\documentclass[11pt]{article}
\usepackage{jheppub}
\usepackage{xparse, physics, graphicx, dcolumn, bm, amsmath, amssymb, mathtools, cancel, dsfont, enumitem, braket, tikz, color}
\usepackage[normalem]{ulem}
\usepackage[linktocpage=true]{hyperref}
\usepackage[all]{hypcap}
\usepackage{cleveref}
\usepackage{siunitx}

\newcommand{\MPl}{M_\text{Pl}}
\newcommand{\msoft}{m_{\text{soft}}}
\NewDocumentCommand{\so}{m!O{}}{\mathfrak{so}(#1)_{\text{#2}}}
\NewDocumentCommand{\su}{m!O{}}{\mathfrak{su}(#1)_{\text{#2}}}
\NewDocumentCommand{\SO}{m!O{}}{\textup{SO}(#1)_{\text{#2}}}
\NewDocumentCommand{\SU}{m!O{}}{\textup{SU}(#1)_{\text{#2}}}
\NewDocumentCommand{\U}{m!O{}}{\textup{U}(#1)_{\text{#2}}}
\NewDocumentCommand{\Z}{O{}}{\mathbb{Z}_{#1}}
\newcommand{\Lag}[1][]{\mathcal{L}_\text{#1}}
\newcommand{\defeq}{\equiv}

\newcommand{\cmt}[1]{}

\newcommand{\GammaNGBGen}{\Gamma^\text{NGB}_\text{wall}(w)}
\newcommand{\GammaNGBWall}{\Gamma^\text{NGB}_\text{wall}(w_i)}
\newcommand{\GammaWall}{\Gamma_\text{wall}}
\newcommand{\GammaFatStr}{\Gamma_\text{fat str}}
\newcommand{\GammaStr}{\Gamma_\text{str}}
\newcommand{\GammaNGBStr}{\Gamma^{\text{NGB}}_\text{str}}

\newcommand{\GammaTot}{\Gamma_\text{tot}}
\newcommand{\GammaGen}{\Gamma_\text{tot}}
\newcommand{\vwall}{{\sigma^{1/3}}}
\newcommand{\vstr}{{\mu^{1/2}}}

\newmuskip\pFqmuskip

\newcommand*\pFq[6][8]{%
	\begingroup % only local assignments
	\pFqmuskip=#1mu\relax
	\mathcode`\,=\string"8000
	\begingroup\lccode`\~=`\,
	\lowercase{\endgroup\let~}\pFqcomma
	{}_{#2}F_{#3}{\left[\genfrac..{0pt}{}{#4}{#5};#6\right]}%
	\endgroup
}
\newcommand{\pFqcomma}{\mskip\pFqmuskip}

\let\deltafunc\delta
\DeclareDocumentCommand\delta{}{\trigbraces{\deltafunc}}
\let\thetafunc\Theta
\DeclareDocumentCommand\Theta{}{\trigbraces{\thetafunc}}

\title{Crescendo Beyond the Horizon: \\ More Gravitational Waves from Domain Walls Bounded by Inflated Cosmic Strings}

\author[a]{Yunjia Bao}
\emailAdd{yunjia.bao@uchicago.edu}
\author[a,b]{Keisuke Harigaya}
\emailAdd{kharigaya@uchicago.edu}
\author[a]{Lian-Tao Wang}
\emailAdd{liantaow@uchicago.edu}

\affiliation[a]{Department of Physics, Enrico Fermi Institute, and Kavli Institute for Cosmological Physics, University of Chicago, Chicago, IL 60637, USA}
\affiliation[b]{Kavli Institute for the Physics and Mathematics of the Universe (WPI), The University of Tokyo Institutes for Advanced Study, The University of Tokyo, Kashiwa, Chiba 277-8583, Japan}

\abstract{
Gravitational-wave (GW) signals offer a unique window into the dynamics of the early universe.
GWs may be generated by the topological defects produced in the early universe, which contain information on the symmetry of UV physics.
We consider the case in which a two-step phase transition produces a network of domain walls bounded by cosmic strings. Specifically, we focus on the case in which there is a hierarchy in the symmetry-breaking scales, and a period of inflation pushes the cosmic string generated in the first phase transition outside the horizon before the second phase transition. 
We show that the GW signal from the evolution and collapse of this string-wall network has a unique spectrum, and the resulting signal strength can be sizeable. In particular, depending on the model parameters, the resulting signal can show up in a broad range of frequencies and can be discovered by a multitude of future probes, including the pulsar timing arrays and space- and ground-based GW observatories. As an example that naturally gives rise to this scenario, we present a model with the first phase transition followed by a brief period of thermal inflation driven by the field responsible for the second stage of symmetry breaking. The model can be embedded into a supersymmetric setup, which provides a natural realization of this scenario. 
In this case, the successful detection of the peak of the GW spectrum probes the soft supersymmetry breaking scale and the wall tension.
}

\begin{document}
\maketitle

%%%%%%%%%%     Sec 1     %%%%%%%%%%
\section{Introduction and Main Result}
Gravitational-wave (GW) signals offer a unique probe into the dynamics of the early universe. In particular, they can carry information about the period during inflation after the large-scale structure modes exit the horizon and the period after the end of inflation and before the Big Bang nucleosynthesis (BBN), which is difficult to probe by other means. Many GW observations are planned, such as pulsar timing arrays (PTAs)~\cite{Janssen:2014dka, NANOGrav:2023gor, EPTA:2023fyk, Antoniadis:2022pcn, Zic:2023gta, Weltman:2018zrl}, Laser Interferometer Space Antenna (LISA)~\cite{Baker:2019nia, Caldwell:2019vru}, Deci-hertz Interferometer Gravitational Wave Observatory (DECIGO)~\cite{Kawamura:2020pcg, Isoyama:2018rjb}, Big Bang Observer (BBO)~\cite{Corbin:2005ny, Harry:2006fi}, TianQin~\cite{TianQin:2015yph, TianQin:2020hid}, Taiji~\cite{Hu:2017mde, Luo:2021qji}, Advanced LIGO-Virgo-KAGRA network~\cite{LIGOScientific:2014pky, LIGOScientific:2016wof}, Einstein Telescope~\cite{Punturo:2010zz, Maggiore:2019uih}, Cosmic Explorer~\cite{LIGOScientific:2016wof, Reitze:2019iox}, galaxy survey data~\cite{Moore:2017ity, Garcia-Bellido:2021zgu}, as well as smaller-scale experiments probing higher-frequency GW signals (see, \textit{e.g.}, Ref.~\cite{Aggarwal:2020olq} for a review on high-frequency GW detection). 

One of the most promising sources of the GW signal is the topological defects in the early universe~\cite{Vilenkin:2000jqa}. These signals may be generated by cosmic strings~\cite{Vilenkin:1981iu, Vilenkin:1981zs, Hogan:1984is, Damour:2000wa, Sousa:2013aaa, Blanco-Pillado:2017oxo, Ringeval:2017eww, Cui:2017ufi, Cui:2018rwi, Cui:2019kkd, Gouttenoire:2019kij, Auclair:2019wcv, Blasi:2020mfx, Ellis:2020ena, Sousa:2020sxs, Co:2021lkc, Gorghetto:2021fsn, Buchmuller:2021mbb, Chang:2021afa, Gouttenoire:2021wzu, Gouttenoire:2021jhk}, metastable domain walls~\cite{Vachaspati:1984gt, Gleiser:1998na, Hiramatsu:2010yz, Hiramatsu:2013qaa, Kawasaki:2011vv, Kamada:2015iga, Nakayama:2016gxi, Ferreira:2022zzo, Bai:2023cqj, Ge:2023rce, Kitajima:2023cek}, and string-bound monopoles~\cite{Martin:1996cp, Babichev:2004gy, Dunsky:2021tih, Lazarides:2022jgr, Roshan:2024qnv}. However, the richness of this class of signals is far from being fully explored.

\begin{figure}[t]
    \centering
    \includegraphics[width=0.9\textwidth]{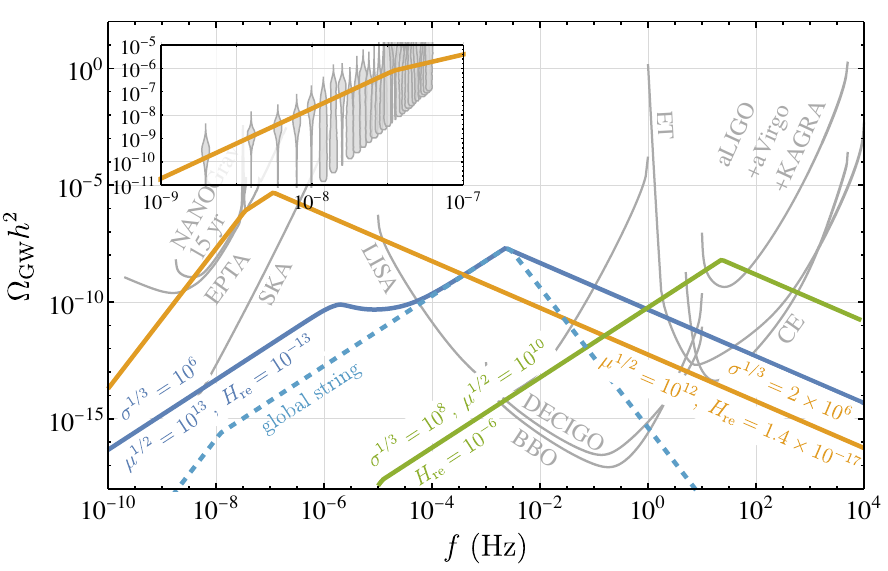}
    \caption{Various benchmarks to show that inflated string-bounded wall networks can potentially produce GW signals across a wide frequency band as observed today, from $10^{-10}$ Hz to $10^{4}$ Hz. Three GW benchmark spectra from walls bounded by inflated gauge strings are shown (solid lines), and one benchmark spectrum from walls bounded by inflated global strings is provided (dashed blue line) to be compared with its gauge-string counterpart (solid blue line). The string tension scale $\vstr$, the wall tension scale $\vwall$, and the Hubble size at string re-entry $H_\text{re}$ of these benchmarks are shown next to the curves in units of GeV. A zoomed-in panel is shown to compare the signal of the low-frequency benchmark to the stochastic background observed by NANOGrav 15-year data release \cite{NANOGrav:2023hvm, NANOGrav:2023gor}. }
    \label{fig:multibenchmarks}
\end{figure}

\begin{figure}[t]
	\centering
	\includegraphics[width=0.9\textwidth]{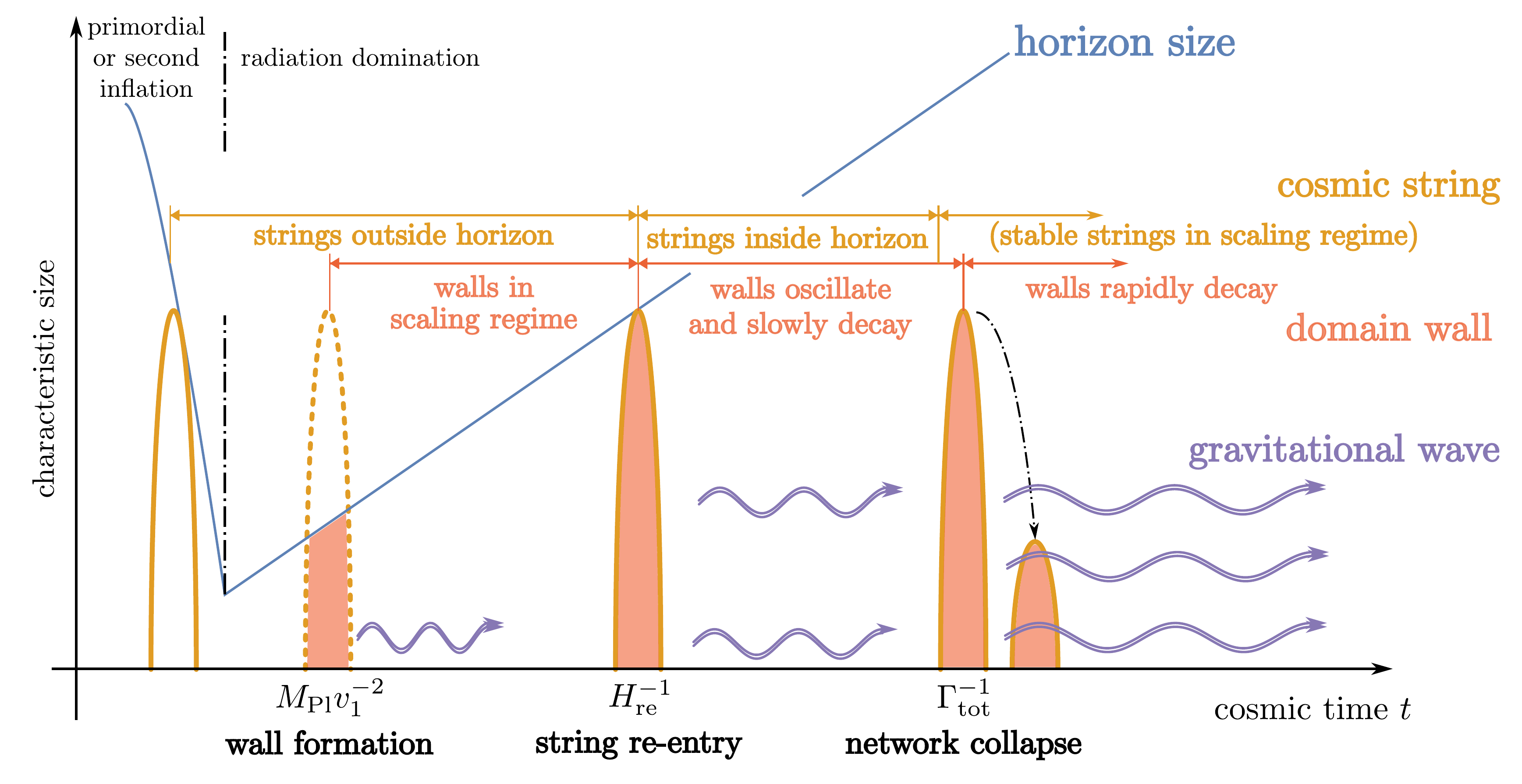}
	\caption{
        Sketch for the evolution of the string-wall network against the horizon size: The blue line denotes the horizon scale, while the orange ellipses show the size of cosmic strings that bound the domain walls. These strings are produced by a phase transition before or during inflation and are frozen outside the horizon. Once inflation ends, the universe enters a radiation-domination epoch. When the radiation bath cools below the energy scale $v_1$ for another phase transition, walls (pink-shaded regions) are produced and enter the scaling regime. The network decouples from the Hubble flow when the strings re-enter the horizon at $t \approx H_\text{re}^{-1}$, and the walls oscillate until their size exponentially decays around $t \approx \GammaGen^{-1}$, where $\GammaGen$ parameterizes the total decay rate of the domain wall. The dynamics of domain walls and the re-entry of cosmic strings are crucial for our mechanism (pink and orange words), while stable strings in the scaling regime (orange words in parenthesis) are a particular feature of our benchmark model and are not required for inflated string-bounded walls. We considered gravitational waves (violet curves) produced at three stages of the inflated string-bounded wall network.
    }
	\label{fig:scalingSketch}
\end{figure}

In this paper, we consider the possibility that there were two phase transitions in the early universe. In the first phase transition, cosmic strings are formed, and subsequently, a brief period of inflation takes place to inflate the strings outside the horizon. Then, a second phase transition occurs after inflation ends to form domain walls. Those domain walls collide with each other and reduce their number to maintain typically one domain wall per horizon volume, and the domain wall network appears to evolve in a scaling solution~\cite{Leite:2011sc, Leite:2012vn, Martins:2016ois}. These collisions produce gravitational waves~\cite{Gleiser:1998na, Hiramatsu:2010yz, Kawasaki:2011vv, Hiramatsu:2013qaa}. Without a difference in the energy among different vacua, the wall network is expected not to collapse and lead to a wall-dominated universe, which is inconsistent with the standard cosmology.%
\footnote{However, see, for example, ref.~\cite{Bai:2023cqj} for how a brief period of the wall-domination epoch can be incorporated in a model.}
However, in our case, the wall can be unstable; once the strings re-enter the horizon, the string-wall network can annihilate. It was considered in the previous literature how boundary defects (here, cosmic strings) and bulk defects (here, domain walls) can interact and produce novel gravitational-wave signatures~\cite{Dunsky:2021tih}. However, the signal is typically less pronounced for general parameter spaces. In our case, the brief period of inflation makes these signals much more conspicuous, as illustrated in \cref{fig:multibenchmarks}. Roughly speaking, the time at which the network starts to annihilate is controlled by the Hubble scale when the inflated strings re-enter the horizon, denoted as $H_\text{re}$, and the final collapse of the system is controlled by a decay rate which we will denote as $\GammaGen$. A sketch of the evolution of the size of the string-wall network is shown in \cref{fig:scalingSketch}. More comparison with Ref.~\cite{Dunsky:2021tih} and explanations for why inflation is advantageous, if not unavoidable, for this scenario is provided in \cref{sec:CompareWithPrev,sec:InevitableInflation}.

Before we investigate the details of the evolution of the defect network, we would like to remark on the generality of the model. First, a cascade of phase transitions with the production of topological defects is quite common in UV models.\footnote{See, for example, fig.~1 of ref.~\cite{Dunsky:2021tih}.} It is natural to expect that phase transitions can happen in hierarchically different scales. If so, there would be enough room for additional dynamics, such as a period of inflation, to happen in between the two phase transitions. An epoch of vacuum domination, so long as it ends before BBN, does not necessarily contradict current cosmological observations and can be consistently incorporated into the cosmological timeline. In our case, the period of inflation after the formation of cosmic strings could be either within the primordial cosmic inflation that seeds the large-scale structure fluctuations or due to a second inflation after the primordial one. If the primordial cosmic inflation is responsible for inflating the strings, this would require a phase transition during inflation, which can be achieved by, for example, inflaton-dependent mass terms for the string-producing field. As the inflaton may traverse a distance $\sim \order{1} \MPl$, it is possible that a phase transition for some field during inflation can be triggered \cite{Sugimura:2011tk, Jiang:2015qor, Ashoorioon:2015hya, Wang:2018caj, Ashoorioon:2020hln, An:2022cce, An:2023jxf}. On the other hand, if a second inflation is responsible for inflating the strings, the phase transitions can be triggered by the decreasing temperature of the thermal bath from the Hubble expansion after the primordial inflation. The second inflation needs not to be a slow-roll inflation; a thermal inflation can also realize a brief period of inflation~\cite{Yamamoto:1985rd, Lazarides:1985ja, Lyth:1995hj, Lyth:1995ka}, and we will use this mechanism to build a model that offers more stringent parameter constraints. This scenario can be motivated from a different angle. As emphasized in \cref{sec:InevitableInflation}, a stage of inflation before the second symmetry-breaking phase transition is expected if we focus on the models with a sizable gravitational wave signal.

Although we will use a particular benchmark model to make our discussion more concrete, we believe that similar discussion for the evolution of the string-wall network and its gravitational signature is applicable to more generic models, 
and most of our estimation will be presented in a less model-dependent way to reflect this generality. The model-specific features of this general mechanism via more thorough analytical and numerical methods are also worth further investigation. 

The crucial ingredients of our scenario are cosmic strings whose typical size is much larger than the horizon size when domain walls are produced. Such cosmic strings can also be produced even if the first symmetry breaking occurs before the observable cosmic inflation, through the quantum nucleation of cosmic strings during inflation~\cite{Basu:1991ig}, or through the accumulation of the fluctuations of the symmetry breaking field outside the horizon~\cite{Gorghetto:2023vqu}. Our analysis is also applicable to those cases.

In this work, we will derive both the size and the spectrum of gravitational wave signals from the defect network. Our emphasis is on the analytical understanding of general features of the spectral shape. Precise calculation of this requires detailed numerical simulation. 

The paper is organized as follows. \Cref{sec:genPic} discusses how the inflated string-bounded wall network can be produced and evolve. Whether the boundary string is a gauge string or a global string slightly alters the physics. For walls bounded by inflated gauge strings, their gravitational-wave signals are computed in \cref{sec:GWSignature}. In the parameter region of interest discussed in \cref{sec:ParamRegionOfInterest}, the network undergoes three stages of evolution, and the spectral shape of these contributions is evaluated in \cref{sec:GWSpectrum,sec:GWReconnection} and summarized in \cref{sec:GWSummary}. Following a similar method as discussed in \cref{sec:GWSignature}, the gravitational-wave signal from walls bounded by global strings is discussed in \cref{sec:GWGlobalSignature} and summarized in \cref{sec:GWGlobalSummary}. A few benchmarks are provided in \cref{sec:GWBenchmark} to show how this signal can cover a wide range of frequencies (\textit{cf.}~\cref{fig:multibenchmarks}). We also show how inflation is generally preferred if one would like domain walls to produce large gravitational-wave signals in \cref{sec:InevitableInflation}. To further restrict the parameter space, we provide in \cref{sec:ThermalInflation} a concrete model that uses the field producing domain walls as the inflaton of a second inflation. In this model, probing the GW spectrum of inflated string-bounded walls provides a probe to the soft supersymmetry breaking scale and the wall tension. We conclude in \cref{sec:conclusion}.

%%%%%%%%%%     Sec 2     %%%%%%%%%%
\section{Productions and Evolution of the Defect Network: General Picture \label{sec:genPic}}

In this section, we discuss how inflated string-bounded domain walls can be produced, evolve, and eventually collapse. 

Topological defects, such as cosmic strings and domain walls, can be produced during phase transitions, and these defects can be classified by the 0\textsuperscript{th} and 1\textsuperscript{st} homotopy group of the vacuum space. In particular, given a symmetry breaking $G \to H$ in which $G$ is the symmetry group of the full UV theory and $H$ is that of the vacuum, the resulting defects are classified by $\pi_0(G/H)$ for domain walls and $\pi_1(G/H)$ for cosmic strings. During cosmic evolution, such defects will be produced through the Kibble-Zurek mechanism~\cite{Kibble:1976sj, Zurek:1985qw}. 

To anchor our discussion, we consider the following sequence of symmetry breaking $\U{1} \to \Z[2] \to \emptyset$ in a model with two complex scalar fields $\phi_1$ and $\phi_{2}$ that have $\U{1}$ charges of $1$ and $2$, respectively. Then, one may consider the following Lagrangian%
\footnote{We ignore the coupling of the form $|\phi_1|^2 |\phi_2|^2$. This coupling needs to be small to preserve the hierarchy $v_2 \gg v_1$. In \cref{sec:ThermalInflation}, we present a SUSY model in which such a small coupling is technically natural. 
}
\begin{equation}
	\Lag = \abs{D_\mu \phi_1}^2 + \abs{D_\mu \phi_{2}}^2 + \lambda_1 \qty(\abs{\phi_1}^2 - \frac{v_1^2}{2})^2 + \lambda_2 \qty(\abs{\phi_{2}}^2 - \frac{v_2^2}{2})^2 + \mu_m \qty(\phi_{2}^* {\phi_1}^2 + \text{h.c.}),
	\label{eqn:toyModel}
\end{equation}
in which $\lambda_{1,2}$ are dimensionless couplings, $v_{1,2}$ are the VEV of $\phi_{1,2}$ respectively, and $\mu_m$ denotes the mixing parameter. If $v_2 \gg v_1$, phase transition of $\phi_{2}$ field leads to $\U{1} \to \Z[2]$ breaking. The corresponding $\phi_2$ cosmic string will form according to the Kibble-Zurek mechanism. 

The $U(1)$ symmetry may be a gauge symmetry, such as the $B-L$ symmetry, or a global symmetry, such as the Peccei-Quinn symmetry. In \cref{sec:GWSignature,sec:GWGlobalSignature}, we consider a gauged and global $U(1)$ symmetry, respectively. This symmetry breaking leads to $\Z$ cosmic strings. Yet, our proposal can be applicable to other types of cosmic strings, such as unstable $\Z[2]$ strings from breaking of $\SO{10}$ to Standard Model (SM) gauge group. More discussions about unstable cosmic strings will be presented in \cref{sec:GWSummary}.

\begin{figure}[t]
	\includegraphics[width=0.9\textwidth]{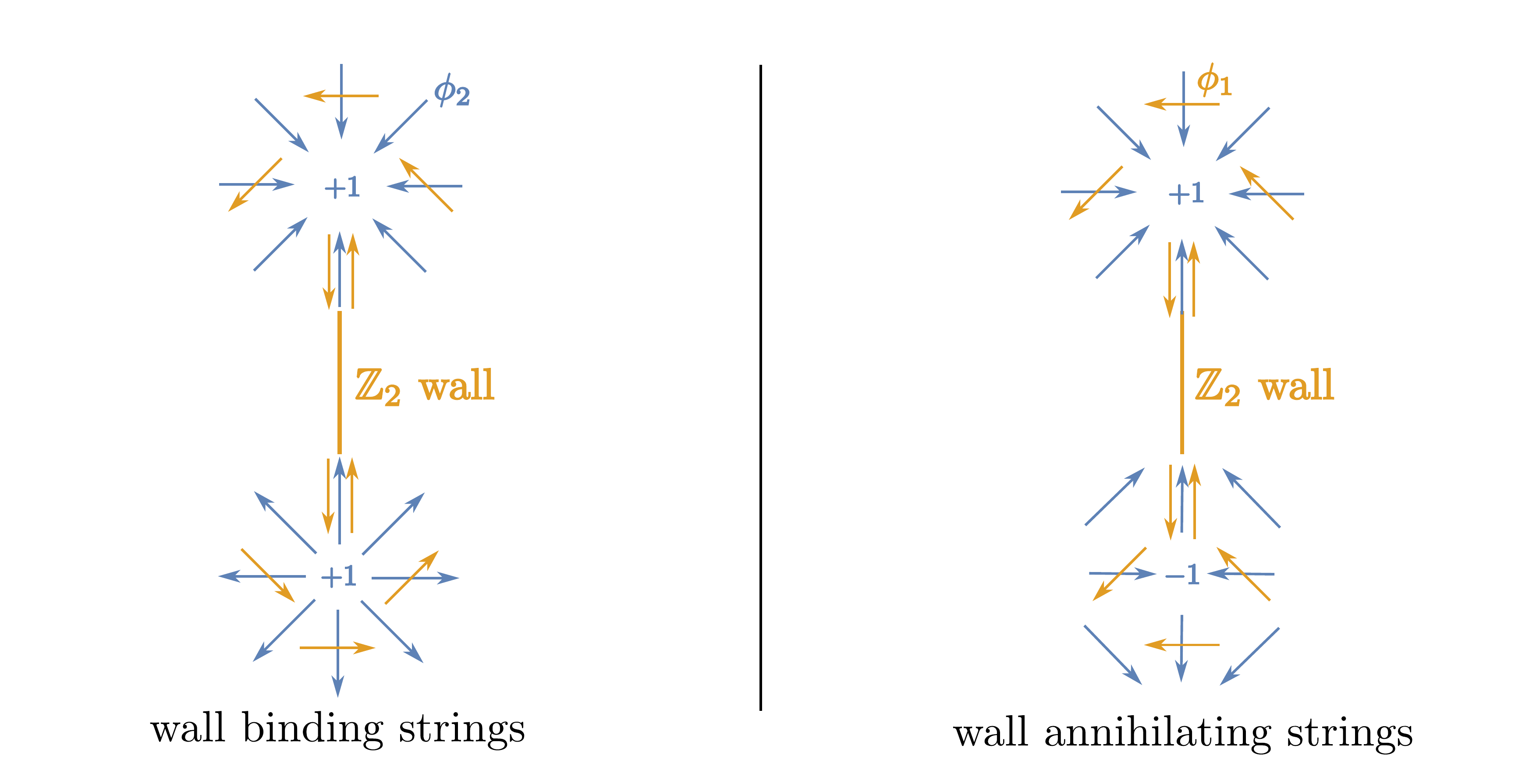}
	\caption{Schematic sketch of the two configurations of the string-bounded walls: For the particular model we consider, there is a stable configuration (left panel) and an unstable configuration (right panel). The phases of two complex scalar fields are indicated by arrows. The first phase transition produces topological defects (strings) of $\phi_2$ fields (blue arrows), and, later, a second phase transition settles $\phi_1$ (orange arrows) to its true vacuum. Due to the trilinear coupling, $\phi_1$ will rotate along the phase set by the phase of $\phi_2$ near the $\phi_2$ strings. Far away from $\phi_2$ strings, $\phi_2$ takes a uniform value so that the phase of $\phi_1$ falls into either vacuum, leading to the formation of $\Z[2]$ walls (orange lines). The left panel shows two $\phi_2$ strings with the same winding number pulled by a wall. As we will show later, wall tension can sufficiently bring these strings together, resulting in a composite string bundle that has the same winding number as the stable gauge string in this particular model. The right panel shows two $\phi_2$ strings with opposite winding numbers. When heavy walls bring two strings together, the two $\phi_2$ strings annihilate.}
	\label{fig:string_wall_winding}
\end{figure}

For the remnant $\Z[2]$ symmetry after the first phase transition, a second phase transition happens to settle $\phi_1$ to its true vacuum, breaking the $\Z[2]$ and producing domain walls. The resulting string-wall bound states are shown in \cref{fig:string_wall_winding}. 
Due to the trilinear interaction 
\begin{equation}
	\Lag \supset \mu_m (\phi_1^2 \phi_{2}^* + \text{h.c.}) \to \frac{\mu_m}{\sqrt{2}} v_1^2 v_2 \cos(2\theta_1 - \theta_2), 
\end{equation}
where we have parameterized $\phi_i = v_i e^{i\theta_i} / \sqrt{2}$, there is a correlation between the winding of $\theta_1$ and that of $\theta_2$

Now, we introduce a stage of inflation before the second phase transition and after strings are formed. The production of the strings can be either during the primordial inflation or followed by a second period of inflation. After its production, the string network may evolve into a scaling solution so that each Hubble patch has $\sim \order{1}$ long cosmic strings.%
\footnote{
It is also possible that the network does not have enough time to evolve into the scaling regime, but this will not be essential to our discussion. If the scaling regime is not reached, the typical string size may no longer be of size $H_i^{-1}$, and the correspondence between $N_{\rm inf}$ and $H_{\rm re}$ should be altered. Nonetheless, inflation still allows us to treat $H_\text{re}$ as a free parameter, which is the only condition we assumed in the remaining text.
}
Hence, the typical distance between strings soon after its production is $\xi \approx H_i^{-1}$ in which $H_i$ denotes the inflationary Hubble size. Due to the subsequent inflation, they will be quickly inflated to super-horizon separations. Then, causality dictates that the co-moving separation between the strings is almost frozen as they exit the horizon. This allows us to estimate the Hubble size when they re-enter the horizon $H_\text{re}$ as
\begin{equation}
	1 \approx \frac{\eval{a H}_\text{exit}}{\eval{a H}_{\text{re-entry}}} \implies H_\text{re} \approx e^{-2N_\text{inf}} H_i,
    \label{eqn:HreEfficientReheat}
\end{equation}
in which $a(t)$ denotes the scale factor. 
Here, we have implicitly assumed that the reheating after this inflation is efficient. When reheating is less efficient, this estimate changes to 
\begin{equation}
    H_\text{re} \approx e^{-2N_\text{inf}} H_i \qty(\frac{T_\text{R}^4}{\rho_\text{inf}})^{1/6}.
\end{equation}
However, the main discussion in \cref{sec:GWSignature} is mostly independent of this assumption. We also provide more discussions on how relaxing this assumption can impact the GW signal in \cref{app:inefficientReheat}.

After the second phase transition, assumed to be after the inflation, a network of stable domain walls is produced following the scaling solution. One might worry that the wall network will dominate the universe. Fortunately, the dynamics change once the inflated strings re-enter the horizon. As the temperature drops below $T \lesssim \sqrt{\MPl H_\text{re}}$, the string-wall network observes the re-entry of boundary strings and starts to collapse. This is to be contrasted with the familiar bias-induced collapse of domain walls~\cite{Vilenkin:1981zs, Gelmini:1988sf, Larsson:1996sp}. In that case, the wall collapses due to the presence of $\Delta V$, a small difference in the energy of the two vacua across the wall, and the annihilation happens around $H \approx \Delta V / \sigma$ assuming that the wall enters the scaling regime~\cite{Hiramatsu:2013qaa, Saikawa:2017hiv, Kitajima:2023cek}. Both $\sigma$ and $\Delta V$ are fixed by the parameters on the wall-producing field $\phi_1$. This usually relates the wall tension with the Hubble scale at wall annihilation and limits the strength of the gravitational-wave signal if one does not carefully tune $\Delta V$. In our case, the network starts to slowly collapse at a scale $H_\text{re}$ controlled by the first phase transition and the inflationary dynamics, both of which are not specific to the dynamics of $\phi_1$. As we will demonstrate in \cref{sec:GWSignature}, this generality also admits sizable gravitational-wave signals. For the scenario that we will consider, although the network decouples from the Hubble flow around $H \approx H_\text{re}$, the network does not necessarily immediately collapse at $H_\text{re}$. We will consider its final collapse due to some decay process controlled by a decay rate $\GammaGen$. Also, the particular model shown in \cref{eqn:toyModel} admits both a stable and an unstable configuration as shown in \cref{fig:string_wall_winding}. Both configurations will lead to the collapse of the wall network, but one of them leaves a stable string defect after the wall collapses. The stable configuration exists since the wall binding two strings with the same $\phi_2$ winding number collapses into a composite string bundle~\cite{Higaki:2016jjh, Long:2018nsl}. These string bundles have the same winding number as the gauged $\U{1}$ string, so we will assume that their evolution will be similar to those gauge strings produced before the second phase transition. On the other hand, when the wall binds two strings with opposite $\phi_2$ winding numbers, the strings annihilate once the wall tugs the boundary defects together. Our analysis of the production of the gravitational wave should be applicable to both cases because their dynamics are similar.

%%%%%%%%%%     Sec 3     %%%%%%%%%%
\section{GW Signal from Network Bounded by Inflated Gauge String \label{sec:GWSignature}}

%%%%%%%%%%    Sec 3.1    %%%%%%%%%%
\subsection{Parameter Region of Interest \label{sec:ParamRegionOfInterest}}

First, we would like to determine the specific parameter region of interest. There are generally two possible hierarchies: (1) $E_\text{str} \gtrsim E_\text{wall}$ or (2) $E_\text{wall} \gtrsim E_\text{str}$, in which $E_\text{str}$ and $E_\text{wall}$ denote the total energy of the string and the wall within the horizon, respectively. This determines which component of the network dominates the dynamics as well as what sources gravitational waves predominantly. The hierarchy $E_\text{str} > E_\text{wall}$ is partially covered in a previous study without assuming inflation between two phase transitions~\cite{Dunsky:2021tih}. Comparisons between this study and the previous one are provided in \cref{sec:CompareWithPrev}, and we will briefly comment on how inflation can modify the GW spectrum from strings in \cref{app:GWString}. Here, we will focus on the hierarchy $E_\text{wall} > E_\text{str}$. As the walls follow the scaling regime, their energy is roughly $E_\text{wall} \approx \pi \sigma R^2 \approx \pi \sigma / H^2$, where we assumed that the walls have characteristic radius $R \approx H^{-1}$ of the horizon size. In contrast, the string on its boundary will have an energy of $E_\text{str} \approx 2\pi \mu R \approx 2\pi \mu / H$. This hierarchy provides a bound on the string re-entry Hubble scale 
\begin{equation}
	E_\text{wall} \gtrsim E_\text{str} \implies H_\text{re} \lesssim \frac{\sigma}{2\mu}.
\end{equation} 
On the other hand, we should avoid wall domination as it will decrease the comoving horizon~\cite{Ipser:1983db} and inflate away the strings bounding the wall network. Domain walls in wall domination remain dynamically stable, causing a domain wall problem in the model. Hence, 
there is also a lower bound on the re-entry Hubble
\begin{equation}
	H_\text{re} \geq H_\text{wd} = \frac{\sigma}{3\MPl^2},
\end{equation}
where $H_\text{wd}$ is the would-be wall-domination Hubble scale \cite{Kibble:1976sj}. 

String loops can also be nucleated on the wall~\cite{Kibble:1982dd, Preskill:1992ck}. The Euclidean bounce action of the string wall system,
\begin{equation}
	S_E = 4\pi \mu R^2 - \frac{4\pi}{3} \sigma R^3,
\end{equation}
has a critical bounce radius of $R_* = 2\mu/\sigma$, and the corresponding probability of nucleating strings per area per Hubble time is 
\begin{equation}
	\frac{P}{L^2 T} \approx \sigma e^{-S_E(R_*)} = \sigma \exp(-\frac{16\pi\mu^3}{3\sigma^2}). 
\end{equation}
This is negligible in our case since even one order-of-magnitude difference between the string tension scale $\mu^{1/2}$ and that of the wall tension $\sigma^{1/3}$ leads to roughly a six order-of-magnitude difference between $\mu^3$ and $\sigma^2$. Such a large exponent suppresses the nucleation rate to be utterly negligible.

In the course of the evolution of the network, there may be other observational bounds on $H_\text{re}$ due to BBN or $\Delta N_\text{eff}$ from cosmic microwave background (CMB), which we will briefly discuss when a set of concrete benchmark parameters are presented in \cref{sec:GWBenchmark}. To recapitulate, the consistent choice of the string re-entry Hubble $H_\text{re}$ must satisfy
\begin{equation}
	\frac{\sigma}{3\MPl^2} \lesssim H_\text{re} \lesssim \frac{2\sigma}{\mu}. 
    \label{eqn:paramHierarchy}
\end{equation}

%%%%%%%%%%    Sec 3.2    %%%%%%%%%%
\subsection{Catalog of Defects in the Network}
\begin{figure}[t]
    \includegraphics[width=0.9\textwidth]{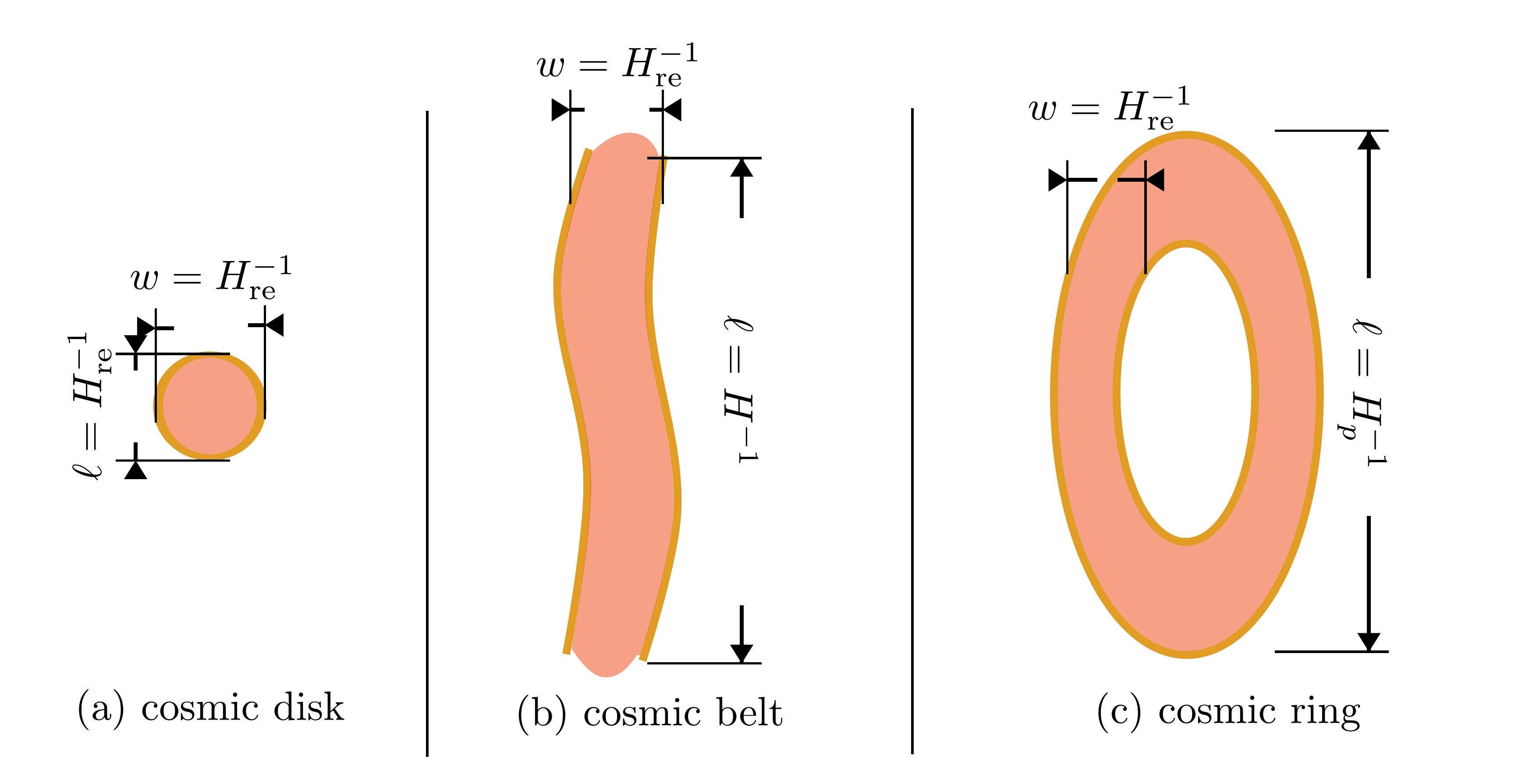}
    \caption{Three types of topological defects in the early universe considered in this paper: Cosmic disks (left) and cosmic belts (middle) are formed mainly due to boundary conditions from inflated cosmic strings. Inflated string loops tend to bound walls to form disks, whereas inflated long strings tend to form belts. When $H \ll H_\text{re}$, the boundary cosmic strings enter the scaling regime by their efficient reconnection. These reconnections may leave behind cosmic rings (right) similar to string loops produced in a pure string network. As long as the walls have not reached their decay time, these belts ``know" both about the string re-entry and the network reconnection, as encoded in their two characteristic sizes. The ring width is typically controlled by the string re-entry Hubble size $w \approx H_\text{re}^{-1}$, but their length (radius) is typically controlled by the Hubble size when they are produced $H_p^{-1}$. }
    \label{fig:defects_catalog}
\end{figure}

In the defect network, there will be three types of defects to consider as illustrated in \cref{fig:defects_catalog}: (a) cosmic disks (walls attached to string loops), (b) cosmic belts (walls attached to long strings), and (c) cosmic rings (annular walls attached to two string loops). Initially, only cosmic strings are present, and they may appear in the form of long strings or string loops. When walls are formed during the second phase transition, some of these walls should eventually terminate on a long string or a string loop. These eventually lead to the production of cosmic disks and belts when the horizon expands sufficiently. Later, as the boundary string of the cosmic belt intercommutes, the length $\ell$ of the cosmic belt remains in the scaling regime $\ell \sim \order{H^{-1}}$ while its width remains $w \sim \order{H_\text{re}^{-1}}$ as walls are still heavy compare to the boundary strings. However, when two long belts reconnect, a residual cosmic ring may be formed. This object will have a typical width of $w \approx H_\text{re}^{-1}$ but a typical radius of $\ell \approx H_p^{-1}$, where $H_p$ denotes the Hubble scale when the two belts intercommute and produce the ring. One may regard the extended cosmic belts as analogous to long strings in a string network, while cosmic rings are more akin to string loops in a string network. 

Here, we distinguish disks and belts from rings. The former two are mainly produced because of the boundary condition set by the inflated cosmic strings, while the latter one is mainly a consequence of the network reconnection. This is reflected by the fact that both disks and belts have characteristic sizes of either $H_\text{re}^{-1}$ or $H^{-1}$ while cosmic rings have widths of $\sim H_\text{re}^{-1}$ but a length of $\sim H_p^{-1}$. Therefore, cosmic disks and belts may oscillate at one characteristic frequency $k \sim H_\text{re}$ while the motion of cosmic rings may have two scales, one set by $H_\text{re}$ and the other set by $H_p$, making modeling GW spectrum from rings more involved.%
\footnote{
Cosmic belts can have motions on the scale of $\sim H$. However, since the typical scale of this motion scales with $H \sim t^{-1}$, it should not be regarded as an oscillation and cannot generate significant GWs, analogous to GW radiation from infinitely long string \cite{Vilenkin:2000jqa}.
}
For cosmic disks and belts, we shall assume that around $t \approx H_\text{re}^{-1}$, almost all Hubble patches are occupied by one such object. This implies that we may estimate their energy density as a function of $H_\text{re}$ and regard these defects as of characteristic size $\sim H_\text{re}^{-1}$. Technically, the configuration of these defects is set by the boundary condition at the horizon exit of boundary strings during inflation. How one precisely evaluates their energy densities around the horizon re-entry may be affected by whether the boundary strings are produced during inflation or have reached scaling before inflation. Additional numerical simulation is required in the future to fully address this. Nonetheless, we expect our parametric estimate to hold on dimensional ground. Detailed discussion will be provided in \cref{sec:GWSpectrum}. On the other hand, when evaluating the gravitational-wave signal from cosmic rings, they can be produced at different $H_p$. Therefore, beyond obtaining their individual gravitational-wave spectrum, we should regard them as following some distribution of sizes due to network reconnection, and the total gravitational-wave signal comes from integrating over this distribution. We will discuss this more in \cref{sec:GWReconnection}.

%%%%%%%%%%    Sec 3.3    %%%%%%%%%%
\subsection{Gravitational Wave from Scaling and Re-entering Defects \label{sec:GWSpectrum}}
Next, we discuss how much this defect network can contribute to stochastic gravitational waves. For the moment, we ignore the network reconnection, in particular, the gravitational-wave signal from cosmic rings. We instead focus on the gravitational-wave spectrum from defects that are already formed before the string re-entry. Cosmic rings are discussed in \cref{sec:GWReconnection}.

As the walls contain most of the energy in the string-wall system, we first focus on the gravitational waves from the dynamics of the walls. It is generally convenient to split the computation into three parts: contribution from (1) walls following the scaling regime before string re-entry, (2) walls oscillating before rapid decay, and (3) the rapidly decaying network. To estimate the gravitational waves emitted from the inflated string-bounded walls, we use the Boltzmann equation for the gravitational-wave energy density
\begin{equation}
	\dot{\rho}_\text{GW} + 4H(t) \rho_\text{GW} = n(t) P(t),
\end{equation}
in which $\rho_\text{GW}$ denotes the energy density of the gravitational waves produced at some cosmic time $t$, $H(t)$ is the Hubble parameter, $P(t)$ denotes the gravitational-waves power emitted by one source, and $n(t)$ denotes the number density of the source. In what follows, we will focus on a qualitative understanding of the signal strength and spectrum. More careful calculations are presented in \cref{app:spectrumComputation}, and the main results are summarized in \cref{sec:GWSummary}.

%%%%%%%%%%   Sec 3.3.1   %%%%%%%%%%
\subsubsection{GW from Scaling Walls ($t \lesssim H_\text{re}^{-1}$) \label{sec:GWScalingWall}}
For walls reaching the scaling regime, previous numerical studies on the case where the walls decay via explicit $\Z[2]$ symmetry breaking~\cite{Kawasaki:2011vv, Hiramatsu:2013qaa} suggest that its gravitational-wave spectrum follows a $k^3$ power law at the IR and falls off like $k^{-1}$ after reaching the peak frequency around the Hubble scale in physical momentum $k \cdot a(t_\text{obs})/a(t_\text{prod})$, in which the ratio of scale factors captures the redshift from the time of gravitational-wave production $t_\text{prod}$ to that of observation $t_\text{obs}$. Here, we use the following parameterization for the power spectrum
\begin{equation}
	\pdv{P}{\ln k} \approx \frac{C \pi \sigma^2}{\MPl^2H^2} \frac{ \qty(k H^{-1} a(t_\text{obs})/a(t_\text{prod}))^3 }{ 1 + \qty(k H^{-1} a(t_\text{obs})/a(t_\text{prod}))^4 },
\end{equation}
in which $C$ denotes some $\sim \order{1}$ to $\order{10}$ dimensionless constant. While the $k^3$ spectrum is a general feature expected from causality~\cite{Caprini:2009fx}, the $k^{-1}$ power law may depend on the specific microscopic physics of wall collisions, which calls for more detailed analytical and numerical studies. As we will see, this UV part of the spectrum is subdominant in comparison with other contributions and will not be observable unless it is shallower than $k^{-1}$. The number density $n(t)$ can be estimated from scaling law, \textit{i.e.}, 
\begin{equation}
	n(t) \sim H^{-3},
\end{equation}
so that each Hubble patch has $\sim\order{1}$ domain walls. 

Now, we may solve the Boltzmann equation by explicit integration. The detailed computation is presented in \cref{app:spectrumComputation}. As it turns out, it is more convenient to consider the fractional energy density at the wall decay time $t = \GammaGen^{-1}$ (to be discussed in more detail in \cref{sec:GWDisks}), and the gravitational-wave spectrum redshifted to $t = \GammaGen^{-1}$ can be approximated as 
\begin{equation}
	\eval{ \pdv{\Omega_\text{GW, scaling}}{\ln k} }_{t = \GammaGen^{-1}} \approx \frac{2\pi \sigma^2 C}{21\MPl^4 H_\text{re}^2} 
	\begin{dcases}
		\qty(\frac{k}{\sqrt{H_\text{re}\GammaGen}})^3, & k \lesssim \sqrt{H_\text{re} \GammaGen}, \\
		\qty(\frac{\sqrt{H_\text{re}\GammaGen}}{k}), & k \gtrsim \sqrt{H_\text{re} \GammaGen}.
	\end{dcases}
\end{equation}
As expected, the GW spectrum follows a power law similar to that of $\pdv*{P}{\ln k}$. Also, the spectrum peaks at $k = \sqrt{H_\text{re}\GammaGen} = H_\text{re} a(H_\text{re}^{-1})/a(\GammaGen^{-1})$ in which $a(H_\text{re}^{-1})/a(\GammaGen^{-1})$ is a redshift factor to evaluate $\Omega_\text{GW}$ at $t = \GammaGen^{-1}$. This is also sensible because the gravitational wave peaks around the horizon scale as dictated by the scaling of domain walls. After the string re-entry, the defect network deviates from scaling solutions, and the rapid wall oscillation produces gravitational waves, as we will discuss next.

%%%%%%%%%%   Sec 3.3.2   %%%%%%%%%%
\subsubsection{GW from Cosmic Disks ($t \gtrsim H_\text{re}^{-1}$) \label{sec:GWDisks}}
%%%%%%%%%%               %%%%%%%%%%
\paragraph{GW from Oscillating Cosmic Disks $(H_\text{re}^{-1} \lesssim t \lesssim \GammaGen^{-1})$ \label{sec:GWOscillatingWall}}
After the strings re-enter the horizon, most of the domain walls reside in the following two configurations: walls attached to long strings (``belts") and walls attached to string loops (``disks"). Domain walls that stretch far outside the horizon are rare, as dictated by causality that forbids the correlation beyond the horizon.
The scaling regime of walls terminates at this point. Therefore, cosmic disks/belts produced from string re-entry are expected to be of radius/width $r \approx H_\text{re}^{-1}$. In the following subsection, we will focus on the disk contribution, and the subdominant contribution from belts is discussed in \cref{sec:GWBelts}.

Because we assume that $E_\text{wall} \gtrsim E_\text{str}$, the dynamics of the network is mainly governed by walls. These walls will oscillate with some characteristic frequency of their size. This characteristic scale is $r_0 \approx H_\text{re}^{-1}$ so that this oscillation of walls with characteristic curvature $r$ can be roughly described by 
\begin{equation}
	r(t) \approx \bar{r}(t) \cos(\frac{t}{\bar{r}(t)}),
\end{equation}
in which $\bar{r}(t)$ is the slowly varying radius of the wall with $\bar{r} \approx r_0$ around $t \approx H_\text{re}^{-1}$. A slightly more sophisticated modeling yields a similar estimate as shown in \cref{app:closedDomainWall}. One may estimate the power radiated into gravitational waves by the quadrupole formula
\begin{equation}
	P= \dv{E}{t} \approx \frac{\ev{\dddot{Q}^2}}{8\pi \MPl^2} \approx \frac{1}{8\pi \MPl^2} \abs{\dv[3]{t}( \sigma \pi \bar{r}^4 \cos[4](\frac{t}{\bar{r}}) )}^2 \approx \frac{\pi \sigma^2 \bar{r}^2}{\MPl^2}. 
	\label{eqn:PowerQuadrupoleGW}
\end{equation}
Therefore, the gravitational-wave radiation damps the domain wall with a characteristic rate of $\GammaWall \approx \pi \sigma / \MPl^2$ so that 
\begin{equation}
	\dv{(\sigma \bar{r}^2)}{t} \approx -\GammaWall \sigma \bar{r}^2. 
\end{equation}
In the discussion to follow, we will use $\GammaGen$ to parameterize the total decay rate of the domain wall. As we average over oscillations with a period $\approx H_\text{re}^{-1}$, we generally expect $\GammaGen \approx \Delta t^{-1} \lesssim H_\text{re}$ to maintain consistency. Although the specific initial condition and geometry of the string dictate the precise spectrum, we will approximate the emission as if all the gravitational-wave power is emitted in the fundamental mode on the wall, \textit{i.e.}, $k \approx H_\text{re}$.%
\footnote{
It is possible that a more complicated mechanism, such as disk self-intersection, can dissipate disks' energy. Unless the self-intersection is very frequent, it should only affect our estimation of the GW spectrum by an $\sim \order{1}$ factor. Whether self-intersection significantly dissipates the disks' energy and alters the GW spectrum calls for further numerical simulations. Also, the presence of higher harmonics may affect the gravitational-wave spectral shape, but the computational technique to obtain the spectrum should be similar. We provide a discussion about higher harmonics in \cref{sec:GWSummary}.
}
This implies that
\begin{equation}
	\pdv{P(t)}{\ln k} \approx \frac{\pi \sigma^2}{\MPl^2H_\text{re}^2} \frac{a(\GammaGen^{-1})k}{a(t)} \delta(\frac{a(\GammaGen^{-1})k}{a(t)} - H_\text{re}). 
	\label{eqn:PowerGWOscillatingWall}
\end{equation}
We expect there to be $\sim \order{1}$ number of disks in a volume of $H_\text{re}^{-3}$ when the strings re-enter the horizon. The gravitational-wave signal can be estimated as follows. The energy fraction of each disk radiated into the gravitational wave by time $t$ is roughly $\approx \GammaWall t$, the energy density of the oscillating wall is approximately $\approx \sigma H_\text{re}$, and the redshift dilution for the number density of walls $\propto a^{-3}$. Denoting the temperature and Hubble scale at the emission of gravitational waves during the oscillating stage of the disks as $T$ and $H \approx T^2 / \MPl$ respectively, the fractional energy density of gravitational waves around $t = \GammaGen^{-1}$ is
\begin{equation}
    \begin{aligned}
	   \eval{\pdv{\Omega_\text{GW, osc.}}{\ln k}}_{t = \GammaGen^{-1}} \approx& \frac{1}{T^4} \qty(\sigma H_\text{re}) \frac{\GammaWall}{H} \qty(\frac{T}{T_\text{re}})^3
	   \approx \frac{\sigma^2 H_\text{re}}{\MPl T_\text{re}^3} \frac{1}{T^3} \\
        \implies & \eval{\pdv{\Omega_\text{GW, osc.}}{\ln k}}_{t = \GammaGen^{-1}} \approx  \frac{\sigma^2}{\MPl^4 H_\text{re}^{1/2} \GammaGen^{3/2}} \qty(\frac{k}{H_\text{re}})^3 ,
    \end{aligned}
    \label{eqn:OmegaGWOscillatingWall}
\end{equation}
where we have used the adiabatic invariant $k(t=\GammaGen^{-1})/T_{\GammaGen} = H_\text{re}/T$, and $T_{\GammaGen} \approx \sqrt{\GammaGen \MPl}$ denotes the temperature of the bath around $t \approx \GammaGen^{-1}$. This agrees with the explicit solution of the Boltzmann equation as shown in \cref{app:spectrumComputation}.

Around $k \approx \sqrt{H_\text{re} \GammaGen}$, $\Omega_\text{GW, osc.}$ has a similar parametric dependence as the peak amplitude of the gravitational waves from walls in the scaling regime, 
\begin{equation}
	\eval{\Omega_\text{GW, osc.}(k=\sqrt{H_\text{re} \GammaGen})}_{t = \GammaGen^{-1}} \approx \frac{2\pi \sigma^2}{3 \MPl^4 H_\text{re}^2} = \frac{7}{C} \eval{\Omega_\text{GW, scaling}(k = \sqrt{H_\text{re}\GammaGen})}_{t = \GammaGen^{-1}}. 
\end{equation}
A more detailed numerical simulation is needed to determine the precise dynamics during the transition from the scaling regime to the oscillating regime and the gravitational-wave spectrum produced by it. Nonetheless, this transition is likely to be smooth enough without producing striking features, such as sharp discontinuous jumps, on the gravitational-wave spectrum; hence, we will match the two contributions with $C = 7$ to obtain a continuous spectrum.

%%%%%%%%%%              %%%%%%%%%%
\paragraph{GW from Collapsing Cosmic Disks $(t \gtrsim \GammaGen^{-1})$ \label{sec:GWCollapsingWall}}

At the last stage of the evolution of the disks, rapid collapse happens, and the network annihilates. Because of this, we may approximate the frequency of the gravitational waves emitted at this stage as if they are all produced around $t = \GammaGen^{-1}$. The average radius $\bar{r}(t)$ starts to decay from $r_0=H_\text{re}^{-1}$ since $t = \GammaGen^{-1}$ and is described by 
\begin{equation}
	\bar{r}(t) \sim H_\text{re}^{-1} \exp(-\frac{\GammaGen t}{2}). 
\end{equation}
Following a strategy similar to that gives rise to \cref{eqn:PowerGWOscillatingWall}, the power spectrum here can be estimated as 
\begin{equation}
	\pdv{P(t)}{\ln k} \approx \frac{\pi \sigma^2 \bar{r}^2(t)}{\MPl^2} k \delta(k - \frac{1}{\bar{r}(t)}),
    \label{eqn:PowerGWCollapsingWall}
\end{equation}
in which we dropped the redshift dependence on the frequency. The estimate for the number density of the network is still $\sim H^3$. The gravitational-wave energy density is
\begin{equation}
    \eval{\pdv{\Omega_\text{GW, col.}}{\ln k}}_{t = \GammaGen^{-1}} 
    \approx \frac{1}{T_{\GammaGen}^4} \sigma H_{\text{re}} \left( \frac{T_{\GammaGen}}{T_{\text{re}}}\right)^3 (\bar{r} H_{\text{re}})^2 
    \approx \frac{\sigma^2}{ \MPl^4 H_\text{re}^{1/2} \GammaGen^{3/2}} \qty( \frac{H_{\text{re}}}{k} )^2,
    \label{eqn:OmegaGWCollapsingWall}
\end{equation}
where we have used $k=\bar{r}^{-1}$, following \cref{eqn:PowerGWCollapsingWall}. 

It is worth remarking that the microscopic physics of the wall collapse could potentially change the power-law dependence of the UV part. For instance, when the disk size is $\lesssim \mu/\sigma$ so that the string energy dominates, one expects that the gravitational-wave spectrum transitions from $\sim k^{-2}$ to $\sim k^{-1}$, which is the typical UV tail of the GW spectrum from a cosmic string loop. This, however, should be in the deep UV as we assumed $H_\text{re} \ll \sigma / \mu$.%
\footnote{
Another example of a change in the GW spectrum due to microscopic physics could be the inter-string interaction, which can potentially compete with the wall tension as the boundary gauge strings are pulled by walls. However, for a string separation larger than that of the string core size, this interaction is exponentially suppressed. When $v_1 \approx \vstr \gg \vwall$, this competition is never important when the wall is large and contributes significantly to the energy of the defect network.
} 
Another potential source of modification to the UV part of the GW spectrum comes from the finiteness of the collapse time. We assumed that the string-wall network collapses sufficiently quickly so that the GW spectrum is produced at $t = \GammaGen^{-1}$ almost instantaneously. However, a finite collapse time for the network gives additional logarithmic dependence on $k$ because of the redshift of $k$ during the collapse process. A dedicated numerical study is required to fully determine the details of the spectrum.

%%%%%%%%%%   Sec 3.3.2   %%%%%%%%%%
\subsubsection{GW from Long Belts \label{sec:GWBelts}}

Now, we direct our attention to the gravitational-wave signal from cosmic belts. These defects may reconnect and enter a scaling regime by breaking off daughter defects. Here, we focus on how long belts (mother defects) evolve and produce additional gravitational-wave signals; we will discuss the production of gravitational waves from the daughter defects in \cref{sec:GWReconnection}.

We assume that the boundary defects enter the scaling regime efficiently. For sufficiently small $H_\text{re}$ considered here, the belts' energy comes from the walls stretching between cosmic strings. These belts have an energy density that scales as 
\begin{equation}
    \rho_\text{belts} \approx \sigma \ell w H^3 \approx \frac{\sigma H^2}{H_\text{re}},
    \label{eqn:rho_belts}
\end{equation}
in which we assumed that the belt has a width $w$ set by $H_\text{re}^{-1}$ and a length $\ell$ set by $H^{-1}$. This energy density redshifts as $\propto a^{-4}$ in radiation domination. When belts collide and reconnect, some of their energy is converted into kinetic energy, which is why belts redshift more than disks. Also, kinks and cusps at the intersection lead to the production of relativistic particles and radiation that may further dissipate the energy stored in long belts. Two analogous mechanisms during reconnection give rise to the scaling regime of cosmic strings, and the scaling regime explains why gravitational-wave radiation from string loops is generally more significant than those emitted by long strings \cite{Vilenkin:2000jqa, Hindmarsh:2008dw}. While we expect this analogy between belts and strings to hold, it is interesting to check whether this expectation is valid in numerical simulations when domain wall energy is larger than string energy.

Let us now estimate the decay rate of cosmic belts into gravitational waves. Here, the gravitational-wave signal mainly comes from the rapid oscillation of domain walls with frequency $k \approx H_\text{re}$. There is no dynamical reason for this motion to be coherent on a scale of the horizon size $\ell \approx H^{-1}$. Hence, when using the quadrupole formula to estimate the power of gravitational radiation, we should use an incoherent sum over patches of size $w \times w$ on an object of size $\ell \times w$, \textit{i.e.},
\begin{equation}
    P_\text{GW} 
    \approx \frac{\ev{\dddot{Q}^2}}{\MPl^2} 
    \approx \frac{1}{\MPl^2} \qty(\sum_{w\times w\text{ patches}} \frac{\sigma w^4}{w^3})^2
    \approx \frac{\sigma^2 w^2}{\MPl^2} \frac{\ell}{w}.
\end{equation}
Cross terms of $\dddot{Q}$ between different patches should vanish on average due to incoherent oscillation. Note that $P_\text{GW} \propto \ell$ instead of $\propto \ell^2$ is a tell-tale feature of this incoherent sum. This leads to $\Gamma_\text{belt} = \GammaWall \approx \sigma/\MPl^2$. Since $\Gamma_\text{belt} = \GammaWall$ and the energy density of the belts redshifts faster than that of the disks does, the gravitational waves from the belts are subdominant in comparison with those from disks, as we confirm explicitly below.

To estimate the gravitational-wave emission before belts collapse, one may consider
\begin{equation}
    \eval{\pdv{\Omega_\text{GW, belt, osc.}}{\ln k}}_{t = \GammaGen^{-1}} 
    \approx \frac{1}{T^4} \qty(\frac{\sigma H^2}{H_\text{re}}) \frac{\GammaWall}{H} 
    = \frac{\sigma^2}{\MPl^4 H_\text{re} \GammaGen} \qty(\frac{k}{H_\text{re}})^2, \quad \sqrt{\GammaGen H_\text{re}} \lesssim k \lesssim H_\text{re}.
    \label{eqn:OmegaGWOscillatingBelts}
\end{equation}
Here, the change in power law $\sim k^2$ in comparison with that of the disks $\sim k^3$ comes from the $\propto a^{-4}$ redshift of belts' energy density in the scaling regime. As anticipated, the belts produce less GW signals than the disks, as can be seen from the maximal abundance $\eval{\Omega_\text{GW, belt}}_{t = \GammaGen^{-1}}^{k = H_\text{re}} \approx (H_\text{re} / \GammaGen)^{1/2} \eval{\Omega_\text{GW, disk}}_{t = \GammaGen^{-1}}^{k = H_\text{re}}$; see \cref{eqn:OmegaGWOscillatingBelts,eqn:OmegaGWOscillatingWall}. As the network rapidly decays at around $t \approx \GammaGen^{-1}$, belts are pulled by domain walls to either annihilate or form stable string bundles as well. Then, the gravitational-wave spectrum can be estimated as 
\begin{equation}
    \eval{\pdv{\Omega_\text{GW, belt, col.}}{\ln k}}_{t = \GammaGen^{-1}} 
    \approx \frac{1}{T_{\GammaGen}^4} \qty(\frac{\sigma \GammaGen^2}{H_\text{re}}) \qty(w H_\text{re})
    = \frac{\sigma^2}{\MPl^4 \GammaGen H_\text{re}} \qty(\frac{H_\text{re}}{k}), \quad k \gtrsim H_\text{re}
    \label{eqn:OmegaGWCollapsingBelts}
\end{equation}
in which we assumed that the gravitational wave is predominantly emitted with a frequency $k \approx w^{-1} \approx H_\text{re}$ controlled by its width while its length $\ell \approx H^{-1}$ remains in the scaling regime. This agrees with the approach by solving the Boltzmann equation as discussed in \cref{app:spectrumComputation}.

%%%%%%%%%%    Sec 3.4    %%%%%%%%%%
\subsection{Gravitational Wave from Network Reconnection \label{sec:GWReconnection}}
In this section, we discuss contributions to the gravitational-wave signal from defects produced during reconnections. These computations generally involve two steps: identifying the GW spectrum from each individual defect and summing over all possible defect sizes. This strategy can reproduce the parametric form of the well-studied gravitational-wave spectrum of gauge cosmic strings as demonstrated in \cref{app:GWString}. Here, we focus on the contribution from cosmic rings that are particular to the inflated string-bounded wall network. 

Each cosmic ring is produced from its mother defects at $t \approx H_p^{-1}$. They should have a typical radius $\ell \approx H_p^{-1}$ and width $w \approx H_\text{re}^{-1}$. The assumption that $\ell \approx H_p^{-1}$ is motivated by similar properties of horizon-sized string loops in the scaling regime \cite{Kibble:1984hp, Vanchurin:2005pa, Olum:2006ix, Blanco-Pillado:2011egf, Blanco-Pillado:2013qja}, but it is currently under debate whether string loops much smaller than horizon size can be amply produced and impact the scaling solution \cite{Ringeval:2005kr, Martins:2005es, Polchinski:2007rg, Auclair:2019zoz}. Different from defects previously discussed, cosmic rings can have two oscillation modes, one from the overall coherent oscillation of the rings with $k \approx H_p$ (hereafter ``string mode") and the other from the rapid incoherent oscillation of the heavy walls with $k \approx H_\text{re}$ (hereafter ``wall mode"). Because the walls are mostly transverse to the string, one may assume that the frequency of the wall mode $k \approx H_\text{re}$ remains mostly unchanged during reconnection. As cosmic belts reconnect later, they break off longer cosmic rings that produce GWs with lower frequencies.

%%%%%%%%%%   Sec 3.4.1   %%%%%%%%%%
\subsubsection{Estimating Spectrum from Cosmic Ring of Fixed Size \label{sec:GWRingInd}}

\begin{figure}[t]
    \centering
    \includegraphics[width=0.9\textwidth]{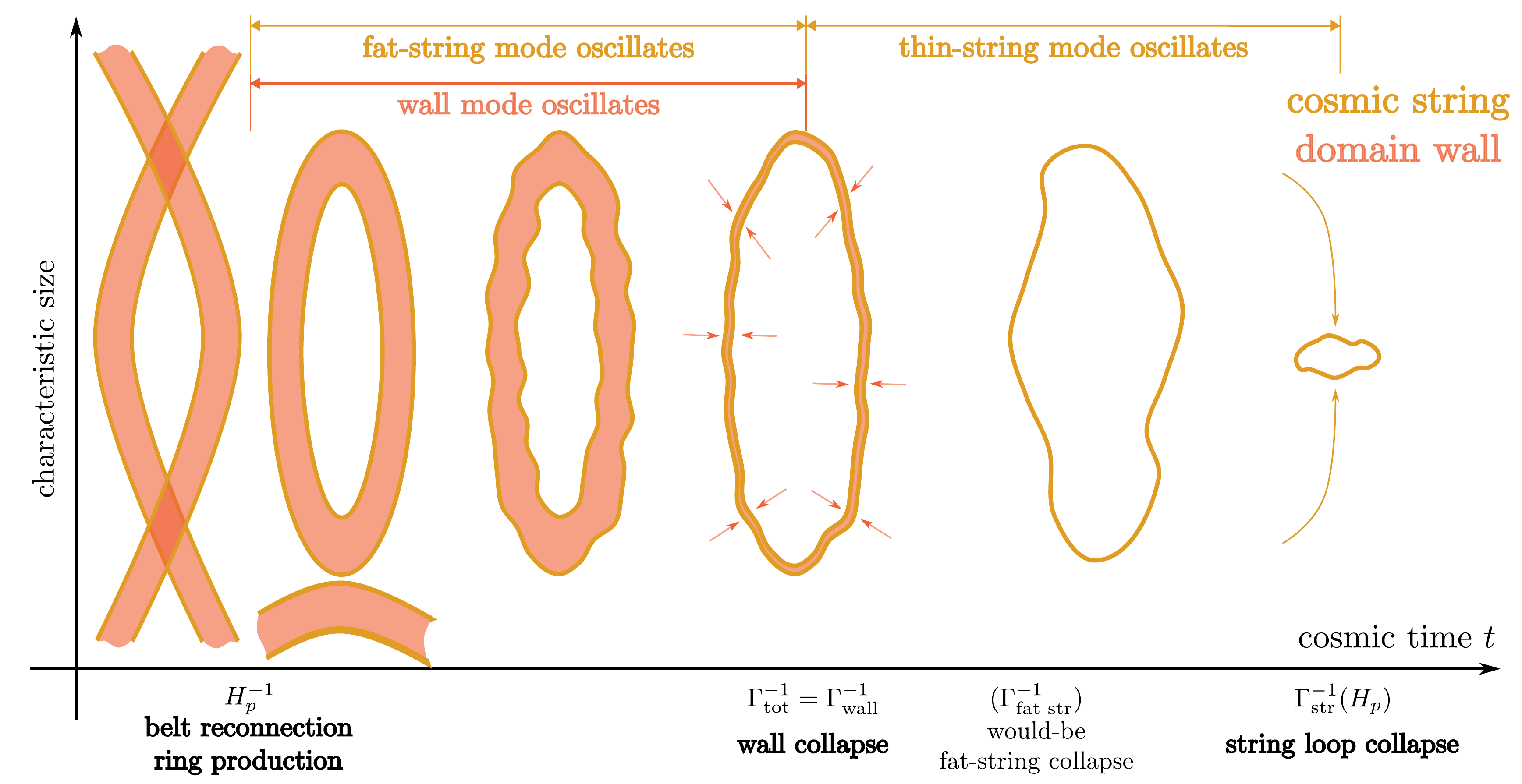}
    \caption{
        Schematic sketch for the evolution of the individual cosmic ring: Cosmic belts meet and reconnect to form cosmic rings around $t \approx H_p^{-1}$. Isolated cosmic rings then redshift like matter and emit gravitational waves in two frequencies: (1) the string mode (yellow) with frequency $k \approx H_p$ and (2) the wall mode (pink) with frequency $k \approx H_\text{re}$. During the wall mode oscillations, the width of cosmic rings remains $w \approx H_\text{re}^{-1}$. This also enhances the string mode signal as the effective tension of the string model $\mu_\text{eff} \defeq \sigma / H_\text{re}$ is much higher than the string tension $\mu$. Until much later around $t \approx \GammaGen^{-1}$, all walls rapidly collapse, and all cosmic rings slim down. Unstable string-wall configuration (\textit{cf.}~\cref{fig:string_wall_winding}) is annihilated at this point while the stable configuration is bound into a loop of stable string bundles. Each thin string loop continues to oscillate and finally collapses much later at their respective $t \approx \GammaStr^{-1}$.
    }
    \label{fig:ringSketch}
\end{figure}

%%%%%%%%%%               %%%%%%%%%%
\paragraph{Wall Mode ($k \approx H_\text{re}$) GW Spectrum} We first estimate the wall mode spectrum for cosmic rings produced around $H \approx H_p$. The ring density around its production is 
\begin{equation}
    \rho_\text{ring} \approx \sigma w \ell H_p^3 \approx \frac{\sigma H_p^2}{H_\text{re}} \approx \rho_\text{belt}(H = H_p),
\end{equation}
comparable to the energy density of the long belt. However, as illustrated in \cref{fig:ringSketch}, after separating from the mother long belt, it becomes an isolated object with $w \approx H_\text{re}^{-1}$ and $\ell \approx H_p^{-1}$, redshifting like matter. During the oscillating stage of the wall mode, the GW abundance may be estimated as 
\begin{equation}
    \begin{multlined}
        \eval{\pdv{\Omega_\text{GW, ring, wall osc.}}{\ln k}{\ln H_p}}_{t = \GammaGen^{-1}} 
        \approx \frac{1}{T^4} \qty(\frac{\sigma}{H_\text{re}} H_p^2) \qty(\frac{T}{T_p})^3 \frac{\GammaWall}{H} \\
        \approx \qty(\frac{H_p}{H_\text{re}})^{1/2} \frac{\sigma^2}{\MPl^4 H_\text{re}^{1/2} \GammaGen^{3/2}} \qty(\frac{k}{H_\text{re}})^3, \quad k \lesssim H_\text{re}. 
    \end{multlined}
    \label{eqn:OmegaGWRingOscillatingWall}
\end{equation} 
The $\sim k^3$ power-law dependence is similar to that of oscillating cosmic disks or string loops. Compared to cosmic disks (\textit{cf.}~\cref{eqn:OmegaGWOscillatingWall}), these cosmic rings are produced later and have ``narrower" walls ($w < \ell$). Because rings first redshift as radiation as part of the long belt until their later breakoff, they necessarily make up a smaller fraction of the energy density than cosmic disks. Consequently, these rings produce gravitational-wave signals that are $\sim (H_p / H_\text{re})^{1/2}$ smaller than those from disks. During the collapsing stage, the width of these rings shrinks, resulting in a decrease in energy $\propto w H_\text{re}$, and produces a $\sim k^{-1}$ spectrum similar to long cosmic belts. Hence, 
\begin{equation}
    \eval{\pdv{\Omega_\text{GW, ring, wall col.}}{\ln k}{\ln H_p}}_{t = \GammaGen^{-1}} 
    \approx \qty(\frac{H_p}{H_\text{re}})^{1/2} \frac{\sigma^2}{\MPl^4 H_\text{re}^{1/2} \GammaGen^{3/2}} \qty(\frac{H_\text{re}}{k}), \quad k \gtrsim H_\text{re}. 
\end{equation}
When $H_p \ll H_\text{re}$, both the oscillating and collapse stage of the wall mode are obscured underneath the disks' GW spectrum because of the $(H_p / H_\text{re})^{1/2}$ suppression. However, when $H_p \approx H_\text{re}$, the collapsing stage of cosmic rings is less rapid than that of disks and produces a shallower UV spectrum $\propto k^{-1}$. This $\sim k^{-1}$ part of the spectrum should be one of the main features in the GW signal from the reconnection of the inflated string-bounded wall network.

\begin{figure}[t]
    \centering
    \includegraphics[width=\linewidth]{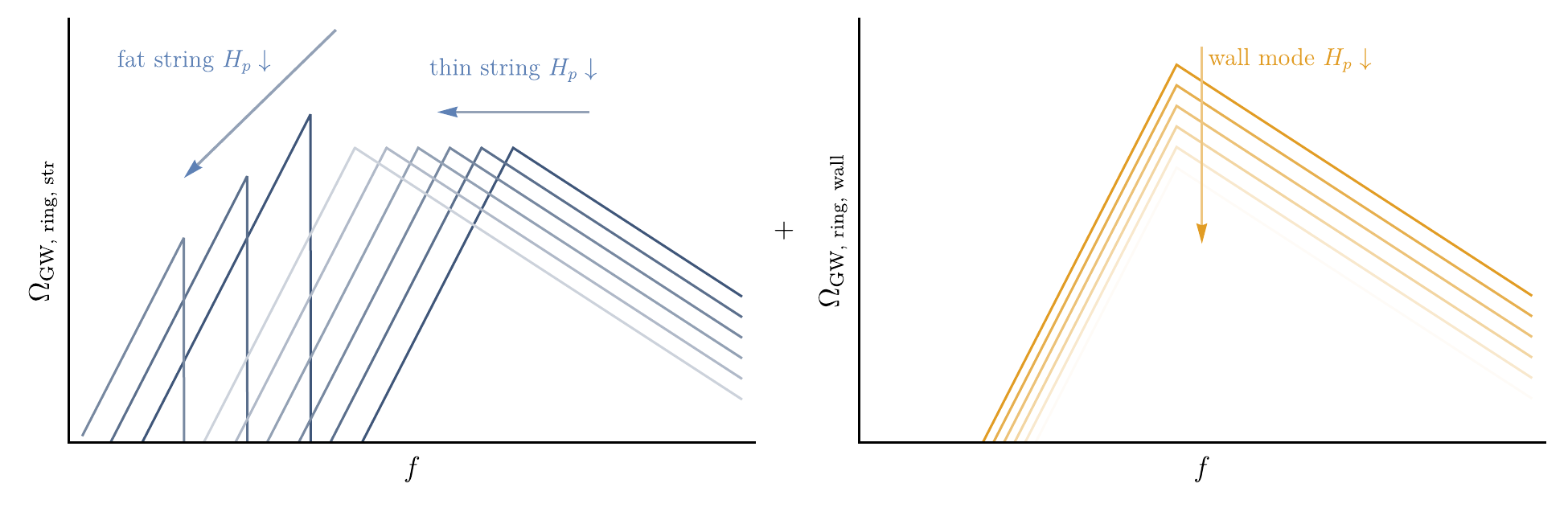}
    \includegraphics[width=0.75\linewidth]{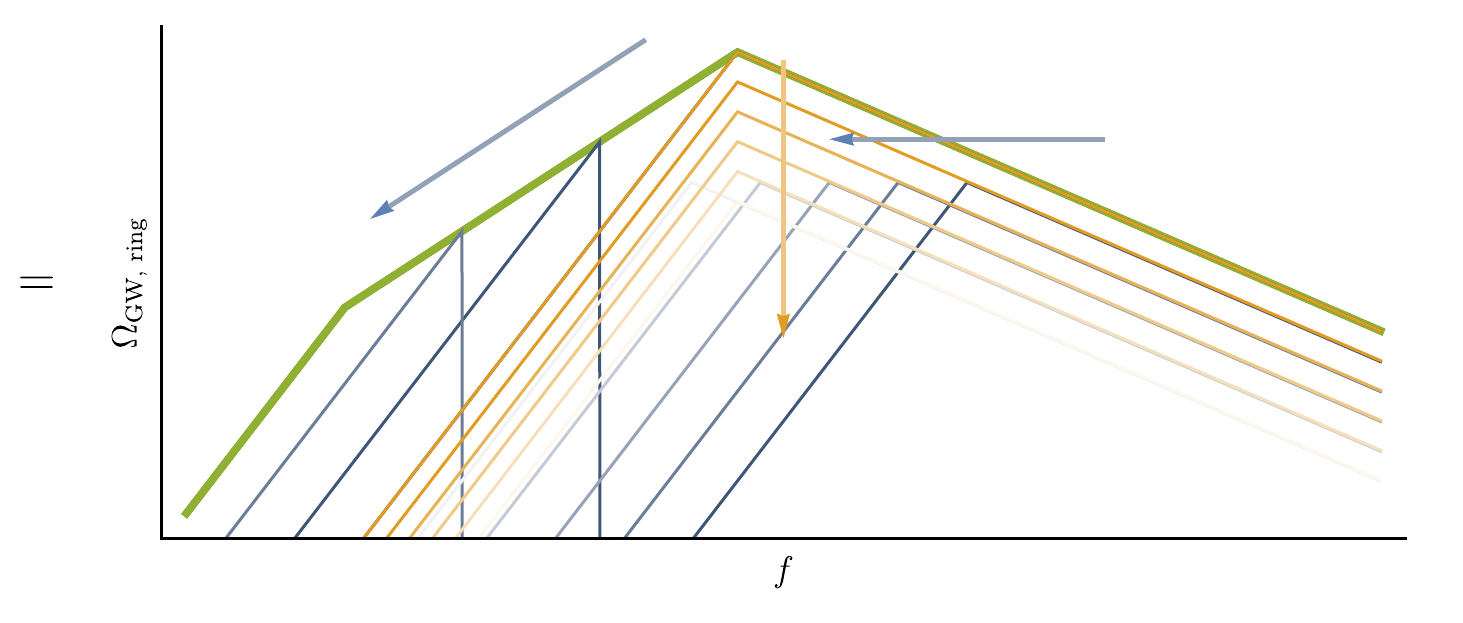}
    \caption{Decomposition of the gravitational-wave spectrum from cosmic rings on a logarithmic scale. Generally, three peaks appear. Two peaks are from the string mode oscillating with a frequency $k \approx H_p$: one from the usual thin string with tension $\mu$, and the other from the fat string with effective tension $\mu_\text{eff} \defeq \sigma / H_\text{re}$. The last peak comes from the wall mode oscillating with frequency $k \approx H_\text{re}$. \textbf{Top left:} As $H_p$ decreases (lighter color), the rings are produced later and radiate GWs of lower frequencies. The GW abundance of thin-string mode remains unchanged. However, that of the fat-string mode decreases and traces out the $\propto f^{3/2}$ power law. The fat-string peak vanishes for $H_p \lesssim \GammaGen$ since domain walls have collapsed when cosmic rings are produced. In other words, almost all cosmic rings (belts) have effectively become string loops (long strings). \textbf{Top right:} The GW spectrum for the wall mode of cosmic rings of decreasing $H_p$ (lighter color). While the wall mode constantly radiates at $k \approx H_\text{re}$, the GW abundance of cosmic rings produced later decreases and is subdominant. \textbf{Bottom:} The total GW spectrum of cosmic rings produced due to network reconnection is obtained by summing over the GW spectrum produced at all possible $H_p$. This is approximated by the envelope of all the spectra (thick green line).}
    \label{fig:ringDecomp}
\end{figure}

%%%%%%%%%%               %%%%%%%%%%
\paragraph{String Mode ($k \approx H_p$) GW Spectrum} The string mode on individual cosmic ring spectrum has more features. We start by noticing that $\rho_\text{ring}$ takes a similar form to that of string loops $\rho_\text{loop} \approx \mu H_p^2$ if we define an effective string tension $\mu_\text{eff} \defeq \sigma / H_\text{re}$. The presence of walls, therefore, makes the cosmic ring appear like a ``fat string".%
\footnote{The word ``fat string" in this work refers to the low-frequency oscillating mode of cosmic rings with frequency $k \sim H_p$ and effective tension $\sim \sigma / H_\text{re}$ before the wall decays. This term should not be confused with the ``fat string" used in numerical simulations of cosmic strings that refer to simulated strings with artificially enlarged string cores for better dynamics resolutions.} 
This means that the string mode of the cosmic rings will have a decay rate
\begin{equation}
    \GammaFatStr \approx \frac{P_\text{GW, ring}}{E_\text{ring}} 
    \approx \frac{\sigma}{\MPl^2} \frac{H_p}{H_\text{re}}, 
\end{equation}
in which we used 
\begin{equation}
    P_\text{GW, ring} \approx \frac{\ev{\dddot{Q}^2}}{\MPl^2} \approx \frac{1}{\MPl^2} \qty(\frac{E_\text{ring} \ell^2}{\ell^3})^2 \approx \frac{\sigma^2}{\MPl^2 H_\text{re}^2}, \quad 
    E_\text{ring} \approx \sigma w \ell.
\end{equation}
Note that the fat-string decay rate $\GammaFatStr$ is always smaller than the wall decay rate into gravitational wave $\GammaWall \lesssim \GammaTot$, as sketched in \cref{fig:ringSketch}, because $H_p < H_\text{re} < \sigma / \mu$. Yet once the wall mode decays, the ring width rapidly shrinks so that rings behave like cosmic string loops with their usual tension $\mu$. Then, much later, these string loops collapse and decay into gravitational waves. It is then helpful to call the string mode with $\mu_\text{eff}$ before $t \lesssim \GammaGen^{-1}$ the fat-string mode and call the string mode with $\mu$ after $t \gtrsim \GammaGen^{-1}$ the thin-string mode. The fat-string mode stays in the oscillating stage so that its gravitational-wave spectrum can be estimated as 
\begin{equation}
    \begin{multlined}
        \eval{\pdv{\Omega_\text{GW, ring, fat str}}{\ln k}}_{t = \GammaGen^{-1}} 
        \approx \frac{1}{T^4} \qty(\frac{\sigma H_p^2}{H_\text{re}}) \qty(\frac{T}{T_p})^3 \frac{\GammaFatStr}{H} \\
        \approx \qty(\frac{H_p}{H_\text{re}})^{3/2} \frac{\sigma^2}{\MPl^4 H_\text{re}^{1/2} \GammaGen^{3/2}} \qty(\frac{k}{H_p})^3, \quad k \lesssim H_p.
    \end{multlined}
\end{equation}
At $t = \GammaGen^{-1}$, domain walls rapidly collapse. Because tension changes suddenly from $\mu_\text{eff} = \sigma / H_p$ to its true tension $\mu$ within $t \approx \GammaGen^{-1}$, the gravitational-wave spectrum of fat-string mode sharply cuts off at $k \approx H_p$,%
\footnote{
Here, we assumed that $H_p \ll H_\text{re}$ so that the typical string mode frequency $k \approx H_p$ cannot resolve the dynamics on a scale $w^{-1} \approx H_\text{re}$. Technically, the spectrum may not be sharply cut off when $H_p \approx H_\text{re}$, and $\sim k^{-1}$ UV part of the GW spectrum from the fat string collapse may be resolved. Nonetheless, this contribution is comparable to that of the wall mode with $H_p \approx H_\text{re}$ and does not introduce a shallower power-law dependence to the total GW spectrum. Hence, we will not further discuss this subtlety.
}
and the cosmic ring, now a thin string loop, remains in its oscillating stage. 

The string mode then continues to behave just like a usual string loop until it decays around $t^{-1} \approx \GammaStr \defeq \mu H_p / \MPl^2$ and matches to the well-studied gauge string GW spectrum. The spectral peak of thin strings should be around $k \approx H_p$ at $t \approx \GammaStr^{-1}$ or, equivalently, $k \approx a(\GammaStr^{-1}) H_p / a(\GammaGen^{-1})$ at $t \approx \GammaGen^{-1}$. Also, it should exhibit $k^{3} \to k^{-1}$ power-law dependence around this peak. A more elaborated computation for this contribution is provided in \cref{app:GWString} with results given in \cref{appeqn:OmegaGWLoopOsc,appeqn:OmegaGWLoopCol}. To sum up, the full GW spectrum from the string mode of cosmic rings should look like
\begin{equation}
    \eval{\pdv{\Omega_\text{GW, ring, str.}}{\ln k}{\ln H_p}}_{t = \GammaGen^{-1}} \approx 
    \begin{dcases}
        \qty(\frac{H_p}{H_\text{re}})^{3/2} \frac{\sigma^2}{\MPl^4 H_\text{re}^{1/2} \GammaGen^{3/2}} \qty(\frac{k}{H_p})^3, & k \lesssim H_p, \\
        \frac{\sqrt{\mu}}{\MPl} \qty(\frac{a(\GammaGen^{-1})}{a(\GammaStr^{-1})} \frac{k}{H_p})^{3}, & H_p \lesssim k \lesssim \frac{a(\GammaStr^{-1})}{a(\GammaGen^{-1})} H_p, \\
        \frac{\sqrt{\mu}}{\MPl} \qty(\frac{a(\GammaGen^{-1})}{a(\GammaStr^{-1})} \frac{k}{H_p})^{-1}, & k \gtrsim \frac{a(\GammaStr^{-1})}{a(\GammaGen^{-1})} H_p. 
    \end{dcases}
\end{equation}
A sketch of example spectra from the string mode of individual cosmic rings with different $H_p$ is shown in the top left panel of \cref{fig:ringDecomp}. 

For our next discussion on the total gravitational-wave spectrum from rings, the important observation here is that the string mode has two peaks, one at $k\approx H_p$ and another around the usual peak of thin string loops. The thin-string mode behaves similarly to the well-studied GW spectrum produced by gauge strings as illustrated in \cref{app:GWString}. Thus, when reporting the string spectrum in \cref{sec:GWSummary}, we will use the spectrum from previous studies that treated the reconnection more carefully with support from detailed numerical simulations and will not distinguish the thin-string contribution from that from a usual cosmic string. The novel contribution is that from the fat string, which will be our main focus in \cref{sec:GWRingTot}, and we will defer further discussion on the thin-string mode until \cref{sec:GWSummary}. 

%%%%%%%%%%   Sec 3.4.2   %%%%%%%%%%
\subsubsection{Summing Contributions from Cosmic Rings of All Sizes \label{sec:GWRingTot}}
Once the GW signal from individual cosmic rings is determined, the total gravitational-wave spectrum can be obtained by summing over spectra of rings of various sizes. Reconnection of the string-wall network produces cosmic rings at different $H_p$. Thus, the total spectrum can be obtained by 
\begin{equation}
    \eval{\pdv{\Omega_\text{GW, loop}}{\ln k}}_{t = \GammaGen^{-1}} = \int_{H_{p, \text{min}}}^{H_{p, \text{max}}} \dd \ln H_p\; f(\ln H_p) \, \eval{\pdv{\Omega_\text{GW, loop}}{\ln k}{\ln H_p}}, 
\end{equation}
in which $f(\ln H_p)$ denotes the distribution of cosmic rings produced with size $\ell \approx H_p^{-1}$ at cosmic time $t \approx H_p^{-1}$. Maintaining a scaling regime for the network implies that this distribution is roughly scale-invariant, and $f(\ln H_p) \sim \text{const.} \approx 1$. This integral can be estimated by to approximating it as envelope of the integrand as we change $H_p$ within the integration bound as illustrated in the bottom panel of \cref{fig:ringDecomp}. 

When $H_p \approx H_\text{re}$, both the wall mode and string mode produce comparable peaks at $k \approx H_\text{re}$ with maximal GW abundance $\approx \sigma^2 / \qty(\MPl^{4} H_\text{re}^{1/2} \GammaGen^{3/2})$. Coincidentally, when domain walls decay mainly into gravitational waves, the $\sim k^{-1}$ falloff from the wall mode spectrum matches parametrically with the $\sim k^{-1}$ spectrum from the thin-string mode (or the usual cosmic string GW spectrum). That is, 
\begin{equation}
    \eval{\pdv{\Omega_\text{GW, ring, wall, col.}}{\ln k}}_{t = \GammaWall^{-1}}^{k = a(\GammaStr^{-1}) H_\text{re}/ a(\GammaWall^{-1})} = \frac{\sqrt{\mu}}{\MPl}.
\end{equation}
As $H_p$ decreases, the maximum abundance of both string and wall modes decreases. However, the (fat-) string mode now oscillates at a lower frequency $H_p$ than $H_\text{re}$. This leads to a $\sim k^{3/2}$ envelope due to the $\sim (H_p / H_\text{re})^{3/2}$ factor. Therefore, by integrating $H_p$ from $\GammaGen$ to $H_\text{re}$, we find that the cosmic rings produce a gravitational-wave spectrum of the form
\begin{equation}
    \eval{\pdv{\Omega_\text{GW, ring}}{\ln k}}_{t = \GammaGen^{-1}} 
    \approx \frac{2\pi \sigma^2}{3 \MPl^4 H_\text{re}^{1/2} \GammaGen^{3/2}} 
    \begin{dcases}
        \qty(\frac{k}{\GammaGen})^3 \qty(\frac{\GammaGen}{H_\text{re}})^{3/2}, & k \lesssim \GammaGen, \\
        \qty(\frac{k}{H_\text{re}})^{3/2}, & \GammaGen \lesssim k \lesssim H_\text{re}, \\
        \qty(\frac{H_\text{re}}{k}), & k \gtrsim H_\text{re},
    \end{dcases}
    \label{eqn:powerLawRing}
\end{equation}
in which we dropped the contribution from the thin-string peak.

%%%%%%%%%%    Sec 3.5    %%%%%%%%%%
\subsection{Summary \label{sec:GWSummary}}

We have computed the gravitational-wave spectrum due to the three-stage evolution of the inflated string-bounded wall network. Taking $\GammaTot = \GammaWall$, the full wall spectrum, including disks, belts, and rings, observed at around $t = \GammaWall^{-1}$, is 
\begin{equation}
	\eval{\pdv{\Omega_\text{GW}}{\ln k}}_{t = \GammaWall^{-1}} \approx \frac{2\pi \sigma^{1/2}}{3 \MPl H_\text{re}^{1/2}}
    \begin{dcases}
        \qty(\frac{k}{\GammaWall})^3 \qty(\frac{\GammaWall}{H_\text{re}})^{3/2}, & k \lesssim \GammaWall, \\
		\qty(\frac{k}{H_\text{re}})^{3/2}, & \GammaWall \lesssim k \lesssim H_\text{re}, \\
		\qty(\frac{H_\text{re}}{k}), & k \gtrsim H_\text{re}. 
	\end{dcases}
	\label{eqn:powerLaw}
\end{equation}
This spectrum is shared among any inflated string-bounded wall networks. Redshifted to today, this provides a peak around 
\begin{equation}
    \begin{aligned}
        \eval{f_\text{peak}}_{T_0} 
        \approx& \SI{2E-3}{\Hz} \qty(\frac{106.75}{g_{*,s}(T_{\GammaGen})})^{1/3} \qty(\frac{g_{*,\rho}(T_{\GammaGen})}{106.75})^{1/4} \qty(\frac{H_\text{re}}{\SI{E-13}{\GeV}}) \qty(\frac{\SI{5.3E-19}{\GeV}}{\GammaGen})^{1/2} \\
        \approx& \SI{2E-3}{\Hz} \qty(\frac{106.75}{g_{*,s}(T_{\GammaGen})})^{1/3} \qty(\frac{g_{*,\rho}(T_{\GammaGen})}{106.75})^{1/4} \qty(\frac{H_\text{re}}{\SI{E-13}{\GeV}}) \qty(\frac{\SI{E6}{\GeV}}{\vwall})^{3/2},
    \end{aligned}
    \label{eqn:kPeakT0}
\end{equation}
in which the 2\textsuperscript{nd} equality assumes that $\GammaGen = \GammaWall$. The fractional energy density of the gravitational wave around this peak, as observed today, is 
\begin{equation}
    \begin{aligned}
        \eval{\Omega_\text{GW} h^2}_{T_0, f_\text{peak}} \approx& 2\times10^{-8} \qty(\frac{\vwall}{\SI{E6}{\GeV}})^6 \qty(\frac{\SI{5.3E-19}{\GeV}}{\GammaGen})^{3/2} \qty(\frac{\SI{E-14}{\GeV}}{H_\text{re}})^{1/2} \\
        =& 2\times10^{-8} \qty(\frac{\vwall}{\SI{E6}{\GeV}})^{3/2} \qty(\frac{\SI{E-14}{\GeV}}{H_\text{re}})^{1/2}. 
    \end{aligned}
    \label{eqn:Omegah2GWT0}
\end{equation}

In the model discussed in \cref{sec:genPic}, topologically stable strings remain after the annihilation of domain walls. The string contribution (both scaling string after $t \gtrsim \GammaGen^{-1}$ and thin-string mode of cosmic rings) will be part of the gravitational-wave spectrum. It has been shown that the gravitational-wave spectrum from gauge string as observed today is approximately \cite{Cui:2018rwi} 
\begin{equation}
    \eval{\pdv{\Omega_{\text{GW, str}} h^2}{\ln f}}_{T_0} \approx 4\times 10^{-11} \qty(\frac{\vstr}{\SI{E13}{\GeV}}) 
    \begin{dcases}
        \qty(\frac{f}{f_\text{str, eq}})^{3/2}, & f \lesssim f_\text{str, eq}, \\
        1, & f_\text{str, eq} \lesssim f \lesssim \frac{a(\GammaGen^{-1}) k_\text{str}}{2\pi}, \\
        \frac{a(\GammaGen^{-1}) k_\text{str}}{2\pi f}, & f \gtrsim \frac{a(\GammaGen^{-1}) k_\text{str}}{2\pi}.
    \end{dcases}
\end{equation}
Here, $f_\text{str, eq}$ is defined as the typical frequency of  GWs emitted by strings around matter-radiation equality 
\begin{equation}
    \frac{2a(t_\text{eq})}{f_\text{str, eq}} = \gamma \frac{\mu}{\MPl^2} \frac{1}{H(t_\text{eq})} \implies f_\text{str, eq} \approx \SI{2E-6}{\Hz} \qty(\frac{\SI{E13}{\GeV}}{\vstr})^2,
\end{equation}
in which $t_\text{eq}$ denotes the cosmic time at equality, and $\gamma \approx 50$ is a dimensionless parameter governing the decay efficiency of gauge strings. The two IR contributions are familiar from cosmic strings in the scaling regime during a matter-dominated and radiation-dominated epoch. The novel signature is the UV falloff. Unlike the usual case where this falloff is controlled by the symmetry-breaking scale, our falloff is controlled by the string re-entry Hubble as well. 

Note that we ignore the GW emission from higher harmonic modes, especially from possible cuspy structures on the domain wall, in this work. The higher modes may potentially modify the UV part of the spectrum. This is analogous to cusps modifying the UV part of the GW spectrum of cosmic strings. It is known that the $n$\textsuperscript{th} harmonic mode on strings may be excited and emit GWs with a power $P_n \sim n^{-q} $ with spectral index $q = 4/3$ for cusp-dominated string configuration \cite{Vachaspati:1984gt, Auclair:2019wcv}. Assuming that the UV spectrum from the fundamental mode follows a power law $\Omega_\text{GW} \propto (k/k_c)^{-n_\text{UV}}$, the higher-order harmonics excited by cusps can lead to a shallower spectral shape $\Omega_\text{GW} = \sum_{n=1}^{\infty} \Omega_{\text{GW}, n} \propto k^{1-q}$ in the UV when $q < n_\text{UV} + 1$ \cite{Blasi:2020mfx}. This modifies the UV tail of the GW spectrum of cosmic strings from $\sim f^{-1}$ to $\sim f^{-1/3}$. Similar effects may appear for the wall oscillation and change the UV part of the GW spectrum, but the investigation of such an effect requires dedicated numerical simulations and is beyond the scope of this work.

\begin{figure}[t]
	\centering
	\includegraphics[width=0.9\textwidth]{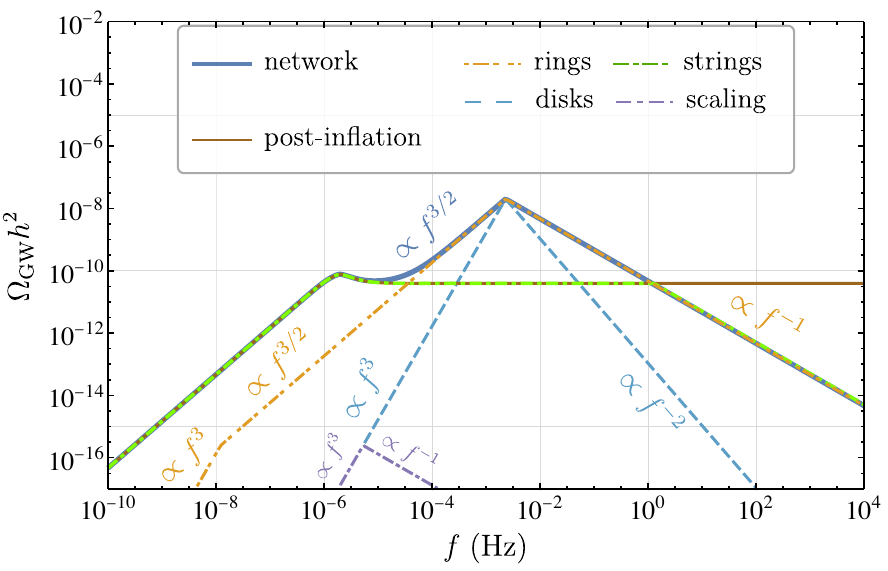}
	\caption{
        Decomposition of the gravitational-wave signal of the defect network bounded by inflated \textit{gauge} strings with the string tension $\vstr = \SI{E13}{\GeV}$, the wall tension $\vwall = \SI{E6}{\GeV}$, the string re-entry Hubble parameter $H_\text{re} = \SI{E-13}{\GeV}$, and the decay rate of domain walls $\GammaGen = \GammaWall \approx \SI{5.3E-19}{\GeV}$. The dominant GW signal is from domain walls and comes in three parts: rings (dot-dash-dotted yellow line), disks (dashed light-blue line), and scaling (dot-dashed violet line) with the characteristic power-law behavior labeled on the figure. Stable gauge strings in the model also produce GW (dot-dashed green line). The full GW signal is the sum of the two spectra (solid dark-blue line). The string spectrum for the inflated string-wall network has a UV falloff because the strings can produce GWs only after $t = H_\text{re}^{-1}$ due to inflation. If there remain no stable gauge strings after the annihilation of walls, only GW from walls will be observed. The same benchmark model with the production of both gauge strings and domain walls after inflation will produce a GW spectrum only from scaling gauge strings (solid brown line), with no sharp feature due to the annihilation of walls on the spectrum.
 }
	\label{fig:benchmarkSignal}
\end{figure}

As a concrete demonstration of this power-law dependence, we consider a benchmark point with $\vstr = \SI{E13}{\GeV}$, $\vwall = \SI{E6}{\GeV}$, $H_\text{re} = \SI{E-14}{\GeV}$, and $\GammaWall \approx \SI{5.3E-19}{\GeV}$, whose spectrum is shown in \cref{fig:benchmarkSignal}. Various contributions from cosmic disks, cosmic rings, scaling strings, and scaling walls are decomposed with the characteristic frequency dependence labeled near each curve. Both cosmic disks and cosmic rings contribute significantly to the GW spectrum near its peak. But because cosmic rings of longer radius $\ell$ are produced from the reconnection of cosmic belts, their contribution dominates the IR part of the GW signal. If the thin string is sufficiently long-lived, this contribution may eventually overtake the IR part of the ring contribution, leaving a string-like spectrum in the lower frequency. 

We note that the presence of the string spectrum (dot-dashed green line) is model-dependent. It is possible that cosmic strings produced in the first phase transition are all topologically unstable and annihilated by walls produced in the second phase transition. For instance, unstable $\Z[2]$ strings can usually arise when breaking an $\SO{10}$ grand unified theory to the Standard Model. Unlike our model with $\pi_1(\U{1}) = \Z$ showing the stability of boundary strings, $\pi_1(\SO{10}/(\SU{3}\times \SU{2} \times \U{1})) = 0$ implies that these strings are not protected by topology. Due to this topological instability, all strings must attach to domain walls, and no string remains after the walls collapse. This may remove the flat string spectrum. Nonetheless, due to inflation, walls (rings and disks) bounded by unstable strings are capable of providing some striking signal.%
\footnote{
However, it should be noted that the unstable string does not affect the dynamics of the network until much after the collapse of the walls. This makes extracting information about the string, such as the string tension, from the GW spectrum quite challenging in general.
}

Here, it is also opportune to compare our scenario (thick blue curve in \cref{fig:benchmarkSignal}) with a period of inflation between phase transitions to the typical post-inflationary production of both cosmic strings and domain walls (solid brown curve labeled ``post-inflation"). For the post-inflationary production scenario, the typical energy of domain walls is initially smaller than that of strings.\footnote{See, however, the discussion in Section VII of ref.~\cite{Dunsky:2021tih} and our remarks in \cref{sec:CompareWithPrev}.} As strings scale with the Hubble size, walls attached to strings shall also grow until $H \approx \sigma / \mu$. Then, domain walls become important and can annihilate some strings (see \cref{fig:string_wall_winding}), analogous to the $N_\text{DW} = 1$ axion strings annihilated by domain walls due to QCD potential. A previous study suggests that wall-driven string annihilation contributes within $\sim \order{1}$ of the GW spectrum from that of the string in scaling regime~\cite{Gorghetto:2021fsn}. On top of this contribution from annihilating strings of opposite winding numbers, the gauge string remains stable due to the nontrivial 1\textsuperscript{st} homotopy group of $\U{1}$ and will continue to produce a scaling spectrum. It is, thus, reasonable to assume that the wall-driven annihilation process almost does not produce significant features on top of the scaling spectrum from the gauge string. This is why we choose to plot the ``post-inflation" curve in \cref{fig:benchmarkSignal} to match the spectrum of a scaling string without domain walls. It is, however, expected that some small $\sim \order{1}$ enhancement of GW spectrum in the IR can appear for the post-inflationary case since the stable string is bundled up from lower-tension strings, and increased string tension generally augments GW abundance. Comparing the thick blue line with the brown line, we see that the wall annihilation delayed by re-entered strings produces a distinctive spectral shape above the usual flat GW spectrum of strings. This shows how the inflated string-bounded walls produce more gravitational-wave signals. Also, in the UV, the flat string spectrum is modified as no earlier strings are present before re-entry. This shows how inflated string-bounded walls produce more spectral features than the usual cosmic strings. We stress that the peak frequency of the spectrum needs not to be around $\approx \SI{E-4}{\Hz}$, and this will be demonstrated with more benchmarks in \cref{sec:GWBenchmark}.

%%%%%%%%%%     Sec 4     %%%%%%%%%%
\section{From Boundary Gauge String to Boundary Global Strings \label{sec:GWGlobalSignature}}
In this section, we will shift our focus to finding the gravitational-wave signature when global strings, instead of gauge strings, bound the walls. This can be achieved by demoting the $\U{1}$ gauge symmetry to a global symmetry for our model shown in \cref{eqn:toyModel}. Different from the scenario discussed in \cref{sec:GWSignature}, the model with global strings has an extra Nambu-Goldstone boson (NGB). Strings may radiate NGBs and open up another channel to dump the energy of the defect network. This can suppress its gravitational wave signal.

%%%%%%%%%%    Sec 4.1    %%%%%%%%%%
\subsection{NGB Radiation from Boundary Strings: an Illustrative Toy Computation \label{sec:GlobalToyExample}}
Before computing the actual GW spectrum from various defects in the network, it is helpful to clarify how NGB radiation changes the dynamics and GW emission of defects with a toy computation. In this computation, crucial features introduced by boundary global strings are demonstrated with an artificially chosen defect shape that is not part of the network. Let us consider a rectangular domain wall ``ribbon" of width $w$ and length $\ell$ (with $w<\ell$) bounded by a global string on its rim. We will also assume that the ribbon oscillates quickly with frequency $k \approx w^{-1}$.

\begin{figure}[t]
    \centering
    \includegraphics[width=0.95\textwidth]{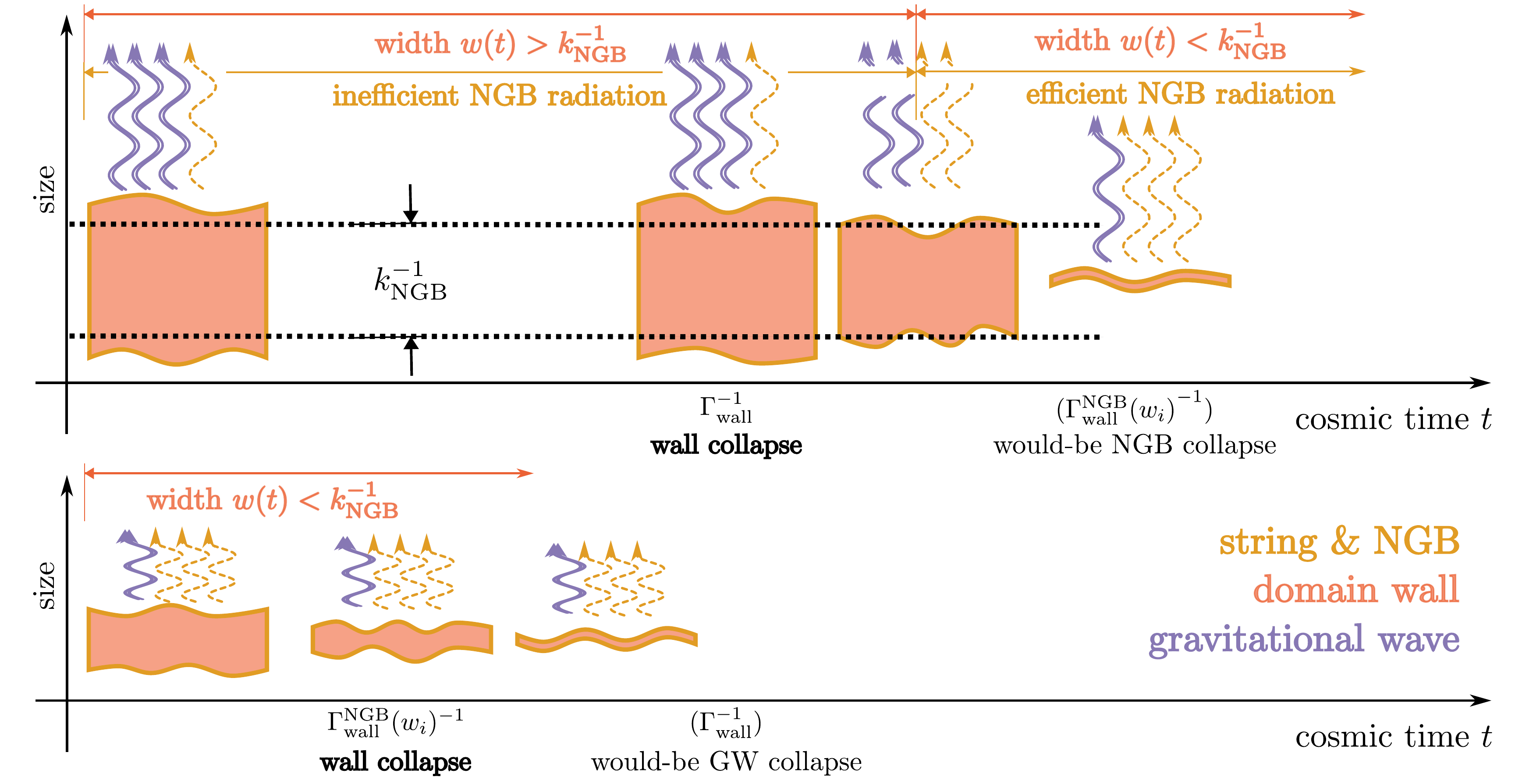}
    \caption{Schematic sketch of the evolution of string-wall ``ribbon" used as a toy model in \cref{sec:GlobalToyExample}: \textbf{Top}: When the ribbon's initial width is larger than $k_\text{NGB}^{-1}$, Nambu-Goldstone boson (NGB) radiation (wavy yellow lines) is not efficient compared to gravitational-wave (GW) emission (wavy violet lines). The ribbon's energy is mostly dumped into GWs, and the wall collapses. However, as the wall width decreases below $k_\text{NGB}^{-1}$, the NGB emission becomes efficient and takes most of the ribbon's energy away in the UV, modifying the GW spectrum. \textbf{Bottom}: When the ribbon's initial width is smaller than $k_\text{NGB}^{-1}$, NGB radiation is efficient compared to that of GW, and the wall mode collapses at $t \approx \GammaNGBWall^{-1}$ which is earlier than its would-be collapse timescale triggered by the GW emission. There is generally less GW emitted since (1) the ribbon decays away earlier, and (2) the ribbon mostly emits NGBs instead of GWs. }
    \label{fig:NGBSketch}
\end{figure}

%%%%%%%%%%   Sec 4.1.2   %%%%%%%%%%
\subsubsection{Changes to Evolution of String-wall System}
At this point, it is worth reviewing the difference between global strings and gauge strings. On the one hand, the string tension has a logarithmic enhancement from NGB modes
\begin{equation}
    \mu \approx \pi v_2^2 \ln(v_2 d),
\end{equation}
where $d$ denotes the typical distance between cosmic strings, and we will estimate the logarithm to be $\sim \order{60}$. Therefore, the condition that $E_\text{wall} \gtrsim E_\text{str}$ translates to a rough bound on $H_\text{re}$ as 
\begin{equation}
    H_\text{re} \lesssim \frac{2\sigma}{\mu} \approx \frac{\sigma}{90 v_2^2}.
\end{equation}
On the other hand, global strings radiate NGBs with power 
\begin{equation}
    P_\text{NGB} \approx \gamma_a v_2^2 k \ell, 
    \label{eqn:PowerNGB}
\end{equation}
in which $\gamma_a \approx 60$ is a dimensionless number controlling the radiation efficiency into Nambu-Goldstone modes \cite{Vilenkin:1986ku, Battye:1993jv}.%
\footnote{
We note that the NGB couples to cosmic strings only, but as topology dictates that all open walls are attached to cosmic strings at their boundaries, the oscillation of a heavy domain wall will drag the boundary cosmic string, force it to radiate NGBs, and decrease the wall's energy. This mechanism is slightly different from the usual NGB radiation from strings that oscillate by their own tension. It is, thus, possible that $\gamma_a \approx 60$ is modified when heavy walls drag strings, but the parametric dependence of $P_\text{NGB}$ should remain the same.
}
The power radiated by a one-dimensional object should scale linearly with its size so long as the rapid oscillation is incoherent on scales larger than its wavelength (see discussion in \cref{sec:GWBelts}), and this motivates the $\propto \ell$ dependence in our parameterization. Without the rapid oscillating wall, the string typically oscillates at a frequency $k \approx \ell^{-1}$ so that the power loss into NGB emission is roughly independent of the string size \cite{Vilenkin:1986ku}. However, when $k \approx w^{-1} > \ell^{-1}$, the rapid oscillation enhances the power radiated by the string-wall defect. 

Here, we will also assume that the wall does not emit NGBs efficiently. This can be justified by two reasons. First, the domain wall in our two-field model mainly consists of the angular field of $\phi_1$. This is nearly orthogonal to the light global NGB mode mainly in $\phi_2$. Also, from the effective action perspective, coupling of NGBs to a string worldsheet involves a lower-dimensional operator than that to a wall's worldvolume. Therefore, the emission rate of NGBs from oscillating walls is suppressed compared to that from their boundary strings. More concrete discussion on this suppression is provided in \cref{app:suppressedCoupling}

By assuming $E_\text{wall} \gtrsim E_\text{str}$, we find that the string-wall system's energy evolves as
\begin{equation}
    \dv{(\sigma w \ell)}{t} \approx - \GammaWall \sigma w \ell - \gamma_a v_2^2 \frac{\ell}{w} = -\qty(\GammaWall + \GammaNGBGen) \sigma w \ell, 
\end{equation}
in which we defined 
\begin{equation}
    \GammaNGBGen \defeq \frac{\gamma_a v_2^2}{\sigma w^2}.
\end{equation}
A special scale is obtained by comparing the decay rates in the two channels, \textit{i.e.},
\begin{equation}
    k_\text{NGB} \defeq \frac{\GammaWall^{1/2} \sigma^{1/2}}{\gamma_a^{1/2} v_2}. 
    \label{eqn:kNGB}
\end{equation}
By estimating that $w \approx k^{-1}$, it is also helpful to relate $\GammaNGBGen$ to $\GammaWall$ by
\begin{equation}
    \GammaNGBGen \approx \GammaWall \qty(\frac{k}{k_\text{NGB}})^2, \quad k \gtrsim k_\text{NGB}.
    \label{eqn:GammaNGBk2}
\end{equation}
This equation highlights that when $w^{-1} \gtrsim k_\text{NGB}$, more energy from the string-wall ribbon will be dumped into NGBs relative to GWs, which changes the GW spectrum. 

The crucial point here is that there is a competition of the decay timescale between the NGB radiation determined by $\GammaNGBWall$ and the GW radiation determined by $\GammaWall$. The larger one of them determines the collapse time of the network. Fortunately, both scales are completely determined by the initial width of the ribbon $w_i$ and independent of its length. Hence, the most important decay channel can be determined by comparing $w_i^{-1}$ with $k_\text{NGB}$. 

%%%%%%%%%%   Sec 4.1.2   %%%%%%%%%%
\subsubsection{Changes to Gravitational-wave Spectrum}
Let us now briefly comment on the GW spectrum for this rectangular string-wall ribbon. When $w_i^{-1} \gtrsim k_\text{NGB}$, NGB radiation can efficiently collapse the string-bounded wall ribbon and set its decay rate. In other words, if the string-wall rectangle's initial width $w_i$ is small enough, the total decay width $\GammaTot$ is roughly determined by
\begin{equation}
    \GammaNGBWall \approx \frac{\mu}{\pi \sigma} w_i^{-2},
\end{equation}
in which we estimated $\pi \gamma_a v_2^2 \approx \mu$ as illustrated in the bottom panel of \cref{fig:NGBSketch}. Note that the $\GammaNGBWall$ defined above is a fixed parameter independent of the width $w(t)$ and should be contrasted with $\GammaNGBGen$ that depends on the $w(t)$. There are two modifications compared to our estimation for a wall bounded by gauge strings: (1) the maximal abundance is affected since the decay time $t \approx \GammaNGBWall^{-1} < \GammaWall^{-1}$, and (2) the UV tail of the spectrum should be altered to reflect the enhancement of NGB radiation while the width decreases as shown in \cref{eqn:GammaNGBk2}. The first change can be addressed by doing the same estimation as \cref{eqn:OmegaGWRingOscillatingWall} with a replacement $\GammaGen \to \GammaNGBWall$ as the total decay rate $\GammaTot$ is almost entirely $\GammaNGBWall$. The second change enters the computation in the form of a branching ratio, \textit{i.e.} 
\begin{equation}
    \frac{\GammaWall}{\GammaTot} 
    = \frac{\GammaWall}{\GammaWall + \GammaNGBGen} 
    \approx \frac{\GammaWall}{\GammaNGBWall} \frac{\GammaNGBWall}{\GammaNGBGen}
    \approx \frac{\GammaWall}{\GammaNGBWall} \qty(w_i k)^{-2},
    \label{eqn:NGBBRSuppressionw<}
\end{equation}
in which $k\sim w^{-1}$ denotes the GW frequency due to the rapidly shrinking width observed at $t \approx \GammaNGBWall^{-1}$. This extra $\sim k^{-2}$ dependence will further suppress the UV part of the GW spectrum from $\sim k^{-n}$ to $\sim k^{-n-2}$. Similar to the wall mode of a cosmic ring, the UV part of the ribbon GW spectrum should have been scaled like $\sim k^{-1}$. However, with the NGB radiation taking over the energy loss process, the UV part of the ribbon GW spectrum scales like $\sim k^{-3}$ instead. 

On the other hand, if initially $w_i^{-1} \lesssim k_\text{NGB}$, the string-bounded rectangular wall still mainly collapses due to energy loss from the wall, \textit{i.e.}, $\GammaTot \approx \GammaWall$, but NGB emission eventually dominates when $w^{-1} \gtrsim k_\text{NGB}$ in the collapse stage as illustrated in the top panel of \cref{fig:NGBSketch}. This means that while the peak of the GW spectrum remains unaffected, we should see a $k^{-n} \to k^{-n-2}$ power-law change in the UV part of the GW spectrum. This effect kicks in when $w^{-1} \approx k \gtrsim k_\text{NGB}$ and, analogous to \cref{eqn:NGBBRSuppressionw<}, the branching fraction is modified to 
\begin{equation}
    \frac{\GammaWall}{\GammaTot} 
    \approx \frac{\GammaGen}{\GammaNGBGen} 
    \approx \frac{\Gamma_\text{GW}}{\GammaWall} \qty(\frac{k_\text{NGB}}{k})^2.
    \label{eqn:NGBBRSuppressionw>}
\end{equation}
Unlike the previous case ($w_i^{-1} \gtrsim k_\text{NGB}$) in which the domain wall oscillation is prematurely terminated by NGB radiation, this case allows the wall to oscillate fully and decay predominantly into GWs while giving an interesting transition in spectral shape in the UV.

%%%%%%%%%%    Sec 4.2    %%%%%%%%%%
\subsection{Gravitational-wave Signal: Spectrum and Strength}
With the previous toy computation in mind, we now compute the gravitational-wave spectrum from the inflated string-bounded wall network. We will focus on the novel features from the gauge string case instead of reiterating already familiar computations. A summary of GW spectra is provided in \cref{sec:GWGlobalSummary}.

%%%%%%%%%%   Sec 4.2.1   %%%%%%%%%%
\subsubsection{GW from Cosmic Disks \label{sec:GWGlobalDisks}}
Cosmic disks generate gravitational-wave signals quite analogous to the scenario discussed in \cref{sec:GlobalToyExample} with $r^{-1} \approx \ell^{-1} \approx w_i^{-1} \approx H_\text{re}$. If $k_\text{NGB} \gtrsim H_\text{re}$, they still undergo three-stage evolution. Both the scaling regime ($t \lesssim H_\text{re}^{-1}$) and the oscillation stage ($H_\text{re}^{-1} \lesssim t \lesssim \GammaTot^{-1}$) remains unaffected. Only the UV part of the gravitational-wave spectrum, which comes from the collapsing stage of cosmic disks, is altered. The estimation for this stage should be 
\begin{equation}
    \begin{multlined}
        \eval{\pdv{\Omega_\text{GW, col.}}{\ln k}}_{t = \GammaWall^{-1}} 
        \approx \frac{1}{T_{\GammaWall}^4} \qty(\sigma H_\text{re}) \qty(\frac{T_{\GammaWall}}{T_\text{re}})^3 \frac{\GammaWall}{\GammaTot} \qty(r H_\text{re})^2 \\
        \approx \frac{\sigma^2}{ \MPl^4 H_\text{re}^{1/2} \GammaWall^{3/2}} 
        \begin{dcases}
            \qty( \frac{H_{\text{re}}}{k} )^2, & H_\text{re} \lesssim k \lesssim k_\text{NGB}, \\
            \qty( \frac{H_{\text{re}}}{k} )^2 \qty(\frac{k_\text{NGB}}{k})^2, & k \gtrsim k_\text{NGB}, 
        \end{dcases}
    \end{multlined}
\end{equation}
following \cref{eqn:NGBBRSuppressionw>}. This means that the UV spectrum of the gravitational wave during the collapse stage is modified such that a steeper falloff $\sim k^{-4}$ may appear. This is corroborated with a more sophisticated computation using the Boltzmann equation as discussed in \cref{app:spectrumComputation}.

When we take $k_\text{NGB} \lesssim H_\text{re}$, the three-stage evolution of walls is altered slightly to (1) scaling regime, (2) oscillating regime that is prematurely terminated by NGB radiation, and (3) rapid decay into predominantly NGBs. This means that the oscillating spectrum remains valid until $t \approx \GammaNGBWall^{-1}$, \textit{i.e.},
\begin{equation}
    \eval{\pdv{\Omega_\text{GW, osc.}}{\ln k}}_{t = \GammaNGBWall^{-1}} \approx \frac{\sigma^2}{\MPl^4 H_\text{re}^{1/2} \qty(\GammaNGBWall)^{3/2}} \qty(\frac{k}{H_\text{re}})^3.
\end{equation}
Then, the radiation into NGBs becomes efficient, and the network rapidly collapses. The gravitational-wave spectrum becomes much steeper because of the enhancement of the decay rate into NGBs from \cref{eqn:NGBBRSuppressionw<}, and its UV part should be 
\begin{equation}
    \eval{\pdv{\Omega_\text{GW, col.}}{\ln k}}_{t = \GammaNGBWall^{-1}} \approx \frac{\sigma^2}{\MPl^4 H_\text{re}^{1/2} \qty(\GammaNGBWall)^{3/2}} \qty(\frac{H_\text{re}}{k})^4.
\end{equation}
While the spectral shape still is interesting, this case with $k_\text{NGB} \lesssim H_\text{re}$ leads to a smaller peak in $\Omega_\text{GW}(k)$, hence yielding typically a smaller GW signature. 

\subsubsection{GW from Cosmic Rings \label{sec:GWGlobalRing}}
Another significant contribution from the string-wall network to the GW spectrum comes from cosmic rings. It is worth reiterating that cosmic rings have a wall mode $(k \approx H_\text{re} \approx w^{-1})$ and a string mode $(k \approx H_p \approx \ell^{-1})$. Each of them gives different decay rates, as summarized in \cref{tab:DecayRateSummary}.
\begin{table}[t]
    \centering
    \begin{tabular}{>{$}c<{$\;}|*{2}{>{$\displaystyle\;}c<{$\;}}}
        \hline\hline
        \Gamma & \text{wall} & \text{string} \\
        \hline\\[-10pt]
        \text{GW} & \frac{\sigma}{\MPl^2} & \frac{\sigma}{\MPl^2} \frac{H_p}{H_\text{re}} \\[12pt]
        \text{NGB} & \frac{\mu}{\pi \sigma} H_\text{re}^2 & \frac{\mu}{\pi \sigma} H_\text{re} H_p \\[12pt]
        \hline\hline
    \end{tabular}
    \caption{Summary table of decay rates of two oscillation modes of cosmic rings into various channels}
    \label{tab:DecayRateSummary}
\end{table}
Most of these have been computed in the previous section, and the new contribution, coming from the NGB radiation of string mode (lower-right entry of \cref{tab:DecayRateSummary}), can be evaluated using the NGB radiation formula (\cref{eqn:PowerNGB}) and dividing the power by the energy of cosmic rings.%
\footnote{
We implicitly assumed that the cosmic rings are in the stable configuration instead of the unstable configuration as shown in \cref{fig:string_wall_winding}. This allows us to treat the NGB radiation of the string mode as if it is emitted from a string. The unstable configuration, however, contains strings that wind oppositely, and the NGBs radiated from the string mode can have a parametrically smaller energy scale $\sim H_p$ than the inverse wall width $\sim H_\text{re}$. Radiated NGB at such a long wavelength typically cannot resolve the winding of individual boundary strings. Hence, the NGB radiation from the string mode of the unstable configuration should be further suppressed by some small parameter controlled by $\sim H_p / H_\text{re}$, analogous to the electric quadrupole radiation as a subleading effect to the electric dipole radiation in classical electrodynamics. 
}
\Cref{tab:DecayRateSummary} tells us that the wall mode dissipates more power and controls the dominant decay rate. It is also worth mentioning that by assuming that $H_\text{re} \lesssim \sigma/\mu$, it is guaranteed that the decay of the walls by NGB radiation can happen only after the string re-entry, \textit{i.e.},
\begin{equation}
    H_\text{re}^{-1} < \frac{\sigma}{\mu H_\text{re}} H_\text{re}^{-1} \approx \GammaNGBWall^{-1}.
    \label{eqn:CompareHreToGammaNGBWall}
\end{equation}
This matches the intuition that the boundary string does not affect the dynamics too much when the wall energy is large.

We can then repeat the analysis for the wall and string mode similar to that presented in \cref{sec:GWRingTot}. If $k_\text{NGB} \gtrsim H_\text{re}$, cosmic rings are wide enough such that NGB emission is not the dominant decay channel during its oscillating stage. Therefore, the IR part of the GW spectrum from cosmic rings remains the same as that discussed in \cref{sec:GWRingTot}, exhibiting $\sim k^3 \to \sim k^{3/2}$ power law. As the wall starts to collapse, the ring width $w(t)$ decreases below $k_\text{NGB}^{-1}$, and the NGB emission becomes the dominant decay process. Then, according to \cref{eqn:NGBBRSuppressionw>}, an extra $\sim k^{-2}$ suppression in the GW spectrum from the branching ratio becomes important. Aggregating all these observations, we obtain the gravitational-wave spectrum analogous to \cref{eqn:powerLawRing},
\begin{equation}
    \eval{\pdv{\Omega_{\text{GW}}}{\ln k}}_{t = \GammaWall^{-1}} 
    \approx \frac{2\pi \sigma^2}{3 \MPl^4 H_\text{re}^{1/2} \GammaWall^{3/2}}
    \begin{dcases}
        \qty(\frac{k}{\GammaWall})^3 \qty(\frac{\GammaWall}{H_\text{re}})^{3/2}, & k \lesssim \GammaWall, \\
        \qty(\frac{k}{H_\text{re}})^{3/2}, & \GammaWall \lesssim k \lesssim H_\text{re}, \\
        \frac{H_\text{re}}{k}, & H_\text{re} \lesssim k \lesssim k_\text{NGB}, \\
        \frac{H_\text{re}}{k} \qty(\frac{k_\text{NGB}}{k})^2, & k \gtrsim k_\text{NGB}.
    \end{dcases}
\end{equation}

If $k_\text{NGB} \lesssim H_\text{re}$, the dominant decay mode is the NGB radiation instead and $\GammaTot \approx \GammaNGBWall$. The $k^{3} \to k^{3/2}$ power law remains unchanged in the oscillating stage, but the maximal gravitational-wave abundance is changed to $\approx \sigma^2 / \qty(\MPl^4 H_\text{re}^{1/2} \GammaNGBWall^{3/2})$. At wall collapse, NGB emission is already important. Hence, the UV tail in this scenario is $\sim k^{-3}$; see the discussion below \cref{eqn:NGBBRSuppressionw<}. The GW spectrum from rings is of the form
\begin{equation}
    \begin{multlined}
        \eval{\pdv{\Omega_{\text{GW}}}{\ln k}}_{t = \GammaNGBWall^{-1}}
    \approx \frac{2\pi \sigma^2}{3\MPl^4 H_\text{re}^{1/2} \GammaNGBWall^{3/2}} \\
    \times 
    \begin{dcases}
        \qty(\frac{k}{\GammaNGBWall})^3 \qty(\frac{\GammaNGBWall}{H_\text{re}})^{3/2}, & k \lesssim \GammaNGBWall, \\
        \qty(\frac{k}{H_\text{re}})^{3/2}, & \GammaNGBWall \lesssim k \lesssim H_\text{re}, \\
        \qty(\frac{H_\text{re}}{k})^3, & k \gtrsim H_\text{re}. 
    \end{dcases}
    \end{multlined}
\end{equation}

At this point, it may be interesting to ask whether the $ k^{3/2}$ spectrum can also be suppressed by the NGB radiation. This requires $H_\text{re} < \GammaNGBWall$ and is incompatible with our assumption that $E_\text{wall} > E_\text{str}$ as demonstrated in \cref{eqn:CompareHreToGammaNGBWall}. This conclusion is also intuitive. By removing the $\sim k^{3/2}$ spectrum, we effectively demand that the cosmic belts almost do not reconnect before walls collapse so that cosmic rings of various sizes are never produced. This is only possible if cosmic belts are similar to global strings and produce string loops instead of cosmic rings. In other words, the scenario without the $k^{3/2}$ part of the spectrum requires domain walls to be a subdominant component of the energy budget of the network.

%%%%%%%%%%   Sec 4.2.3   %%%%%%%%%%
\subsubsection{GW from Other Defects}
Besides rings and disks, there are other defects in the network, such as cosmic string loops and cosmic belts. Now, we show that their contribution to the GW spectrum is negligible. We will omit these contributions when we report the benchmark gravitational-wave spectrum from walls bounded by inflated global strings. 

First, cosmic belts are already known to be a subdominant source of GWs even in the boundary gauge string case as discussed in \cref{sec:GWBelts}. Its maximal abundance should still be suppressed compared to that from cosmic disks or rings, \textit{i.e.},
\begin{equation}
    \eval{\Omega_\text{GW, belt, peak}}_{t = \GammaTot^{-1}} \approx \qty( \frac{H_\text{re}}{\GammaTot} )^{1/2} \eval{\Omega_\text{GW, ring, peak}}_{t = \GammaTot^{-1}}.
\end{equation}
This claim still holds for the case with boundary global strings with $\GammaTot \approx \max\{ \GammaWall, \GammaNGBWall \}$, and the prefactor still suppresses the belt contribution to the GW spectrum when $\GammaTot \approx \GammaNGBWall$ as shown in \cref{eqn:CompareHreToGammaNGBWall}. Hence, we may safely ignore the cosmic belt contribution regardless of whether NGB emission is significant or not.

For the string spectrum, the NGB radiation is extremely efficient in damping the energy in string loops, and the typical decay rate by the NGB radiation for global strings is roughly 
\begin{equation}
    \dv{(\mu \ell)}{t} \approx -\gamma_a v_2^2 \implies \GammaNGBStr \approx \frac{H_p}{\pi} \sim H_p.
\end{equation}
This should be contrasted with the decay rate of strings into gravitational waves, $\GammaStr \approx \mu H_p / \MPl^2$ (see \cref{appeqn:GammaGWStr}). Then, we may estimate the maximal gravitational-wave abundance emitted by global strings as 
\begin{equation}
    \Omega_\text{GW, str, peak} 
    \approx \frac{\mu H_p^2}{\MPl^2 H_p^2} \frac{\GammaStr}{\GammaNGBStr} 
    \approx \frac{\mu^2}{\MPl^4} \approx 10^5 \qty(\frac{v_2}{\MPl})^4.
\end{equation}
If we take $v_2 \approx \SI{E13}{\GeV}$, we find that the maximum gravitational-wave signal from global strings observed today should be 
\begin{equation}
    \eval{\Omega_\text{GW, global str} h^2}_{T_0} \approx 2\times 10^{-21} \qty(\frac{v_2}{\SI{E13}{\GeV}})^4.
\end{equation}
Our naïve estimate here agrees parametrically with previous studies that implemented numerical simulations of GWs from global strings \cite{Gouttenoire:2019kij, Chang:2021afa, Gorghetto:2021fsn}.%
\footnote{
The precise form of a logarithmic dependence on the horizon size is currently under debate \cite{Klaer:2017qhr, Gorghetto:2018myk, Kawasaki:2018bzv, Vaquero:2018tib, Buschmann:2019icd, Klaer:2019fxc, Gouttenoire:2019kij, Chang:2021afa, Gorghetto:2021fsn, Hindmarsh:2021vih, Hindmarsh:2021zkt}. Nonetheless, it is more or less agreed in the literature that the maximum gravitational-wave abundance observed today is at most $\Omega_\text{GW} h^2 \lesssim 10^{-21} \cdot (v_2 / \SI{E13}{\GeV})^4$.
}
This GW abundance is too small to be observed by current or near-future GW observatories, so we may safely ignore the contribution to GWs from global strings.

%%%%%%%%%%    Sec 4.3    %%%%%%%%%%
\subsection{Summary \label{sec:GWGlobalSummary}}

In recapitulation, when the boundary defect is a global string, NGB radiation becomes efficient for a small enough string-bounded wall. If $H_\text{re} \lesssim k_\text{NGB}$, the string-wall network continues to oscillate until $t = \GammaWall^{-1}$. In this case, the gravitational-wave spectrum is
\begin{equation}
    \eval{\pdv{\Omega_\text{GW}}{\ln k}}_{t = \GammaWall^{-1}} 
    \approx \frac{2\pi \sigma^{1/2}}{3 \MPl H_\text{re}^{1/2}}
    \begin{dcases}
        \qty(\frac{k}{\GammaWall})^3 \qty(\frac{\GammaWall}{H_\text{re}})^{3/2}, & k \lesssim \GammaWall, \\
        \qty(\frac{k}{H_\text{re}})^{3/2}, & \GammaWall \lesssim k \lesssim H_\text{re}, \\
        \frac{H_\text{re}}{k}, & H_\text{re} \lesssim k \lesssim k_\text{NGB}, \\
        \frac{H_\text{re}}{k} \qty(\frac{k_\text{NGB}}{k})^2, & k \gtrsim k_\text{NGB}.
    \end{dcases}
    \label{eqn:powerlawGlobal1}
\end{equation}
The maximum gravitational-wave fractional energy and characteristic peak frequency observed today remain unchanged from \cref{eqn:kPeakT0,eqn:Omegah2GWT0} so long as $H_\text{re} \lesssim k_\text{NGB}$. Here, the characteristic scale for NGB emission is 
\begin{equation}
    k_\text{NGB} \approx \SI{E-13}{\GeV}\, \qty(\frac{\vwall}{\SI{E6}{\GeV}})^3 \qty(\frac{\SI{E13}{\GeV}}{\vstr}), 
\end{equation}
or as observed today,
\begin{equation}
    \eval{f_\text{NGB}}_{T_0} \approx \SI{2E-3}{\Hz} \; \qty(\frac{106.75}{g_{*,s}(T_{\GammaWall})})^{1/3} \qty(\frac{g_{*,\rho}(T_{\GammaWall})}{106.75})^{1/4} \qty(\frac{\vwall}{\SI{E6}{\GeV}})^{3/2} \qty(\frac{\SI{E13}{\GeV}}{\vstr}). 
    \label{eqn:kNGBT0}
\end{equation}

If $k_\text{NGB} \lesssim H_\text{re}$, the maximum GW abundance is suppressed, and fewer power-law changes are present in the GW spectrum. The GW spectrum takes a characteristic power law 
\begin{equation}
    \begin{multlined}
        \eval{\pdv{\Omega_\text{GW}}{\ln k}}_{t = \GammaNGBWall^{-1}}
    \approx \frac{2\pi^{5/2} \sigma^{7/2}}{3\MPl^4 H_\text{re}^{7/2} \mu^{3/2}} \\
    \times 
    \begin{dcases}
        \qty(\frac{k}{\GammaNGBWall})^3 \qty(\frac{\GammaNGBWall}{H_\text{re}})^{3/2}, & k \lesssim \GammaNGBWall, \\
        \qty(\frac{k}{H_\text{re}})^{3/2}, & \GammaNGBWall \lesssim k \lesssim H_\text{re}, \\
        \qty(\frac{H_\text{re}}{k})^3, & k \gtrsim H_\text{re}. 
    \end{dcases}
    \end{multlined}
    \label{eqn:powerlawGlobal2}
\end{equation}
The maximal gravitational-wave fractional energy and characteristic peak frequency observed today can still be estimated similar to \cref{eqn:kPeakT0,eqn:Omegah2GWT0} by replacing $\GammaGen = \GammaNGBWall$ as 
\begin{gather}
    \begin{multlined}
        \eval{f_\text{peak}}_{T_0} 
        = \eval{f_\text{NGB}}_{T_0} \\
        \approx \SI{2E-3}{\Hz} \; \qty(\frac{106.75}{g_{*,s}(T_{\GammaTot})})^{1/3} \qty(\frac{g_{*,\rho}(T_{\GammaTot})}{106.75})^{1/4} \qty(\frac{\vwall}{\SI{E6}{\GeV}})^{3/2} \qty(\frac{\SI{E13}{\GeV}}{\vstr}), 
    \end{multlined}\\
    \eval{\Omega_\text{GW} h^2}_{T_0, f_\text{peak}} 
    \approx 10^{-11} \qty(\frac{\vwall}{\SI{E6}{\GeV}})^{21/2} \qty(\frac{\SI{E13}{\GeV}}{\vstr})^{3} \qty(\frac{\SI{E-12}{\GeV}}{H_\text{re}})^{7/2}.
\end{gather}

\begin{figure}[t]
	\centering
	\includegraphics[width=0.9\textwidth]{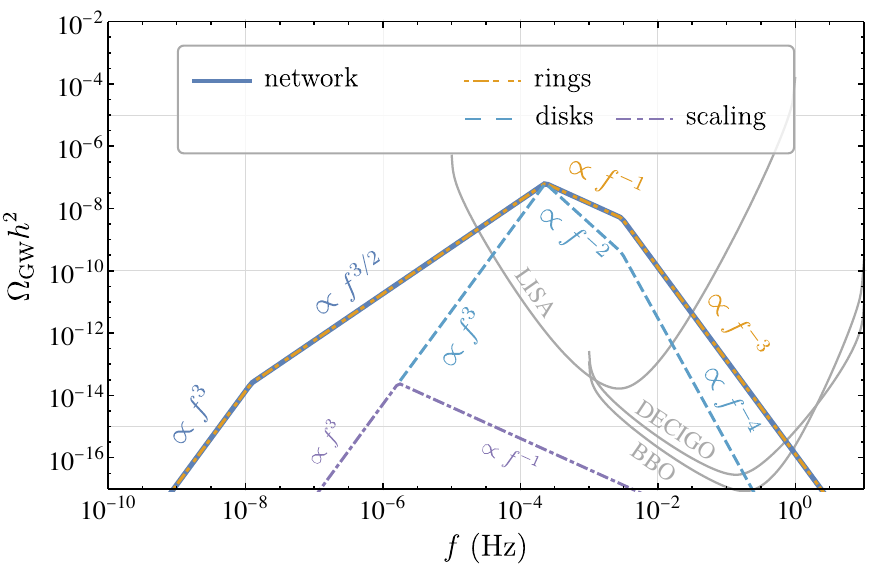}
	\caption{Decomposition of the gravitational-wave (GW) signal of walls bounded by inflated \textit{global} strings for the scale related to string tension $\vstr = \SI{E13}{\GeV}$, the scale related to wall tension $\vwall = \SI{E6}{\GeV}$, the string re-entry Hubble parameter $H_\text{re} = \SI{E-14}{\GeV}$, for which the network decay width $\GammaTot = \GammaWall \approx \SI{5.3E-19}{\GeV}$. The wall contribution to the GW spectrum still comes from three parts: rings (dot-dash-dotted yellow line), disks (dashed light-blue line), and scaling (dot-dashed violet line). The new feature, in comparison with gauge strings, is the quicker falloff in both the disk and belt spectrum in the UV. They appear because global strings can radiate Nambu-Goldstone bosons (NGB) and can further accelerate the collapse of the string-wall network when walls are sufficiently small. The full GW signal is the envelope of all these spectra (solid dark blue line). The global string considered here has a low tension and produces a GW signal that is too small to be visible by current and near-future GW observatories. Nonetheless, the change in power-law dependence in the UV part of the GW spectrum comes from the NGB emission of the boundary strings and can be used to infer the string tension.}
	\label{fig:benchmarkSignalGlobal}
\end{figure}

Now, we take a slightly modified benchmark from that in \cref{sec:GWSummary} with $H_\text{re}$ changed from $\SI{E-13}{\GeV}$ to $\SI{E-14}{\GeV}$. All other parameters, such as $\vstr = \SI{E13}{\GeV}$, $\vwall = \SI{E6}{\GeV}$, and $\GammaTot = \GammaWall \approx \SI{5.3E-19}{\GeV}$, remains unchanged. This leads to a characteristic scale of NGB emission $k_\text{NGB} \approx \SI{E-13}{\GeV}$. This scale was coincidentally close to the peak frequency (\textit{cf.} \cref{eqn:kPeakT0}) for the benchmark in \cref{sec:GWSummary}. Thus, for clarity, we choose to decrease $H_\text{re}$ by one order of magnitude so that $\eval{f_\text{peak}}_{T_0} \approx \SI{2E-4}{\GeV}$. This shows the entire $\sim k^{-1} \to \sim k^{-3}$ spectrum as shown \cref{fig:benchmarkSignalGlobal}. 

Interestingly, the scale $k_\text{NGB}$ observed today is independent of $H_\text{re}$. If a $\sim k^{-1} \to \sim k^{-3}$ transition is observed in the GW spectrum, the prominent peak can be used to infer both $\vwall$ and $H_\text{re}$ while the transition frequency, related to $k_\text{NGB}$, may be used to further extract the string tension. In other words, although the low-lying gravitational-wave signature from the strings is not detectable, solely using the wall spectrum is sufficient to determine the property of boundary strings in this scenario. On the other hand, even if $\sim k^{3/2} \to \sim k^{-3}$ spectrum is observed, hinting at $\GammaTot = \GammaNGBWall$, if the global NGB is massless or has a small mass, one may also expect it to be a component of dark radiation as well. In this scenario, while the GW spectrum near its peak cannot provide more information about the boundary string, $\Delta N_\text{eff}$ and other dark radiation searches provide a complementary reach for these inflated topological defects that decay into either GWs or very light NGBs, offering an alternative probe to the string tension. 

When the global NGB has a mass and is stable, it can be an axion-like particle dark matter. Then, the inflated string-bounded wall considered in this work could lead to another way to produce axion dark matter beyond the standard misalignment mechanism or production from defects in the minimal case, allowing for a smaller decay constant than what the minimal case predicts. Note that unlike Refs.~\cite{Baratella:2018pxi, Redi:2022llj, Harigaya:2022pjd}, where domain walls are made from the axion, the domain walls in our setup are made from a heavy field, and hence, the domain-wall energy density can be larger and produce larger GW signals without overproducing axion dark matter.

%%%%%%%%%%     Sec 5     %%%%%%%%%%
\section{GW Spectrum Benchmarks and Comparison \label{sec:GWBenchmark}}

In this section, we provide a few more benchmarks to show the generality of this mechanism in producing interesting gravitational-wave signals across a wide range of frequencies. We also compare the signal of our setup with the case without intermediate inflation and other possible stochastic GW sources.

%%%%%%%%%%    Sec 5.1    %%%%%%%%%%
\subsection{Signals from Nanohertz to Kilohertz}

\begin{table}[t]
    \centering
    \begin{tabular}{*{4}{r}|*{2}{r}}
        \hline\hline
        $\vstr\;(\text{GeV})$ & $\vwall\;(\text{GeV})$ & $H_\text{re}\;(\text{GeV})$ & $H_\text{wd}\;(\text{GeV})$ & $\eval{f_\text{peak}}_{T_0}\;(\text{Hz})$ & $\eval{\Omega_\text{GW} h^2}_{T_0, f_\text{peak}}$\\[3pt]
        \hline
        $10^{13}$ & $10^{6}$ & $10^{-13}$ & $6\times 10^{-20}$ & $2\times 10^{-3}$ & $2\times 10^{-8}$ \\
        $10^{7}$ & $2\times 10^{6}$ & $1.4\times 10^{-17}$ & $6\times 10^{-23}$ & $1.2\times 10^{-7}$ & $5\times 10^{-6}$ \\
        $10^{11}$ & $10^{8}$ & $10^{-6}$ & $6\times 10^{-14}$ & $20$ & $6 \times 10^{-9}$ \\
        $10^{13}$ & $10^{6}$ & $10^{-14}$ & $6\times 10^{-20}$ & $2\times 10^{-4}$ & $6\times 10^{-8}$ \\
        \hline\hline
    \end{tabular}
    \caption{Benchmarks of the model parameters $\{\vstr, \vwall, H_\text{re}\}$ adopted in \cref{fig:benchmarkSignal,fig:benchmarkSignalGlobal,fig:multibenchmarks}. The would-be wall domination scale $H_\text{wd}$ is also shown as a comparison with the string re-entry Hubble. We also provide the peak GW frequency and abundance as observed today for these benchmarks.}
    \label{tab:benchmarkParams}
\end{table}

Before discussing particular benchmarks, we comment on possible observational or phenomenological constraints from $\Delta N_\text{eff}$ and BBN. First, the dominant light decay products, GWs for gauge-string-bounded walls and possibly NGBs for global-string-bounded walls, should not give too much dark radiation beyond the current bound, $\Delta N_\text{eff} \lesssim 0.17$ at $1\sigma$ level from CMB and BBO \cite{Planck:2018vyg}. Second, the defects should not significantly disrupt BBN. While it is not strictly required when the domain wall energy fraction is small, $T_{\GammaTot} \approx \sqrt{\GammaTot \MPl} \gtrsim T_\text{BBN}$ guarantees that the wall network decays before BBN. It is also worth commenting that so long as the GW emission is the dominant decay channel of the string-wall network, demanding the relic gravitational-wave density to be below the $\Delta N_\text{eff}$ bound also limits the maximal abundance of these defects around BBN, assuming a radiation-dominated background from BBN to recombination. Therefore, it suffices to check the $\Delta N_\text{eff}$ bound for all benchmarks we considered below because they all predominantly decay into gravitational waves. 

Let us now revisit the scenario with $\vstr = \SI{E13}{\GeV}$, $\vwall = \SI{E6}{\GeV}$, $H_\text{re} = \SI{E-13}{\GeV}$, and $\GammaTot = \GammaWall \approx \SI{5.3E-19}{\GeV}$ that we considered in \cref{sec:GWSummary}. In this case, the wall-domination Hubble scale is around $H_\text{wd} \approx \SI{6E-20}{\GeV}$, which is 6 orders of magnitude smaller than $H_\text{re}$. This benchmark point also satisfies the observational and phenomenological constraints. First, the $\Delta N_\text{eff}$ bound is satisfied because of the small fractional energy density of the resulting GW $\eval{\Omega_\text{GW}}_{t = \GammaTot^{-1}} \approx 2\times 10^{-4}$. Second, the wall network decays around $T_{\GammaWall} \approx \SI{1}{\GeV}$, well above the typical BBN temperature $\sim \order{\SI{10}{\MeV}}$. 

Now, we turn to the potential signature of this benchmark. As shown in \cref{fig:multibenchmarks}, this benchmark (solid blue curve) produces a gravitational-wave signal across a wide range of frequency bands due to stable gauge strings, and a sharp peak is present in the spectrum due to the wall collapse after string re-entry. This spectrum is widely visible in many future gravitational-wave observatories from pulsar timing array measurements, such as Square Kilometer Array (SKA) \cite{Janssen:2014dka, Weltman:2018zrl}, to space-based observatories, such as LISA \cite{Baker:2019nia, Caldwell:2019vru}, DECIGO \cite{Kawamura:2020pcg, Isoyama:2018rjb}, and BBO \cite{Corbin:2005ny, Harry:2006fi}, to 3\textsuperscript{rd}-generation ground-based observatories, such as Einstein Telescope (ET) \cite{Punturo:2010zz, Maggiore:2019uih}, and Cosmic Explorer (CE) \cite{LIGOScientific:2016wof, Reitze:2019iox}. The power-law-integrated sensitivity curves for these observatories \cite{Schmitz:2020syl, NANOGrav:2023ctt} are shown in gray to compare with signals shown in colors. Even when NGB radiation is present to compete with the gravitational-wave production, domain walls bounded by inflated global strings (dashed blue line) can still produce gravitational waves that are detectable and have intriguing spectral shapes encoding the string tension scale as well.

By choosing different model parameters, the gravitational-wave signal in this scenario can show up in different observations. 
In \cref{fig:multibenchmarks}, we provide a few more parameter choices with the peaks of the gravitational-wave spectra centered at vastly different frequencies as shown in \cref{tab:benchmarkParams}. These parameters all satisfy the desired hierarchy as shown in \cref{eqn:paramHierarchy}, and for the benchmark with the latest string re-entry (orange line with the smallest $H_\text{re}$), we have checked that the walls decay around $T\approx \SI{35}{\MeV}$ well before BBN starts. It is interesting that this benchmark also matches decently with the observed GW spectrum by NANOGrav 15-year data release \cite{NANOGrav:2023hvm, NANOGrav:2023gor} and provides another explanation of this spectrum based on new physics.

%%%%%%%%%%    Sec 5.2    %%%%%%%%%%
\subsection{Comparison with Other Typical Stochastic GW Spectra \label{sec:otherGWSpectra}}

It is worth comparing the gravitational-wave spectrum produced by inflated string-bounded walls with that from other possible sources. For simplicity, we will use $\sim f^{3} \to \sim f^{3/2} \to \sim f^{-1}$ from walls bounded by \textit{gauge} strings as the benchmark for this discussion. It is not hard to compare the GW spectrum for the boundary \textit{global} string case. As the global string case may provide more power-law transitions, its spectrum is more distinguishable from those in the gauge string case. This is not a proof that our scenario is unique in producing such a signal. Instead, we will demonstrate that it is different from the signals from often considered benchmark scenarios.

\begin{table}[t]
    \centering
    \begin{tabular}{rcl}
		\hline\hline
		source & spectral shape & ref(s) \\
		\hline 
		gauge str. + inf. + wall & $f^{3} \to f^{3/2} \to f^{-1}$ & \cref{eqn:powerLaw} \\
        global str. ($w_i \gtrsim k_\text{NGB}$) + inf. + wall & $f^{3} \to f^{3/2} \to f^{-1} \to f^{-3}$ & \cref{eqn:powerlawGlobal1} \\
        global str. ($w_i \lesssim k_\text{NGB}$) + inf. + wall & $f^{3} \to f^{3/2} \to f^{-3}$ & \cref{eqn:powerlawGlobal2} \\
		\hline
		primordial metric perturbation & $f^{n_T} \to f^{n_T - 2}$ & \cite{Kuroyanagi:2014nba} \\
        secondary GW (log-normal $P_\zeta$) & $f^{3}\ln^2 f \to $ cutoff & \cite{Yuan:2021qgz} \\
		secondary GW (Dirac delta $P_\zeta$) & $f^{2}\ln^2 f \to $ cutoff & \cite{Yuan:2021qgz} \\
        secondary GW ($k^{n_\text{IR}} \to k^{-n_\text{UV}}$) & $f^3 \ln^2 f \to f^{-2n_\text{UV}}$ & \cite{Domenech:2021ztg} \\
        phase transition, turbulence, analytical & $f^{3} \to f^{-7/2}$ & \cite{Gogoberidze:2007an} \\
		phase transition, turbulence, numerical & $f^{1} \to f^{-8/3}$ & \cite{RoperPol:2019wvy} \\
        phase transition, sound wave & $f^9 \to f^{-3}$ & \cite{Hindmarsh:2019phv} \\
        domain wall & $f^3 \to f^{-1}$ & \cite{Hiramatsu:2013qaa} \\
        cosmic gauge string & $f^{3/2} \to f^0 \to f^{-1}$ & \cite{Cui:2018rwi} \\
        gauge string in kination domination & $f^1 \to f^{-1/3}$ bump & \cite{Co:2021lkc, Gouttenoire:2021wzu, Gouttenoire:2021jhk} \\
        supermassive black hole binary & $f^{2/3}$ & \cite{Phinney:2001di} \\
		\hline\hline
	\end{tabular}
    \caption{Summary table of a few other GW spectral shapes in comparison with this work: Various sources and their power-law dependence of GW spectra are provided with references. More discussions are offered in \cref{sec:otherGWSpectra}}
    \label{tab:comparisonWithOtherSpectra}
\end{table}

A summary of other benchmark GW spectral shapes is given in \cref{tab:comparisonWithOtherSpectra}.
Generally, cosmological sources of gravitational waves fall into four broad categories: (1) primordial tensor perturbation, (2) scalar-induced (secondary) gravitational wave from curvature perturbation, (3) phase transition, and (4) early-universe topological defects. One feature of the GW spectrum generated by inflated string-bounded walls is that their frequency dependence is rather different from these standard scenarios.

Primordial tensor perturbation typically has some small spectral tilt $n_T \sim 0$. However, due to reheating, its spectrum shape typically takes the form $f^{n_T} \to f^{n_T - 2}$ \cite{Kuroyanagi:2014nba}, distinct from our $f^3 \to f^{3/2} \to f^{-1}$ spectrum. Of course, the primordial tensor perturbations are also vanishingly small and not accessible by future gravitational wave detectors unless the inflation scale is near the current upper bound. 

Another benchmark is scalar-induced gravitational waves. The particular spectral shape depends on the choice of the primordial curvature perturbation $P_\zeta$. For instance, for a log-normal distribution $P_\zeta(k)$, the GW spectrum $\Omega_\text{GW}$ takes roughly the power law $f^3 \ln^2 f \to \text{cutoff}$; for delta-function-distributed $P_\zeta(k)$, the GW spectrum is roughly $f^2 \ln^2 f \to \text{cutoff}$; for a broken-power-law-distributed $P_\zeta(k) \sim k^{n_\text{IR}} \to k^{-n_\text{UV}}$ with $0 < n_\text{UV} < 4$ and $n_\text{IR} > 3/2$, the GW spectrum is approximately $f^3\ln^2 f \to f^{-2n_\text{UV}}$ \cite{Yuan:2021qgz, Domenech:2021ztg}. For $0 < n_\text{IR} < 3/2$, one does expect the IR tail to behave as $\sim f^{2n_\text{IR}}$ from the naïve counting of the curvature power spectrum. $n_{\rm IR}=3/4$ and $n_{\rm UV}=1/2$ can mimic the $f^{3/2} \rightarrow f^{-1}$ transition from walls bounded by strings.

As for phase transitions, its turbulence phase may produce a GW spectrum of the form $f^3 \to f^{-7/2}$\cite{Gogoberidze:2007an}\footnote{We should remark that an updated numerical study seems to provide a different power-law dependence $f^1 \to f^{-8/3}$ due to the turbulence \cite{RoperPol:2019wvy}.} and its sound waves have a GW spectral shape of $f^9 \to f^{-3}$ \cite{Hindmarsh:2019phv}. 

Lastly, various topological defects may produce interesting GW spectra. For instance, a rapid domain wall collapse can produce a GW spectrum that looks like $f^3 \to f^{-1}$ \cite{Hiramatsu:2013qaa}%
\footnote{The simulation done in Ref.~\cite{Hiramatsu:2013qaa} does not include a bias in the potential. However, a similar spectral shape is seen in a numerical study that incorporates a bias in the potential \cite{Kitajima:2023cek}.} 
while a gauge cosmic string typically has a flat spectrum ($\sim f^0$) with an IR roll-off of $f^{3/2}$ around matter-radiation equality \cite{Cui:2018rwi}. Due to the long lifetime of gauge strings, previous studies also proposed ideas to use the change in their gravitational spectral shape as a probe of the early universe dynamics \cite{Cui:2017ufi, Gouttenoire:2019kij, Chang:2021afa}. For instance, if a period of early matter and kination domination occurs, which is common from axion rotation~\cite{Co:2019wyp}, a bump of the form $f^1 \to f^{-1/3}$ on top of the flat spectrum may appear as discussed in Refs.~\cite{Co:2021lkc, Gouttenoire:2021wzu, Gouttenoire:2021jhk}. Nonetheless, these spectral shapes are distinct from our benchmark spectra. 

In addition to the cosmological sources, the supermassive black hole binary merger induces a stochastic gravitational-wave background with a power law $f^{2/3}$ \cite{Phinney:2001di}, which also differs from the IR part of our benchmark GW signal.

%%%%%%%%%%    Sec 5.3    %%%%%%%%%%
\subsection{Comparison with a Previous Study \label{sec:CompareWithPrev}}
Ref.~\cite{Dunsky:2021tih} is an enlightening study on how the interaction between boundary defects and bulk defects may affect their GW spectrum. The scenario considered in this paper is akin to the ``walls eating strings" case discussed in section VII of Ref.~\cite{Dunsky:2021tih}. The crucial difference between the previous study and this work is that we introduce a period of inflation after string formation and before wall formation. When both strings and walls are produced after inflation, the typical wall size at its formation is much smaller than the critical scale $\sim \mu / \sigma$. This means that the dominant contribution to $\Omega_\text{GW}$ comes from the boundary string. The main role of these small walls is to grow along with the horizon-sized scaling strings and pull strings together after the horizon size $H^{-1}$ exceeds the critical scale $\sim \mu / \sigma$. This is how the IR roll-off of the GW spectrum is determined by the wall dynamics in Ref.~\cite{Dunsky:2021tih}. In our case, domain walls are larger than the critical size because the boundary strings are inflated away. Domain walls overtake the network energy and produce a large gravitational-wave signal. This feature is most clearly demonstrated by comparing the blue curve and the brown curve of \cref{fig:benchmarkSignal}, showing the striking difference between this work and post-inflationary production.

It is also worth stressing that the brown curve of \cref{fig:benchmarkSignal}, labeled as ``post-inflation", is not identical to the post-inflationary production considered in Ref.~\cite{Dunsky:2021tih}. This is because whether the string spectrum is terminated by walls depends on the particular UV model as discussed in \cref{sec:GWSummary}. For the $\SO{10}$ grand unified theory considered in Ref.~\cite{Dunsky:2021tih}, cosmic strings are unstable, \textit{i.e.}, $\pi_1(\SO{10}/(\SU{3} \times \SU{2} \times \U{1}) = 0$. Thus, as walls pull boundary strings together, the entire defect network must annihilate completely, hence walls ``eating" the GW spectrum of strings. On the other hand, for the scenario considered in this work, $\pi_1(\U{1}) = \Z$ implies that a stable string configuration is possible. Hence, after walls collapse and pull in boundary strings, the formation of stable string bundles is allowed (\textit{cf.}~\cref{fig:string_wall_winding}), and they can still follow the scaling regime to produce a flat GW spectrum. The IR roll-off of the GW spectrum for the particular symmetry-breaking pattern in this work is not determined by the wall dynamics, even if both strings and walls are produced after inflation. This is another subtle difference between this work and Ref.~\cite{Dunsky:2021tih}.

At this point, one may ask if a large-wall scenario ($E_\text{wall} \gtrsim E_\text{str}$ at wall formation) can be realized without inflation. Ref.~\cite{Dunsky:2021tih} also discusses this possibility, in which the walls are formed later and have a characteristic size larger than $\sim \mu / \sigma$. In this case, there is an enhanced GW signal (see, \textit{e.g.}, fig.~16 of Ref.~\cite{Dunsky:2021tih}). This large-wall scenario is analogous to our contribution from cosmic disks, having a $\sim f^{3}$ part in the IR and a UV falloff similar to that of a string.%
\footnote{
The spectral shape in the UV $\sim f^{-1/3}$ in Ref.~\cite{Dunsky:2021tih} is shallower than our $\sim f^{-1}$ because they considered the possible enhancement of the GW spectrum in the higher harmonics from cusps on the string. The precise contribution from higher-order harmonics is beyond the scope of this work, so we drop this potential contribution when evaluating the GW spectrum. See \cref{sec:GWSummary} for more discussions.
}
However, this enhanced spectral peak appears at very high frequency $\sim \SI{E11}{\Hz}$ 
in the parameter region with an appreciable amount of GWs. With a period of inflation, cosmic strings are ``diluted" so that the network produces, somewhat counterintuitively, a larger gravitational-wave signal. The characteristic size of the network is set by $H_\text{re}$ which is less constrained than the scenario considered in Ref.~\cite{Dunsky:2021tih}. In addition, our study incorporated an improved analysis of the network reconnection effects from cosmic rings. We will also argue in the next section that inflation is not only preferred but also almost inevitable if domain walls are to produce large gravitational-wave signals.

%%%%%%%%%%    Sec 5.4    %%%%%%%%%%
\subsection{Inevitability of Inflation for Large GW Signal from Walls \label{sec:InevitableInflation}}
In this section, we will show that in order to produce large GW signals from domain walls during a radiation-dominated epoch, a period of inflation before its production is inevitable. Here, we will assume that 1) the network almost entirely collapses into gravitational waves $\GammaTot = \GammaWall$ at $T=T_{\GammaTot}$ to maximize gravitational-wave production, and 2) the defects are in a radiation-dominated background from their formation to decay.

Since all energy density stored in the wall network $\rho_{\rm wall}$ decays into gravitational waves, the maximal gravitational-wave abundance equals to the maximum fractional energy density of domain walls, \textit{i.e.}, $\max_{k}\{ \Omega_\text{GW} \} \approx \max_{t}\{ \Omega_\text{wall} \}$. Then, we may estimate the maximum GW abundance as 
\begin{equation}
    \Omega_\text{GW, peak} 
    \approx \frac{\sigma H_p}{\MPl^2 \GammaWall^2} \qty(\frac{T_{\GammaWall}}{T_p})^{3} 
    \approx \qty(\frac{v_1}{\MPl})^{1/2} \frac{\sqrt{m_1 v_1}}{T_p},
    \label{eqn:OmegaGWPeakForInevitability}
\end{equation}
in which we used $\GammaWall \approx \sigma / \MPl^2$ (predominant GW decay), $H_p \approx T_p^2/\MPl$ (production during radiation domination), $T_{\GammaWall} \approx \sqrt{\GammaWall \MPl}$ (collapse during radiation domination), $\sigma \approx m_1 v_1^2$ (typical wall tension as a function of the VEV $v_1$ and mass $m_1$ of the field $\phi_1$ comprising the walls), and $\rho_{\text{wall}}\propto T^3$ (walls bounded by strings dilute as matter). Unless $v_1 \sim \MPl$, the first factor typically suppresses the GW spectrum. It is, therefore, necessary to demand that 
\begin{equation}
    T_p \ll \sqrt{m_1 v_1}
\end{equation}
in order to obtain a large GW signal. However, the typical energy density of $\phi_1$ before the domain wall production should be $ \rho_1 \sim m_1^2 v_1^2$.%
\footnote{
Here, we implicitly assumed that there is no additional mechanism to dial the potential energy of $\phi_1$ so that its energy density at $\phi_1 =0$ is at least from its potential energy $\rho_1 \sim m_1^2 v_1^2$. The cancellation of $\rho_1$ will generically require fine tuning.
} 
This means that demanding a large gravitational-wave signal from domain walls implies $\rho_\text{rad} \sim T_p^4 \ll m_1^2 v_1^2 \sim \rho_{1}$; $\phi_1$ needs to have a much larger energy density than the radiation bath, implying inflation by vacuum-energy domination. If $v_1 \sim \MPl$ instead, $\rho_1$ may be smaller than $\rho_{\rm rad}$, and inflation does not seem to be required for large GW signals. But the reheating by the dissipation of $\rho_1$ is generically not efficient because of the Planck-suppressed dissipation rate of $\phi_1$. As a result, there will be a prolonged period of matter domination by $\phi_1$, violating our assumption that defects are formed and decay in a radiation-dominated background, and their GW signal will be diluted. In the next section, we will present a concrete model of inflation and revisit this point on the inevitability of inflation.

%%%%%%%%%%     Sec 6     %%%%%%%%%%
\section{Model Realization: Thermal Inflation \label{sec:ThermalInflation}}
So far, we have discussed the evolution of an inflated string-bounded wall network. This mechanism is generally applicable so long as a brief period of inflation appears between two phase transitions. Note that $H_\text{re}$, $\vstr$, and $\vwall$ are essentially free parameters coming from different physics. In this section, we further restrict parameters in the mechanism considered and propose a concrete model relating $H_\text{re}$ to the mass of the field producing domain walls. Interestingly, measuring the gravitational-wave signal provides a novel probe to model parameters, such as the soft supersymmetry (SUSY) breaking scale.%
\footnote{
There are other ideas to use gravitational-wave signals to probe the soft SUSY breaking scale, such as those discussed in Refs.~\cite{Takahashi:2008mu, Kamada:2014qja}.
}

We consider a scenario involving thermal inflation \cite{Yamamoto:1985rd, Lazarides:1985ja, Lyth:1995hj, Lyth:1995ka} driven by inflaton $\phi_{1}$, which is also the field producing domain walls. The crucial mechanism here is the interplay between a flat potential with a tachyonic instability near $\phi_1 = 0$ and the thermal mass of $\phi_1$. Specifically, let us consider the following potential
\begin{equation}
	V(\phi_1) = - m_1^2 \abs{\phi_1}^2 + V_{\text{lift}}(\phi_1),
\end{equation}
in which $m_1$ is a small mass parameter and $V_\text{lift}(\phi_1)$ denotes some lifting potential. Taking $m_1^2 >0$, $\phi_1$ has a tachyonic mass at the origin, but the potential is eventually stabilized by $V_{\text{lift}}(\phi_1)$ at $\phi_1 = v_1$. The energy difference $\Delta V $ between $\phi_1 = 0$ and $\phi_1 = v_1$ is of order $ m_1^2 v_1^2$. Thus, a large $v_1$ leads to large energy for $\phi_1$ if it is trapped at the origin. To trap $\phi_1$, one can consider some coupling between $\phi_1$ and the thermal bath. Then, thermal correction to the $\phi_1$ potential leads to a positive mass term of the form $V \supset y^2 T^2 \abs{\phi_1}^2$, in which $y$ denotes some dimensionless coupling, and $T$ is the temperature. The thermal mass compensates for the tachyonic mass of $\phi_1$ and traps $\phi_1$ near the origin so that $\phi_1$ can dominate the universe and drive a brief period of inflation. This happens when the radiation density drops below the energy density in $\phi_1$ at a temperature $T_i\approx \sqrt{m_1 v_1}$. On the other hand, cosmic expansion cools the universe, and $\phi_1$ can see its vacuum potential when the temperature drops below $T_f \lesssim m_1 /y$. This terminates the thermal inflation. Hence, the number of $e$-foldings for this thermal inflation can be estimated by
\begin{equation}
	N_\text{inf} \approx \ln(\frac{T_i}{T_f}) \approx \frac{1}{2} \ln(\frac{y^2 v_1}{m_1}). 
\end{equation}
Using \cref{eqn:HreEfficientReheat}, we determine the re-entry Hubble for cosmic strings as 
\begin{equation}
	H_\text{re} \approx e^{-2N_\text{inf}} H_i \approx \frac{1}{y^2} \frac{m_1^2}{\MPl}
\end{equation}
in terms of the mass $m_1$. The wall tension $\sigma$ comes from the trilinear coupling between $\phi_1$ and $\phi_{2}$ and is irrelevant for the thermal inflation dynamics. We will thus treat $\vwall$ as an independent model parameter. Then, it is possible to probe $m_1/y$ and $\vwall$ via its gravitational-wave signature. 

The thermal inflation scenario provides a concrete example of the connection between the a stage of inflation before the second phase transition and the size of the gravitational wave signal. From the discussion in the previous paragraph, we see that a long period of thermal inflation is naturally realized when $T_i \gg T_f$ or $y^2 v_1 / m_1 \gg 1$. Recall from \cref{eqn:OmegaGWPeakForInevitability} that a lower domain-wall production temperature $T_p$ is preferred. In the context of thermal inflation, this occurs at the end of thermal inflation when $\phi_1$ rolls to its true vacuum, \textit{i.e.}, $T_p \sim T_f \sim m_1 / y$. Thus, $T_p \ll \sqrt{m_1 v_1}$ means $y v_1^{1/2} / m_1^{1/2} \gg 1$, which coincides with the condition for a period of inflation before phase transition. On the other hand, the condition $y v_1^{1/2} / m_1^{1/2} \gg 1$ means $v_1 \gg m_1$. A parametrically large $v_1$ typically means that the lifting potential is sufficiently flat. Some examples of a flat lifting potential include higher-dimensional operators and a SUSY Coleman-Weinberg potential. 

A flat potential is generally destroyed by the quartic coupling of $\phi_1$ generated by the renormalization group running $\sim y^4$. To protect the flatness of the potential, we consider the following supersymmetric theory with 4 chiral superfields charged under some $\U{1}$ symmetry. They are $\phi_1$ with charge $1$, $\phi_2$ with charge $2$, $\phi_{-2}$ with charge $-2$, and $\psi$ with charge $-1$. We also introduce two more fields $X$ and $\bar{\psi}$ that are not charged under $\U{1}$. Let us consider the following superpotential
\begin{equation}
	W = X \qty(\phi_2 \phi_{-2} - v_2^2) + \lambda \phi_1^2 \phi_{-2} + y \phi_1 \bar{\psi} \psi. 
\end{equation}
The $F$ term of $X$ fixes $\phi_2$ and $\phi_{-2}$ on the $F$-flat direction $\phi_2 \phi_{-2}= v_2^2$. This breaks the $\U{1}$ symmetry spontaneously and produces cosmic strings. Soft SUSY breaking, parameterized by the scale $\msoft$, lifts the $F$-flat direction and stabilize $\phi_{2}\approx \phi_{-2} \approx v_2$. When the $\U{1}$ symmetry is gauged, the $D$ term potential dominates over the soft SUSY breaking, and it also stabilizes $\phi_{2} \approx \phi_{-2} \approx v_2$. The same SUSY breaking can also radiatively generate a flat potential for $\phi_1$ through the Yukawa coupling $y \sim \order{1}$ to $\psi$ and $\bar{\psi}$ by
\begin{equation}
 V_{\cancel{\text{SUSY}}, \abs{\phi_1}^2} \approx - \msoft^2 \abs{\phi_1}^2 + \frac{y^2 \msoft^2}{\qty(4\pi)^2} \abs{\phi_1}^2 \ln(\frac{\abs{\phi_1}^2}{\Lambda^2}),
\end{equation}
in which $\Lambda$ denotes a reference energy scale. This flat potential gives a VEV of $\phi_1$ much larger than $\msoft$. Therefore, we may treat the VEV of $\phi_1$ as an arbitrary parameter satisfying $\msoft \ll v_1 \ll v_2$. At the same time, when $\phi_1$ is near the origin, $\psi$ becomes a light field in the thermal bath and can give $\phi_1$ a thermal mass term $\approx y^2 T^2 \abs{\phi_1}^2$
that stabilizes $\phi_1$ around the origin at high temperatures. Then, using our conclusions from the previous paragraph, we determine that the string re-entry Hubble scale is 
\begin{equation}
	H_\text{re} \approx \frac{\msoft^2}{\MPl}. 
\end{equation}

As for the wall tension, soft SUSY breaking also induces a trilinear coupling of the form
\begin{equation}
	V_{\cancel{\text{SUSY}}, \text{tri.}} \approx -\msoft \lambda \qty(\phi_1^2 \phi_{-2} + \text{h.c.}) \approx -\msoft \lambda v_1^2 v_2 \cos(\frac{2a_1}{v_1}).
\end{equation}
Then, the angular direction of $\phi_1$ obtains a cosine potential, and domain walls are formed between the two degenerate vacua with a mass $m_a \sim \sqrt{\lambda \msoft v_2}$. This give rises to a wall of tensions $\sigma \sim m_a v_1^2 \sim \sqrt{\lambda \msoft v_2} v_1^2$. Note that while minimizing the potential for $\phi_1$, we ignored the trilinear term; hence, to maintain consistency, we must demand an upper bound on $\lambda \lesssim \msoft/v_2$.%
\footnote{
While the trilinear coupling is numerically small according to this bound, its small value is technically natural. Alternatively, one may consider a theory with $\phi_{1}$ and $\phi_{\pm 3}$ and a higher-dimensional operators $W\supset \phi_1^3 \phi_{-3}$ to generate a naturally small domain wall tension.
}
Instead of $\lambda$, $v_1$, and $v_2$, we use the wall tension scale $\vwall$ and soft SUSY breaking scale $\msoft$ as model parameters in presenting our results.

\begin{figure}[t]
	\centering
	\includegraphics[width=0.9\textwidth]{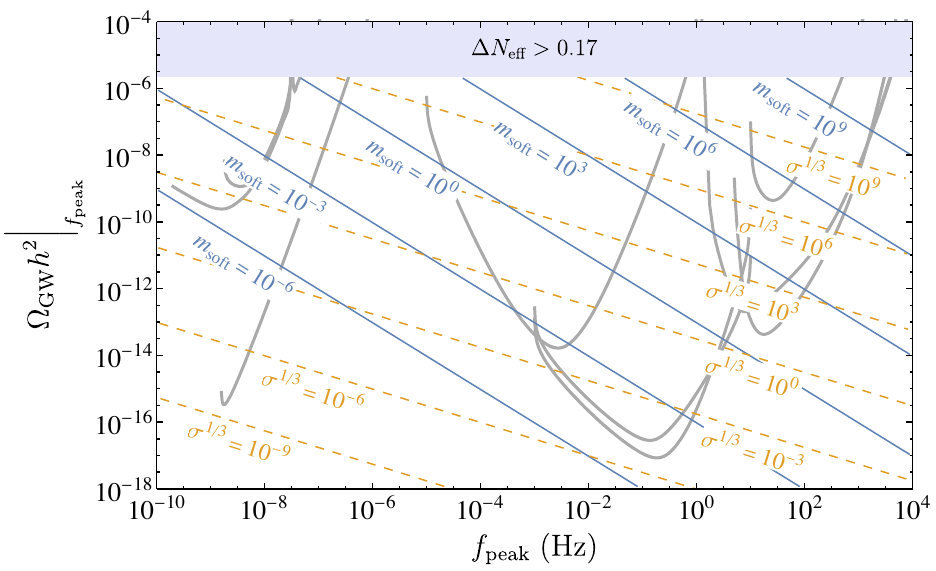}
	\caption{ Maximal gravitational-wave abundance and peak frequency observed today as functions of the model parameters in the thermal inflation model $\msoft$ and $\vwall$ in units of GeV: Blue solid lines show $\eval{(f_\text{peak}, \eval{\Omega_\text{GW} h^2}_{f_\text{peak}})}_{T_0}$ for fixed $\msoft$ with varying $\vwall$, and the orange dashed curve is that for fixed $\vwall$ with varying $\msoft$. The values of $\msoft$ and $\vwall$ are shown near each line. Parameter space with too much gravitational wave is shown as the shaded region. In plotting this figure, we assumed instantaneous reheating. This turns out to be possible for most of the parameter space considered except for a small region. More detailed discussions can be found in \cref{app:inefficientReheat}. 
    }
	\label{fig:thermalInflationParam}
\end{figure}

Let us now focus on the gauged $\U{1}$ case such that walls entirely decay into gravitational waves, \textit{i.e.}, $\GammaTot = \GammaWall(\sigma)$, for which the maximal fractional energy $\Omega_\text{GW} h^2$ and peak frequency $f_\text{peak}$ today are determined by $\msoft$ and $\vwall$. In particular, the dependence of the peak frequency observed today in terms on the thermal inflation parameters is 
\begin{equation}
	\eval{f_\text{peak}}_{T_0} \propto \msoft^2 \sigma^{-1/2},
\end{equation}
whereas the maximal abundance 
\begin{equation}
	\eval{\Omega_\text{GW} h^2}_{T_0, f_\text{peak}} \propto \sigma^{1/2} \msoft^{-1}. 
\end{equation}
Due to the different power-law dependence, upon observing this particular GW spectrum, the peak frequency and maximal abundance point to a particular combination of $\msoft$ and $\vwall$. This is shown in \cref{fig:thermalInflationParam}. Interestingly, the SUSY breaking scale $\msoft$ can be probed by near-future gravitational-wave detectors for reasonably large wall tension. We note that we assumed instantaneous reheating. If a significant matter-dominated epoch follows the thermal inflation, the string re-entry Hubble size $H_\text{re}$ becomes smaller than that estimated here, leading to a larger $\Omega_\text{GW} h^2$. More remarks regarding non-instantaneous reheating are provided in \cref{app:inefficientReheat}.

%%%%%%%%%%     Sec 7     %%%%%%%%%%
\section{Conclusion and Discussions \label{sec:conclusion}}

In this paper, we discussed how domain walls bounded by inflated cosmic strings can produce large gravitational-wave (GW) signals. The mechanism relies on two phase transitions, the first one producing cosmic strings and the second one producing domain walls, and an epoch of inflation between them. This inflation can be either the primordial cosmic inflation or a second stage of inflation. The crucial mechanism here is that while the domain walls may be topologically unstable, they remain dynamically stable so long as no cosmic string terminates their boundaries. In other words, although $G \to H \to K$ is the full symmetry-breaking pattern, because boundary defects are inflated away, each local Hubble patch does not see cosmic strings and cannot appreciate the full breaking pattern. The breaking $H \to K$ then appears to produce stable domain walls until boundary defects associated with $G\to H$ re-enter the horizon. After string re-entry, the full pattern emerges, and the network starts to annihilate. We demonstrated this with a simple two-field model as shown in \cref{eqn:toyModel} that follows a breaking pattern $\U{1} \to \Z[2] \to \emptyset$. 

At first, we gauged $\U{1}$ to suppress the energy loss due to the radiation of Nambu-Goldstone bosons (NGBs) from boundary strings. In particular, we considered the case where the walls' total energy is larger than the strings' energy when the strings re-enter the horizon around $t \approx H_\text{re}^{-1}$. In this case, the string-wall network undergoes three stages of evolution: 1) walls enter a scaling regime while strings are frozen outside the horizon, 2) the string-wall network begins oscillations when cosmic strings re-enter, and 3) the network rapidly decays. We estimated the gravitational-wave spectrum from domain wall dynamics in each of these stages and provided the power-law dependence (\cref{eqn:powerLaw}) and the approximate peak frequency (\cref{eqn:kPeakT0}) of the spectrum, as well as the maximal gravitational-wave abundance (\cref{eqn:Omegah2GWT0}) as observed today. A particular feature of gauging the $\U{1}$ symmetry here is the existence of gauge strings. These gauge strings produce more gravitational-wave signals than their global string counterparts due to the absence of the decay channel into NGBs. This leads to the flat shoulder below the sharp domain wall peak as shown in \cref{fig:benchmarkSignal}. 

We then discussed the case in which the $\U{1}$ symmetry is not gauged and show that it can still lead to novel gravitational-wave signals in \cref{sec:GWGlobalSignature}. The radiation into NGBs changes the spectrum from both the cosmic disks (walls bounded by inflated string loops) and cosmic rings (string-bounded walls formed due to the reconnection of cosmic belts). These can produce extra power-law suppressions to the spectrum, and the frequencies at which the power-law dependence changes (\cref{eqn:powerlawGlobal1,eqn:powerlawGlobal2}) can be used to infer the string tension scale in spite of the low-lying GW spectrum from strings.

We provided a few more benchmark spectra in \cref{sec:GWBenchmark}. Since there needs not be a relation between the wall tension $\sigma$, the string tension $\mu$, and the string re-entry Hubble scale $H_\text{re}$, the gravitational-wave signature from these inflated string-bounded walls can peak at a wide range of frequencies as shown in \cref{fig:multibenchmarks}. We also compared the GW spectrum in our scenario with other GW spectra by various cosmological or astrophysical sources and showed that our GW spectrum has a distinct spectral shape. A comparison between this study and a previous study is made in \cref{sec:CompareWithPrev}. We then argued in \cref{sec:InevitableInflation} that to have large gravitational-wave signals from domain walls, a period of inflation before their production, as considered in this work, seems to be inevitable. 

In \cref{sec:ThermalInflation}, we discussed a concrete model relating $H_\text{re}$ to the mass of the wall-producing field $\phi_1$ by thermal inflation. In this model, only two free parameters $(m_1, \vwall)$ appear, and they can be constrained by the peak frequency and amplitude of the GW signal. This model is more naturally realized with SUSY, and observing this GW signal can probe the soft SUSY breaking scale $\msoft$ along with the wall tension scale $\vwall$ as shown in \cref{fig:thermalInflationParam}.

One of the most exciting directions for future studies is to cross-check the estimated spectra presented in this paper against numerical simulations. In particular, the following points should be checked:
\begin{itemize}[leftmargin=15pt]
    \item Cosmic belts are assumed to reconnect efficiently and follow a scaling law. Is this a reasonable assumption? What are the dominant configurations of the resulting breakoff topological defects such as cosmic rings?
    \item We assumed that string-wall defects mainly oscillate to dissipate energy into gravitational waves or NGBs. However, self-intersection during such oscillations may also provide additional damping on the system. It would be interesting to check if such an effect is important and how it may modify the GW spectrum. 
    \item Given the scaling regime of the cosmic-belt network, we assume that cosmic rings produced from these connections have a size controlled by $H_p^{-1}$, and the distribution of the size is logarithmically even in the range of $ H_\text{re}^{-1} < H_p^{-1} < \GammaTot^{-1}$. Are the size indeed $\sim H_p^{-1}$, and are the upper and lower bounds for the size reasonable?
    \item During reconnection of the inflated string-bounded wall network, small structures on the defects, such as kinks and cusps, may also produce bursts of gravitational waves or NGBs. How significant are these contributions in the scenario proposed in this paper? Also, will small structures on the wall modify the UV part of the GW spectrum as discussed in \cref{sec:GWSummary}? Our naïve model only takes into account GW radiation in the fundamental mode. It would be interesting to see whether higher-order harmonics due to the oscillation of the defect network alter the spectrum significantly. 
    \item Upon reconnection, cosmic strings may bring domain walls very near each other. This may prompt the formation of Y-junction-like structures on the cosmic belt. It is possible that these Y-junction-like structures can zip the cosmic belts. In this study, we implicitly assumed that the violent oscillation of the domain wall prevents this zipping process from destroying the cosmic belt network. Does this emerge in numerical simulations? If so, under what conditions is this a good assumption?
    \item We also assumed that inflated cosmic strings re-enter around the same cosmic scale $H_\text{re}$. Is this true if the phase transition happens during inflation? As for strings reaching the scaling regime before they are inflated away, it is likely that this characteristic scale $H_\text{re}$ follows some distribution. This could smear the spectrum as $H_\text{re}$ takes various possible values in different Hubble patches. In this case, is it still reasonable to talk about some characteristic scale $H_\text{re}$, or is the smearing effect too strong?
\end{itemize}

There are possible future studies besides numerical simulation.
It would be interesting to see if more complicated symmetry-breaking patterns can result in model-dependent features in the GW spectrum. For instance, one of the best-motivated global $\U{1}$ symmetry would be the PQ symmetry, for which the NGB radiation may be affected by the non-zero mass of the QCD axion due to the explicit PQ breaking by QCD anomaly. The axions radiated from the string-wall network may be dark matter. As these walls once occupied a significant portion of the Hubble patch, it would also be interesting to ask how they might impact the matter power spectrum, and whether smaller-scale structures seeded by these large walls can be formed and detected. If these defects couple to the Standard Model, it may also be worth investigating whether current or near-future collider and beam dump experiments can probe these couplings and what novel cosmological signature they may induce such as that proposed in \cite{Long:2014mxa, Long:2014lxa}. If inflation happens between phase transitions that produce other kinds of topological defects, such as monopoles and strings, their GW signal may also beg for further investigation. Lastly, in our mechanism, the causality plays an important role in the evolution of topological-defect networks and producing a feature in the GW spectrum. It, hence, would be enlightening to learn if such mechanism can be applied in other contexts.

%%%%%%%%%%Acknowledgments%%%%%%%%%%
\acknowledgments
We thank David Dunsky and Andrew Long for the enlightening comments on our draft and Daniel Figueroa, Sungwoo Hong, Sung Mook Lee, Hitoshi Murayama, and Jun'ichi Yokoyama for discussions. The work of K.H. was supported by Grant-in-Aid for Scientific Research from the Ministry of Education, Culture, Sports, Science, and Technology (MEXT), Japan (20H01895), by World Premier International Research Center Initiative (WPI), MEXT, Japan (Kavli IPMU), and by the Department of Energy grant DE-SC0025242. Y.B.~and L.T.W.~are supported by the Department of Energy grant DE-SC0013642.

%%%%%%%%%%%%%%%%%%%%%%%%%%%%%%%%%%%
%%%%%%%%%%    Appendix   %%%%%%%%%%
%%%%%%%%%%%%%%%%%%%%%%%%%%%%%%%%%%%
\appendix

%%%%%%%%%%     App A     %%%%%%%%%%
\section{Evolution of Domain Wall from Nambu-Goto Action \label{app:closedDomainWall}}
In this appendix, we discuss how we estimate the domain wall evolution using the Nambu-Goto action and justify our claim that when $E_\text{wall} > E_\text{str}$, the characteristic frequency of the wall dynamics is set by $k \sim r_0^{-1}$ in which $r_0$ is its initial size. For a semi-classical string and wall, its Nambu-Goto action reads
\begin{equation}
	S = - \sigma \int \dd^3 \zeta\; \sqrt{\gamma} - \mu \int \dd^2 \zeta\; \sqrt{-\Upsilon},
\end{equation}
in which $\zeta$ denotes the world-volume coordinates on the wall, $\gamma$ and $\Upsilon$ are the determinants of the induced metric on the wall and string respectively, defined as 
\begin{equation}
	\gamma \defeq \abs{ g_{\mu\nu} \pdv{X^\mu}{\zeta^a} \pdv{X^\nu}{\zeta^b} },
\end{equation}
in which the $g_{\mu\nu}$ is given by the FLRW metric $\dd s^2 = \dd t^2 - a(t)^2 \dd \vb{r}^2$. 

Considering a hemispherical wall bounded by a string with a radius $r$, its worldvolume is most conveniently parameterized as $X^\mu = (t, r(t), \theta, \phi)$ in spherical coordinates with
\begin{equation}
	\pdv{X^\mu}{t} = \qty(1, \dot{r}, 0, 0)^\mu, \quad \pdv{X^\mu}{\theta} = \qty(0, 0, 1, 0)^\mu, \quad \pdv{X^\mu}{\phi} = \qty(0, 0, 0, 1)^\mu, 
\end{equation}
while fixing the gauge of the worldvolume coordinate on the wall to be $\zeta^a = \qty(t, \theta, \phi)^a$. The effective action becomes 
\begin{equation}
	S = -2\pi \int \dd t\; \sigma a^2 r^2 \sqrt{1 - a^2 \dot{r}^2} + \mu a r \sqrt{1 - a^2 \dot{r}^2}. 
\end{equation}
Assuming that $E_\text{str} \ll E_\text{wall}$, we may drop the term proportionate to $\mu$ governing the string dynamics and find that the variation of the action by $r \to r + \delta r$ is 
\begin{equation}
	\qty(-2\pi \sigma)^{-1} \delta S = \int \dd t\; \delta r \qty[2a^2 r \sqrt{1 - a^2 \dot{r}^2}] - \delta \dot{r} \qty[ \frac{a^4 r^2 \dot{r}}{\sqrt{1 - a^2 \dot{r}^2}} ].
\end{equation}
In the flat spacetime limit with $a(t) = 1$, this reduces to an equation of motion for $r(t)$ 
\begin{equation}
	\dv{t}( \frac{r^2 \dot{r}}{\sqrt{1 - \dot{r}^2}} ) = - 2 r \sqrt{ 1 - \dot{r}^2 },
\end{equation}
which can be solved by 
\begin{equation}
	r(t) = r(0) \cdot \text{cd}\qty(\frac{t}{r(0)},  -1),
\end{equation}
in which $\text{cd}(u, m)$ denotes the Jacobi elliptic function. This function has an angular frequency $\omega = 2\pi \qty(4 K(-1) r(0))^{-1} \approx 1.2 r(0)^{-1}$, in which $K(m)$ denotes the complete elliptical integral of the first kind. Hence, one can see that the wall radius oscillates with a characteristic frequency comparable to $k \sim r(0)^{-1}$. 

Alternatively, one may consider a cylindrical wall of height $L$ bounded by a circular string of a radius $r$ at the endpoints such that the wall's world-volume is parameterized as $X^\mu = (t, r(t), \theta, z)$ in cylindrical coordinates. Then, by choosing $\zeta^a = \qty(t, \theta, z)^a$ on the worldvolume, the effective Lagrangian becomes 
\begin{equation}
	S = -2\pi \int \dd t\; \sigma a^2 r L \sqrt{1 - a^2 \dot{r}^2} + \mu a r \sqrt{1 - a^2 \dot{r}^2},
\end{equation}
and the equation of motion on the flat background, 
\begin{equation}
	\dv{t}(\frac{r \dot{r}}{\sqrt{1 - \dot{r}^2}}) = - \sqrt{1 - \dot{r}^2},
\end{equation}
can be solved by 
\begin{equation}
	r(t) = r(0) \cos(\frac{t}{r(0)}),
\end{equation}
demonstrating again that the wall oscillates at a characteristic frequency $k \sim r(0)^{-1}$.

%%%%%%%%%%     App B     %%%%%%%%%%
\section{Details for the Computation of the Gravitational Wave Spectrum \label{app:spectrumComputation}}
In this appendix, we will provide some more details on how the gravitational-wave spectrum was obtained in \cref{sec:GWSignature,sec:GWGlobalSignature}. 

When the wall is still in the scaling regime ($t \lesssim H_\text{re}^{-1}$) as discussed in \cref{sec:GWScalingWall}, we may solve the Boltzmann equation at $t = H_\text{re}^{-1}$ as
\begin{equation}
	\eval{\pdv{\rho_\text{GW, scaling}}{\ln k}}_{t=H_\text{re}^{-1}} = \int_{\MPl v_2^{-2}}^{H_\text{re}^{-1}} \dd t\; \qty(\frac{a(t)}{a(H_\text{re}^{-1})})^4 n(t) \pdv{P(t)}{\ln k}. 
\end{equation}
For simplicity, we will assume that the second phase transition happens sufficiently early compared to the string re-entry time so that the lower bound of the integral can be effectively taken to be zero. Numerically, as long as the two timescales are separated by more than one order of magnitude, the result will not change much as the redshift factor $\propto a^4$ peaks at later time $t\sim H_\text{re}^{-1}$. Then, by assuming that $n(t) \sim \rho_\text{wall}/E_\text{wall} \sim H^{-3}$, one obtains 
\begin{equation}
	\begin{multlined}
		\eval{\pdv{\rho_\text{GW, scaling}}{\ln k}}_{t=H_\text{re}^{-1}} \approx \frac{\pi \sigma^2 C}{\MPl^2} \int_{0}^{H_\text{re}^{-1}} \dd t\; \frac{\qty(H_\text{re} t)^{2}}{t} \frac{\qty(k \sqrt{t/H_\text{re}})^3}{1 + \qty(k \sqrt{t/H_\text{re}})^4} \\
		= \frac{2\pi \sigma^2 C}{7\MPl^2} \qty( \frac{k}{H_\text{re}} )^3 \pFq{2}{1}{1, 7/4}{11/4}{-\frac{k^4}{H_\text{re}^4}}.
	\end{multlined}
\end{equation}
Here, the essential power law can be captured by expanding in small and large $k$ limits, which we find to be 
\begin{equation}
	\begin{gathered}
		\lim_{k \to 0} \eval{\pdv{\rho_\text{GW, scaling}}{\ln k}}_{t=H_\text{re}^{-1}} = \frac{2\pi \sigma^2 C}{7\MPl^2} \qty(\frac{k}{H_\text{re}})^3, \\
		\lim_{k \to \infty}	\eval{\pdv{\rho_\text{GW, scaling}}{\ln k}}_{t=H_\text{re}^{-1}} = \frac{2\pi \sigma^2 C}{3\MPl^2} \qty(\frac{k}{H_\text{re}})^{-1}.
	\end{gathered}
\end{equation}
As the IR spectrum is more universal, we will focus on reproducing the IR spectrum while making a power law approximation for this spectrum as 
\begin{equation}
	\eval{\pdv{\rho_\text{GW, scaling}}{\ln k}}_{t = H_\text{re}^{-1}} \approx \frac{2\pi \sigma^2 C}{7\MPl^2} 
	\begin{dcases}
		\qty(\frac{k}{H_\text{re}})^3, & k \leq H_\text{re}, \\
		\qty(\frac{H_\text{re}}{k}), & k > H_\text{re}.
	\end{dcases}
\end{equation}
After $t \approx H_\text{re}^{-1}$, the wall network becomes a string-wall network and deviates from the scaling regime. This contribution to the gravitational-wave energy density redshifts with radiation. The fractional density at $t = \GammaGen^{-1}$ is 
\begin{equation}
	\eval{ \pdv{\Omega_\text{GW, scaling}}{\ln k} }_{t = \GammaGen^{-1}} \approx \frac{2\pi \sigma^2 C}{21\MPl^4 H_\text{re}^2} 
	\begin{dcases}
		\qty(\frac{k}{\sqrt{H_\text{re}\GammaGen}})^3, & k \leq \sqrt{H_\text{re} \GammaGen}, \\
		\qty(\frac{\sqrt{H_\text{re}\GammaGen}}{k}), & k > \sqrt{H_\text{re} \GammaGen},
	\end{dcases}
\end{equation}
in which we take into account the redshift in frequency. 

Now, we turn our attention to cosmic disks. After the cosmic string re-enters the horizon, the network no longer grows but only oscillates for $H_\text{re}^{-1} \lesssim t \lesssim \Gamma^{-1}$ (see \cref{sec:GWOscillatingWall} for more details). Here, we obtain the gravitational-wave density by integrating 
\begin{equation}
	\eval{\pdv{\rho_\text{GW, osc.}}{\ln k}}_{t = \GammaGen^{-1}} 
    = \int_{H_\text{re}^{-1}}^{\GammaGen^{-1}} \dd t\; \qty(\frac{a(t)}{a(\GammaGen^{-1})})^4 n(t) \pdv{P(t)}{\ln k}, 
    \label{appeqn:rhoGWOsc}
\end{equation}
with the number density $n$ given by 
\begin{equation}
    n(t) \sim H_\text{re}^{3} \qty(\frac{a(H_\text{re}^{-1})}{a(t)})^3 = \qty(\GammaGen H_\text{re})^{3/2} \qty(\frac{a(\GammaGen^{-1})}{a(t)})^3,
    \label{appeqn:nOscillatingWall}
\end{equation}
and power spectrum $\pdv*{P}{\ln k}$ given by \cref{eqn:PowerGWOscillatingWall}. We shall also approximate the power spectrum as if all the gravitational-wave energy is dumped into the fundamental mode, which has a frequency comparable to the oscillation frequency of the wall. This means that the power spectrum will roughly have a Dirac delta peak around this frequency, and integrating this over a finite range of cosmic time $t$ leads to a Heaviside step function that marks the UV and IR cutoff of the spectrum during this period, \textit{i.e.}
\begin{equation}
    \eval{\pdv{\rho_\text{GW, osc.}}{\ln k}}_{t = \GammaGen^{-1}} = \frac{2\pi \sigma^2}{\MPl^2} \sqrt{\frac{\GammaGen}{H_\text{re}}} \qty(\frac{k}{H_\text{re}})^3 \Theta(H_\text{re} - k) \Theta(k - \sqrt{H_\text{re} \GammaGen})
\end{equation}
agreeing parametrically with \cref{eqn:OmegaGWOscillatingWall} from simple estimate.

In the last stage of the evolution ($t \gtrsim \GammaGen^{-1}$), the string-wall network collapses rapidly and dumps most of all of its remaining energy into gravitational waves as presented in \cref{sec:GWCollapsingWall}. The collapse of the disk is reflected in its average radius that exponentially decreases its size as 
\begin{equation}
	\bar{r}(t) \approx H_\text{re}^{-1} \exp(-\frac{\GammaGen}{2} \qty(t - \GammaGen^{-1})). 
    \label{eqn:rBarCollapsingWall}
\end{equation}
Then, the energy density of the gravitational wave can be evaluated as 
\begin{equation}
    \eval{\pdv{\rho_\text{GW, col.}}{\ln k}}_{t = \GammaGen^{-1}} 
	= \int_{\GammaGen^{-1}}^{\infty} \dd t\; n(\GammaGen^{-1}) \pdv{P(t)}{\ln k}
    = \frac{2\pi \sigma^2}{\MPl^2} \sqrt{\frac{\GammaGen}{H_\text{re}}} \qty(\frac{H_\text{re}}{k})^2 \Theta(k - H_\text{re}),
    \label{appeqn:rhoGWCol.}
\end{equation}
in which the number density is still estimated by \cref{appeqn:nOscillatingWall} similar to the oscillating regime while the power spectrum is given by \cref{eqn:PowerGWCollapsingWall} taking into account the rapid shrinking of the network as described by \cref{eqn:rBarCollapsingWall}. After normalizing with respect to the critical density, this leads to \cref{eqn:OmegaGWCollapsingWall}. 

For the evolution of cosmic belts (see \cref{sec:GWBelts}), we assumed that the energy density of these objects follows the scaling law as shown in \cref{eqn:rho_belts}. Then, for $t = t_\text{prod} \lesssim \GammaGen^{-1}$, the power spectrum roughly follows 
\begin{equation}
    \pdv{P(t_\text{prod})}{\ln k} \sim \frac{\pi w \ell}{\MPl^2} \frac{a(t_\text{obs}) k}{a(t_\text{prod})} \delta(\frac{a(t_\text{obs}) k}{a(t_\text{prod})} - \frac{1}{w}),
\end{equation}
with $w \sim H_\text{re}^{-1}$ and $\ell \sim H^{-1}$. The scaling regime also enforces that one observes $\sim\order{1}$ belt per Hubble volume so that the number density can be estimated as $n \sim H^{3}$. Therefore, using \cref{appeqn:rhoGWOsc}, we obtain that 
\begin{equation}
    \eval{\pdv{\rho_\text{GW, osc., belt}}{\ln k}}_{t = \GammaGen^{-1}} = \frac{2\pi \sigma^2}{\MPl^2} \frac{\GammaGen}{H_\text{re}} \qty(\frac{k}{H_\text{re}})^2 \Theta(H_\text{re} - k) \Theta(k - \sqrt{H_\text{re}\GammaGen}), 
\end{equation}
similar to the estimate provided in \cref{eqn:OmegaGWOscillatingBelts}. Here, one may take $t_\text{prod} \approx H^{-1}$ while letting $t_\text{obs} \approx \GammaGen^{-1}$. Note that at the IR boundary, $k = \sqrt{\GammaGen H_\text{re}}$, this density matches that of a disk since this part of the gravitational wave is generated by cosmic strings that first re-enter the horizon. Here, both belts and disks appear of similar sizes. However, as the belt energy density scales with Hubble following $\propto a^{-4}$, the disks' contribution becomes more important. For $t_\text{prod} \gtrsim \GammaGen^{-1}$, the network rapidly decays into gravitational waves whose power spectrum is 
\begin{equation}
    \pdv{P}{\ln k} \approx \frac{\pi \sigma^2 w(t) \ell}{\MPl^2} k \delta(k - \frac{1}{w(t)}),
\end{equation}
in which $w(t)$ follows \cref{eqn:rBarCollapsingWall} and $\ell \sim \GammaGen^{-1}$ is the Hubble length when belts collapses. Computing an integral similar to \cref{appeqn:rhoGWOsc}, we find 
\begin{equation}
    \eval{\pdv{\rho_\text{GW, col., belt}}{\ln k}}_{t = \GammaGen^{-1}} = \frac{2\pi \GammaGen \sigma^2}{\MPl^2 H_\text{re}} \qty(\frac{H_\text{re}}{k}) \Theta(k - H_\text{re}),
\end{equation}
which agrees with \cref{eqn:OmegaGWCollapsingBelts}.

Lastly, when we consider walls bounded by global strings in \cref{sec:GWGlobalSignature}, we argued that the UV spectrum is suppressed by an extra factor of $\sim k^{-2}$ due to NGB radiation. This can also be understood in the formalism discussed in this appendix. We will illustrate this point by considering circular walls bounded by a global string loop in its final collapse stage $t \gtrsim \GammaNGBWall^{-1}$. We will also assume that $H_\text{re} = k_\text{NGB}$ so that the NGB radiation becomes important immediately after the network collapses. This assumption is not necessary for using this formalism, but it keeps the following computation concise and highlights the salient physics. When $H_\text{re} = k_\text{NGB}$, the evolution of the typical radius of the wall during collapse follows approximately
\begin{equation}
    \dv{(\sigma \bar{r}^2)}{t} \approx - \gamma_a v_2^2 \implies \bar{r}(t) \approx \sqrt{r_0^2 - \frac{\gamma_a v_2^2}{\sigma} \qty(t - t_0)} \approx \frac{1}{H_\text{re}} \sqrt{2 - \GammaNGBWall t},
    \label{appeqn:rbarNGBrad}
\end{equation}
in which we used $\GammaGen \approx \GammaNGBWall \approx t_0^{-1}$. Then, using \cref{appeqn:rhoGWCol.,eqn:PowerGWCollapsingWall}, we find that 
\begin{equation}
    \eval{\pdv{\rho_\text{GW, col., global}}{\ln k}}_{t = \GammaGen^{-1}}
    = \frac{2\pi \sigma^2}{\MPl^2} \sqrt{\frac{\GammaGen}{H_\text{re}}} \qty(\frac{H_\text{re}}{k})^4 \Theta(k - H_\text{re}). 
\end{equation}
The extra $\qty(H_\text{re}/k)^{2}$ factors show how the more rapid decay of the network due to NGB radiation appears in the gravitational-wave spectrum. To relax our assumption $H_\text{re} = k_\text{NGB}$, one simply needs to modify $t_0$ in \cref{appeqn:rbarNGBrad} to the appropriate timescale at which the NGB radiation becomes efficient. Changing $t_0$ shifts where the extra $\sim k^{-2}$ suppression appears on the spectrum, but the extra UV suppression persists as it is essentially the same computation as done above.

%%%%%%%%%%     App C     %%%%%%%%%%
\section{Reproducing GW Spectrum from Gauge Strings \label{app:GWString}} 
In addition to the string-bounded walls, there is a stable gauge string for the model in \cref{eqn:toyModel}. We will leave the following discussion general enough so that not only the well-studied gauge string GW spectrum is reproduced, but one can also apply it to the benchmark model considered in this work. We start by estimating the gravitational-wave signal from individual string loops. Because the gauge string loops tend to decay much later than Hubble time when they are produced, they tend to emit more gravitational radiation at a particular frequency than their long-string counterparts \cite{Allen:1991bk, Vilenkin:2000jqa}. To find the spectrum of an individual string loop of an initial radius $\ell \sim H_p^{-1}$, where $H_p$ is the Hubble scale when the loop is produced, we partition its evolution into an oscillating stage and a decay stage again. To connect with our previous estimates, it is convenient to anchor our spectrum as that observed at the wall decay time $t = \GammaGen^{-1}$. Here, the decay rate to string loop of size $H_p^{-1}$ can be estimated by considering the following damping of gauge string due to gravitational interactions
\begin{equation}
	\dv{(\mu \ell)}{t} \approx P_\text{GW} \sim - \frac{\mu^2}{\MPl^2} \implies \GammaStr \approx \frac{\mu H_p}{\MPl^2},
    \label{appeqn:GammaGWStr}
\end{equation}
in which $\GammaStr$ is the characteristic decay rate at which the string of size $H_p^{-1}$ decays efficiently into gravitational waves. Then, in the oscillating stage, the string loop produces radiations with a frequency $k \sim H_p$, and its fractional energy density at $t = \GammaGen^{-1}$ can be estimated by 
\begin{equation}
    \eval{\pdv{\Omega_\text{GW, loop, osc.}}{\ln k}{\ln H_p}}_{t = \GammaGen^{-1}} \approx \frac{1}{T^4} \qty(\mu H_p^2) \qty(\frac{T}{T_p})^3 \frac{\GammaStr}{H}
    = \sqrt{\frac{\mu}{\MPl^2}} \qty(\frac{a(\GammaGen^{-1})}{a(\GammaStr)} \frac{k}{H_p})^3.
    \label{appeqn:OmegaGWLoopOsc}
\end{equation}
in which we used the adiabatic invariant $k/T_{\GammaGen} = H_p/T$ and $\mu^{1/2} H_p^{1/2} / \qty(\MPl \GammaGen^{1/2}) = a(\GammaGen^{-1}) / a(\GammaStr^{-1})$ during radiation domination.%
\footnote{
Similar to the cosmic disk case, we technically need to impose the $k \gtrsim \sqrt{H_p \GammaGen}$ for the oscillation contribution. However, on a large scale, these string loop with length $\ell \sim H_p^{-1}$ provides a white-noise-like fluctuation. One can see this by evaluating the GW generated by large field fluctuation $\Omega_\text{GW, fluc.} \sim \qty(\delta\rho / \rho_c)^2$ around string re-entry and noticing the parametric agreement between the GW abundance from the oscillating stage and the fluctuating stage. These fluctuations will continue the $\sim k^3$ power-law dependence, analogous to the scaling solution of domain walls. Therefore, the $\sim k^{3}$ spectral shape continues to the deep IR.
} 
In the decay stage, the loop energy, which is proportionate to $\ell \propto k^{-1}$, is quickly dumped into gravitational waves, resulting in a UV spectrum of the form
\begin{equation}
    \eval{\pdv{\Omega_\text{GW, loop, col.}}{\ln k}{\ln H_p}}_{t = \GammaGen^{-1}} \approx \frac{1}{T_{\GammaStr}^4} \qty(\mu H_p^2) \qty(\frac{T_{\GammaStr}}{T_p})^3 \qty(\ell H_p)
    = \sqrt{\frac{\mu}{\MPl^2}} \qty(\frac{a(\GammaGen^{-1})}{a(\GammaStr^{-1})} \frac{k}{H_p})^{-1}.
    \label{appeqn:OmegaGWLoopCol}
\end{equation}
Here, we included the appropriate redshift factor $\ell \approx a(\GammaStr^{-1}) / \qty(k a(\GammaGen^{-1}))$ because we chose to evaluate the GW spectrum around the wall decay time $t = \GammaGen^{-1}$ instead of the string decay time $t = \GammaStr^{-1}$. From this computation, we noticed that the individual string loop has a GW spectrum that peaks at $a(\Gamma_\text{str}^{-1}) H_p / a(\GammaGen^{-1})$ and has a maximal abundance of $\sqrt{\mu / \MPl^2}$. The important feature is that the abundance is mostly independent of $H_p$ while the peak frequency changes with $H_p$.

\begin{figure}[t]
    \centering
    \includegraphics[width=0.6\linewidth]{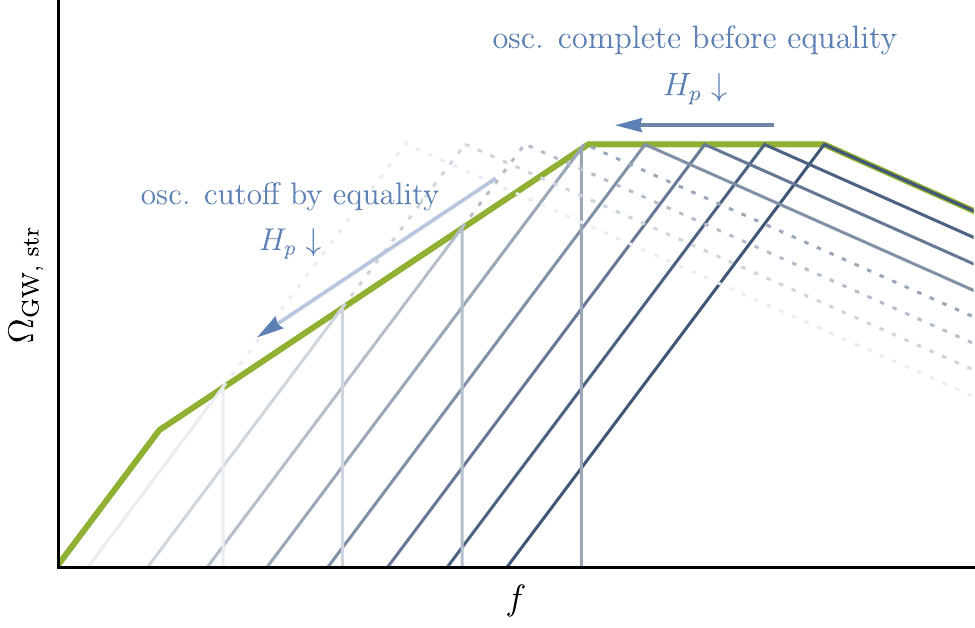}
    \caption{Decomposition of gauge string spectrum: gravitational-wave spectrum from individual string loops of size $\ell \approx H_p^{-1}$ are shown as blue lines. A lighter color indicates a lower $H_p$ or later production of string loop from string reconnection. The envelope (green line) of the individual spectra gives rise to the total GW spectrum of the string network. Some long string loops produced later do not have time to oscillate fully and generate a large enough GW signal around matter-radiation equality. In other words, their decay timescale is longer than that of equality. Thus, their GW spectrum is prematurely terminated around equality, resulting in the $\sim k^{3/2}$ spectrum. Had they oscillated fully, they would generate a GW spectrum with a higher amplitude (dashed line). }
    \label{fig:strDecomp}
\end{figure}

Similar to \cref{sec:GWRingTot}, we will now sum over individual string loop spectrum by finding its envelope as shown in \cref{fig:strDecomp}. At the smallest $H_p$, the IR spectrum looks like $\sim \sqrt{\mu} \MPl^{-1} (k/k_\text{min})^3$; at the largest $H_p$, the UV spectrum looks like $\sim \sqrt{\mu} \MPl^{-1} \MPl (k_\text{max}/k)$. In between, the envelope mainly traces out the peak amplitude, which is mostly flat. However, when $H_p$ becomes more IR, the string may not have time to fully oscillate to its maximum amplitude. Choosing our IR cutoff $H_{p, \text{min}}$ to be the Hubble scale around matter-radiation equality $H_\text{eq}$, the string of size $\ell = H_p^{-1}$ cannot fully oscillate to reach its decay time if 
\begin{equation}
    \eval{\GammaStr}_{H_p} < H_\text{eq} \implies H_p < \frac{\MPl^2}{\mu} H_\text{eq}.
\end{equation}
When the oscillation is incomplete, the $\sim k^{3}$ spectrum is prematurely terminated at $k = H_\text{eq}$. The maximum gravitational-wave abundance generated by these long loops is also suppressed by
\begin{equation}
    \Omega_\text{GW} \sim \frac{1}{T_\text{eq}^4} \qty(\mu H_p^2) \qty(\frac{T_\text{eq}}{T_p})^3 \frac{\GammaStr}{H_\text{eq}}
    = \frac{\sqrt{\mu}}{\MPl} \qty(\frac{\mu}{\MPl^2} \frac{H_p}{H_\text{eq}})^{3/2}.
\end{equation}
More precisely, the spectrum for long loops still follows $\sim k^{3}$ power law and has the form
\begin{equation}
    \eval{\pdv{\Omega_\text{GW}}{\ln k}}_{t = \GammaGen^{-1}} \approx \sqrt{\frac{\mu}{\MPl^2}} \qty(\frac{\mu}{\MPl^2} \frac{H_p}{H_\text{eq}})^{3/2} \qty(\frac{a(\GammaGen^{-1})}{a(t_\text{eq})} \frac{k}{H_p})^{3}, \quad k < \frac{a(t_\text{eq})}{a(\GammaGen^{-1})} H_p.
\end{equation}
This is illustrated as the change in power-law dependence of the envelope (green line) and missing part of the GW spectrum (dashed line) in \cref{fig:strDecomp}. Therefore, integrating over these long loops produces a GW spectrum of the form
\begin{equation}
    \eval{\pdv{\Omega_\text{GW, str}}{\ln k}}_{t = \GammaGen^{-1}} \approx \frac{\sqrt{\mu}}{\MPl} 
    \begin{dcases}
        \qty(\frac{k}{k_\text{min}})^3 \qty(\frac{k_\text{min}}{k_\text{med}})^{3/2}, & k \lesssim k_\text{min}, \\
        \qty(\frac{k}{k_\text{med}})^{3/2}, & k_\text{min} \lesssim k \lesssim k_\text{med}, \\
        1, & k_\text{med} \lesssim k \lesssim k_\text{max}, \\
        \qty(\frac{k_\text{max}}{k}), & k \gtrsim k_\text{max},
    \end{dcases}
\end{equation}
in which we defined 
\begin{equation}
    k_\text{min} \defeq \sqrt{\GammaGen H_\text{eq}}, \quad
    k_\text{med} \defeq \frac{\MPl^2 \GammaGen^{1/2}}{\mu} H_\text{eq}^{1/2}, \quad 
    k_\text{max} \defeq \frac{\MPl \GammaGen^{1/2}}{\mu^{1/2}} H_{p, \text{max}}^{1/2}. 
\end{equation}
Usually, the $\sim k^{3}$ spectral shape is hidden in the deep IR and much below the experimental sensitivity. The remaining $\sim k^{3/2} \to \sim k^0 \to \sim k^{-1}$ power-law dependence agrees with previous studies on GW signal generated by cosmic strings during radiation domination \cite{Blanco-Pillado:2017oxo, Cui:2018rwi, Sousa:2020sxs}.%
\footnote{
Here, our naïve estimate, assuming that cosmic string predominantly radiates in its fundamental mode, produces $\sim k^{-1}$ GW spectrum for strings generated during inflation. This claim agrees with some previous studies~\cite{Guedes:2018afo, Gouttenoire:2019kij}. However, it is currently under debate whether higher-order harmonics of the string may produce a shallower spectrum due to the presence of cusps and kinks~\cite{Cui:2019kkd}.
} 
For the usual gauge string scenario, $H_{p, \text{max}}$ is usually set by either the Hubble scale of the phase transition that produces cosmic strings or the Hubble scale when friction on strings is subdominant and string's scaling regime is reached. For the inflated string-bounded wall discussed in this work, the choice of $H_{p, \text{max}}$ should be the Hubble scale around the wall decay scale $\sim \GammaGen$ because thin strings emerge from reconnecting rings only after all the domain walls have decayed.\footnote{The stable string has a higher winding number in $\phi_2$; therefore, its production during the first phase transition is suppressed.} When one includes both the thin-string mode (\textit{cf.} \cref{sec:GWReconnection}) and the stable string contribution, one may take $H_{p, \text{max}} = H_\text{re}$.

%%%%%%%%%%     App D     %%%%%%%%%%
\section{Remarks on Instantaneous Reheating Assumption \label{app:inefficientReheat}}
In our discussion in \cref{sec:ThermalInflation}, we assumed that the reheating is instantaneous such that a radiation-dominated epoch follows immediately after the inflation that pushes cosmic strings away.
In this appendix, we discuss how non-instantaneous reheating can minimally modify our results and evaluate if the instantaneous reheating for thermal inflation is a reasonable assumption.

Regardless of the details of inflation, the dominant impact of a non-instantaneous reheating is that it introduces a brief period of a matter-domination epoch, which delays the string re-entry. This leads to a smaller re-entry Hubble scale
\begin{equation}
    H_\text{re} \approx e^{-2N_\text{inf}} H_i \qty(\frac{T_\text{R}^4}{\rho_\text{inf}})^{1/6},
\end{equation}
in which $T_\text{R}$ denotes the reheating temperature, which should be smaller than the fourth root of the inflaton energy density $\rho_\text{inf}$. As long as the string re-entry happens after the end of reheating, our estimate in \cref{sec:GWSignature} should hold. This would mean that while keeping the domain wall tension fixed, the gravitational wave will peak at a slightly lower frequency while having a large maximal abundance as shown in \cref{eqn:kPeakT0,eqn:Omegah2GWT0}. This is because a delayed re-entry leads to a larger string radius. This both gives a larger defect network, resulting in a larger gravitational-wave signal and a lower frequency for the normal mode oscillating on the domain wall in its oscillating stage ($H_\text{re}^{-1} < t < \GammaGen^{-1}$).
More specifically, for a second thermal inflation, a non-instantaneous reheating of the thermal inflation can potentially impact our estimate in \cref{sec:ThermalInflation}. In this case, the re-entry Hubble size is modified to
\begin{equation}
    H_\text{re} \approx \frac{m_1^2}{\MPl} \qty(\frac{T_\text{R}^2}{m_1 v_1})^{1/3}
\end{equation}
so that the peak frequency and maximal abundance have a different power-law dependence on model parameters
\begin{equation}
    \eval{f_\text{peak}}_{T_0} \propto m_1^{5/3} \sigma^{-1/2} T_\text{R}^{2/3} v_1^{-1/3}, \quad 
    \eval{\Omega_\text{GW} h^2}_{T_0, f_\text{peak}} \propto m_1^{-5/6} \sigma^{1/2} T_\text{R}^{-1/3} v_1^{1/6}.
\end{equation}

If string re-entry happens during matter domination, the power-law dependence of the frequency for its oscillating phase may be altered due to different redshift factors and could lead to other interesting features on the gravitational-wave spectrum. Nonetheless, as this complication is somewhat tangential to our pursuit in this paper, we leave further discussions on this possibility to a future study.

We now discuss if instantaneous reheating is possible for thermal inflation. Let us consider a coupling of $\phi_1$ to chiral multiplets $\chi$ and $\bar{\chi}$ with superpotential $W= y_\chi \phi_1 \chi \bar{\chi}$. Via a Yukawa-type interaction, $\phi_1$ decays into $\chi \bar{\chi}$ if $y_\chi v_1 < m_1$ with a decay rate $y_\chi^2 m_1/(8\pi) < m_1^3 /(8\pi v_1^2)$. Even if $y_\chi v_1 > m_1$, $\phi_1$ can still decay into light particles that couple to $\chi \bar{\chi}$ with a decay rate $(g_\chi^2/16\pi^2)^2 m_1^3 / (8\pi v_1^2)$, where $g_\chi$ is the coupling of $\chi$ and $\bar{\chi}$ to the light particles. Thus, we may generically parameterize the decay rate as
\begin{align}
    \Gamma_{\phi_1} = \frac{\kappa^2}{8\pi} \frac{m_1^3}{v_1^2},~~\kappa \lesssim 1.
\end{align}
The reheating temperature is 
\begin{equation}
    \frac{\kappa^2}{8\pi} \frac{m_1^3}{v_1^2} \approx \frac{T_\text{R}^2}{\MPl} \implies 
    T_\text{R} \approx \frac{\kappa}{\qty(8\pi)^{1/2}} \frac{m_1^{3/2} \MPl^{1/2}}{v_1}.
\end{equation}
An instantaneous reheating requires $T_\text{R} \gtrsim \sqrt{m_1 v_1}$, which translates to an upper bound on the wall symmetry breaking scale $v_1$,
\begin{equation}
    v_1 \lesssim \SI{E7}{\GeV} \qty(\frac{m_1}{\SI{1}{\TeV}})^{2/3} \qty(\frac{\kappa}{0.1})^{2/3}.
\end{equation}
Recall that the trilinear coupling $\lambda$ is at most $m_1/v_2$ so that the domain wall tension is at most 
\begin{equation}
    \sigma \sim \lambda^{1/2} m_1^{1/2} v_2^{1/2} v_1^2 
    \lesssim m_1 v_1^2;
\end{equation}
thus, to be consistent with an instantaneous reheating, the wall tension scale should be 
\begin{equation}
    \vwall \lesssim \SI{4.6E5}{\GeV}\; \qty(\frac{m_1}{\SI{1}{\TeV}})^{5/9} \qty(\frac{\kappa}{0.1})^{2/9}.
\end{equation}

However, decay of $\phi_1$ is not the most efficient way to terminate this reheating. One may consider dissipation of the inflaton by the scattering with particles in the thermal plasma. In this case, the dissipation rate is given by
\begin{equation}
    \Gamma = \frac{\kappa^2}{8\pi} \frac{T^3}{v_1^2},~~\kappa \lesssim 1.
\end{equation}
The energy dumped into the radiation bath due to scattering of $\phi_1$ within a Hubble time after the end of thermal inflation is given by 
\begin{equation}
    \Delta \rho_\text{rad} \approx \Delta T^4 \approx \Gamma \rho_\phi \Delta t \approx \frac{\kappa^2}{8\pi} \frac{T^3}{v_1^2} \qty(m_1^2 v_1^2) \frac{\MPl}{m_1 v_1} \implies \Delta T \sim \frac{\kappa^2}{32\pi} \frac{m_1 \MPl}{v_1}
\end{equation}
Demanding this process depletes all energy in the inflaton field into radiation implies $\Delta T \gtrsim \sqrt{m_1 v_1}$, and this provides an upper bound on $v_1$
\begin{equation}
    v_1 \lesssim \SI{E11}{\GeV} \; \qty(\frac{m_1}{\SI{1}{\TeV}})^{1/3} \qty(\frac{\kappa}{0.1})^{4/3}.
\end{equation}
This translates to an upper bound on $\vwall$,
\begin{equation}
    \vwall \lesssim \SI{E8}{\GeV} \; \qty(\frac{m_1}{\SI{1}{\TeV}})^{5/9} \qty(\frac{\kappa}{0.1})^{8/9}.
\end{equation}
These constraints are generally satisfied for the parameter space considered in \cref{fig:thermalInflationParam}. Thus, at least for the parameter space of interest, instantaneous reheating is a valid assumption for a second thermal inflation. Also, it is in principle possible to enhance the dissipation rate by the cancellation of the mass of daughter particles between $\phi_1$-dependent contribution and other contributions~\cite{Co:2020jtv}, or by adding many fields.

%%%%%%%%%%     App E     %%%%%%%%%%
\section{Suppressed Coupling between NGB and Domain Wall \label{app:suppressedCoupling}}
In this appendix, we provide a quick parametric estimate for the coupling between the NGB and the domain wall considered in this work in support of our claim made in \cref{sec:GlobalToyExample} that oscillating walls almost do not radiate NGB efficiently. In this section, we will assume, similar to the main text, that $v_2 \gg v_1$. Recall that the Lagrangian considered in our model contains 
\begin{equation}
    \begin{aligned}
        \Lag \supset& \abs{\partial_\mu \phi_1}^2 
        + \qty(\mu_m \phi_1^2 \phi_2^* + \text{h.c.}) - V(\abs{\phi_1}, \abs{\phi_2}) \\
        \to& \qty(1 + \frac{r_1}{v_1})^2 \frac{1}{2} \qty(\partial_\mu a_1)^2 
        + \frac{\mu_m v_1^2 v_2}{\sqrt{2}} \qty(1 + \frac{r_1}{v_1})^2 \qty(1 + \frac{r_2}{v_2}) \cos(\frac{2a_1}{v_1} - \frac{a_2}{v_2}) \\
        \sim& \qty(1 + \frac{2r_1}{v_1}) \frac{1}{2} \qty[\partial_\mu \qty(a_h + \frac{v_1}{2v_2} a_l + \order{v_2^{-2}})]^2 
        + \frac{m_a^2 v_1^2}{4} \qty(1 + \frac{2r_1}{v_1}) \cos(\frac{2a_h}{v_1} + \order{v_2^{-2}}),
    \end{aligned}
\end{equation}
in which we used $\phi_i = \qty(v_i + r_i) \exp(i a_i / v_i) / \sqrt{2}$ and performed a field redefinition 
\begin{equation}
    \begin{dcases}
        a_h = a_1 + \frac{v_1}{2v_2} a_2 + \order{v_2^{-2}}, \\
        a_l = a_2 - \frac{v_1}{2v_2} a_1 + \order{v_2^{-2}}. 
    \end{dcases}
\end{equation}
Here, the field redefinition allows us to identify the heavy angular direction $a_h$ that forms the domain walls and separate it from the massless NGB mode $a_l$. This leads to two interaction terms among $a_l$, $a_h$, and the lightest radial mode $r_1$, 
\begin{equation}
    \Lag \supset \frac{1}{v_2} r_1 \partial_\mu a_h \partial^\mu a_l + \frac{m_a^2 v_1}{2} r_1 \cos(\frac{2a_h}{v_1}). 
\end{equation}
Here, the radial direction $r_1$ typically has a mass $m_1$ that is larger than the angular mode. Then, by integrating out the heavy $r_1$, we obtain an effective Lagrangian that includes interaction between the NGB mode $a_l$ and the heavy domain wall field $a_h$
\begin{equation}
    \Lag[eff] \sim \frac{m_a^2 v_1}{2m_1^2 v_2} \qty(\partial^\mu a_l) \qty(\partial_\mu a_h) \cos(\frac{2a_h}{v_1}) = \frac{2}{m_1^2 v_1 v_2} \qty(\partial^\mu a_l) \qty(\partial_\mu a_h) V(a_h).
\end{equation}
Note that away from the domain wall, both $V(a_h)$ and $\partial_\mu a_h$ terms vanish. Only on domain walls does the heavy angular field $a_h$ have fluctuation that can be coupled to the light axion field $a_l$. Integrating along the direction transverse to the domain wall, we can find that 
\begin{equation}
    \int \dd z\; \partial_z a_h(z) V(a_h(z)) = \frac{m_a^2 v_1^3}{8} = \frac{\sigma^2}{32v_1}.
\end{equation}
The light NGB can then couple to the domain wall by dualizing $\partial_\mu a_l$ to a 3-form field strength $B_{\mu\nu\rho} \propto \epsilon_{\mu\nu\rho\sigma} \partial^\sigma a_l$ that couples to the domain wall worldvolume. Thus, parametrically, we expect the effective action between the domain wall at the light axion is
\begin{equation}
    S_\text{wall} \sim \frac{m_a^2 v_1^2}{m_1^2 v_2} \int \dd \Sigma^{\mu\nu\rho} B_{\mu\nu\rho}.
\end{equation}
This should be compared to the coupling between the cosmic string and NGB mode
\begin{equation}
    S_\text{str} \sim 2\pi v_1 \int \dd \sigma^{\mu\nu} B_{\mu\nu},
\end{equation}
in which $B_{\mu\nu}$ denotes the 2-form potential of the 3-form field strength $B_{\mu\nu\rho}$. Then, one may estimate the power loss due to NGB radiation on a $H_\text{re}^{-1} \times H_\text{re}^{-1}$-sized wall oscillating with $k \approx H_\text{re}$ as 
\begin{equation}
    P_\text{wall} \approx \frac{m_a^4 v_1^4}{m_1^4 v_2^2} \approx \frac{m_a^4}{v_2^2},
\end{equation}
in which we assumed that $m_1 \approx v_1$ in the 2\textsuperscript{nd} equality. This should be contrasted with the power radiated by the boundary string of length $\ell \approx H_\text{re}^{-1}$ and frequency $k \approx H_\text{re}$, which is
\begin{equation}
    P_\text{NGB} \approx \gamma_a v_2^2,
\end{equation}
according to \cref{eqn:PowerNGB}. Indeed, as $m_a \ll v_2$, the power radiated by the wall is highly suppressed. If we take $m_1 \ll v_1$ instead so that thermal inflation can be realized, there is an upper bound on $m_a \sim \sqrt{\lambda m_1 v_2} \lesssim m_1$ so that a large trilinear coupling does not spoil thermal inflation. In this case, the power loss into NGB from the bulk defect is
\begin{equation}
    P_\text{wall} \lesssim \frac{v_1^4}{v_2^2}, 
\end{equation}
which is suppressed compared to the boundary defect. This suppression of direct NGB emission from walls is not too counterintuitive because (1) the wall-producing field is almost orthogonal to the light NGB direction whereas the string consists mainly of the NGB-producing field, and (2) despite the large overall energy of the wall, the wall is locally always lighter than cosmic strings due to $v_1 \ll v_2$.

%%%%%%%%%%  Biblography  %%%%%%%%%%
\bibliographystyle{jhep}
\bibliography{ref_inflated_string_wall}
\end{document}